\documentclass[preprint,3p,12pt]{elsarticle}

\usepackage{mathrsfs}
\usepackage{amsmath}
\usepackage{stmaryrd}
\usepackage{bbding}
\usepackage{dcolumn}
\usepackage{graphicx}
\usepackage{amsfonts}
\usepackage{amssymb}
\usepackage{psfrag}
\usepackage{wrapfig}
\usepackage{subfigure}
\usepackage{makeidx}
\usepackage{bm}
\usepackage{epsf}
\usepackage{epsfig}
\usepackage{setspace}
\usepackage{graphicx}
\usepackage{epstopdf}
\usepackage{psfrag}
\usepackage{subfigure}
\usepackage{color}
\usepackage{comment}
\usepackage{algorithm}
\usepackage{algorithmic}
\usepackage{enumitem}
\newcommand \td {\mathrm{~d}}

\journal{}

\begin{document}

\begin{frontmatter}

  \title{An efficiency and memory-saving programming paradigm for the unified gas-kinetic scheme }
  \author[HKUST1]{Yue Zhang}
  \ead{yzhangnl@connect.ust.hk}
  \author[HKUST1]{Yufeng Wei}
  \ead{yweibe@connect.ust.hk}
  \author[HKUST1]{Wenpei Long}
  \ead{wlongab@connect.ust.hk}
  \author[HKUST1,HKUST2,HKUST3]{Kun Xu\corref{cor1}}
  \ead{makxu@ust.hk}

  \address[HKUST1]{Department of Mathematics, Hong Kong University of Science and Technology, Clear Water Bay, Kowloon, Hong Kong}
  \address[HKUST2]{Department of Mechanical and Aerospace Engineering, Hong Kong University of Science and Technology, Clear Water Bay, Kowloon, Hong Kong}
  \address[HKUST3]{Shenzhen Research Institute, Hong Kong University of Science and Technology, Shenzhen, China}
  \cortext[cor1]{Corresponding author}

  \begin{abstract}
  In recent years, non-equilibrium flows have gained significant attention in aerospace engineering and micro-electro-mechanical systems. The unified gas-kinetic scheme (UGKS) follows the methodology of direct modeling to couple particle collisions and free transport during gas evolution. However, like other discrete-velocity-based methods, the UGKS faces challenges related to high memory requirements and computational costs, such as the possible consumption
   of $1.32$ TB of memory when using $512$ cores for the simulations of the hypersonic flow around an X38-like space vehicle.
  This paper introduces a new UGKS programming paradigm for unstructured grids, focusing on reducing memory usage and improving parallel efficiency. By optimizing the computational sequence, the current method enables each cell in physical space to store only the distribution function for the discretized velocity space, eliminating the need to retain the entire velocity space for slopes and residuals. Additionally, the parallel communication is enhanced through the use of non-blocking MPI.
  Numerical experiments demonstrate that the new strategy in the programming effectively simulates non-equilibrium problems while achieving high computational efficiency and low memory consumption. For the hypersonic flow around an X38-like space vehicle, the simulation, which utilizes $1,058,685$ physical mesh cells and $4,548$ discrete velocity space mesh cells, requires only $168.12$ GB of memory when executed on $512$ CPU cores. This indicates that memory consumption in the UGKS is much reduced.
  This new programming paradigm can serve as a reference for discrete velocity methods for solving kinetic equations.
  \end{abstract}


  \end{frontmatter}



\section{Introduction}
Multiscale flows are commonly encountered in applications of aerospace engineering and micro-electro-mechanical systems (MEMS). For high-speed flying vehicles, the highly compressed gas at the leading edge and the strong expansion wave in the trailing edge can cover the whole flow regimes \cite{bird1994molecular,xu2021unified}. In MEMS, the small size of the structure and the low pressure of the encapsulation result in significant rarefaction effects of gas \cite{senturia1997simulating,wang2022investigation,alexeenko2003numerical,wang2004simulations}. Therefore, developing an efficient and accurate multiscale simulation program is very important.

For multiscale flows across all Knudsen regimes, the description of particle collisions and free streaming are equally important. Accurately capturing non-equilibrium physics requires more degrees of freedom which are provided in the kinetic formulation.
The Boltzmann equation is the fundamental governing equation in rarefied gas dynamics.
Theoretically, it can capture multiscale flow physics in all Knudsen regimes, with the enforcement of resolving the flow physics in the particle mean free path and collision time scale.
Mainstream computational methods of solving non-equilibrium flows can be divided into stochastic and deterministic approaches.
The stochastic methods employ stochastic particles to simulate the statistical behavior of molecular gas dynamics, and the direct simulation Monte Carlo (DSMC) method  \cite{bird1963approach,
bird1998recent,fan2001statistical} is the most popular, which uses probabilistic Monte Carlo simulation to solve the Boltzmann equation.
This method characterizes the non-equilibrium physics through particles within local velocity space, achieving high computational efficiency for hypersonic rarefied flow.
During the gas evolution process, particles' free streaming and collision strictly follow conservation law, guaranteeing high robustness.
However, its stochastic nature of particles  introduce statistical noise.
In low-speed flow simulations, the method requires many particles or multiple statistical averaging to reduce the noise.
Meanwhile, the computational cost of solving intensive partial collisions in the continuum flow regime is very high.
Deterministic methods like the discrete velocity method (DVM) \cite{chu1965kinetic,yang1995rarefied,mieussens2000discrete,tcheremissine2005direct} use the discrete velocity distribution function to solve the Boltzmann equation with different models.
In the deterministic method, the gas evolution flux on the cell interface is often required to be constructed based on the same velocity space.
Therefore, an accurate solution with no statistical noise can be obtained.
However, the global velocity space brings massive memory consumption and computational cost, especially for three-dimensional high-speed flow simulations.
In both stochastic and deterministic methods, when the solution of the kinetic model is split into particle-free streaming and instant collision, numerical dissipation that is proportional to the timestep often becomes inevitable.
Consequently, with the splitting treatment, the mesh size and time step must be smaller than the particle mean free path and mean collision time, respectively.
This ensures that the numerical dissipation in the continuum flow regime does not overshadow the physical dissipation.

In recent years, gas-kinetic methods, such as the DVM-based unified gas-kinetic scheme (UGKS)  \cite{xu2010unified,juan-chen_huang_unified_2012}, the particle-based unified wave-particle (UGKWP) method \cite{liu2020ugkwp,zhu2019ugkwp}, discrete UGKS (DUGKS)  \cite{guo2013dugks} and discrete UGKWP method  \cite{yang2023dugkwp} are proposed. With the coupled particle collisions and free transport in gas evolution, these methods release the constraints on the mesh size and time step for accurate solution in different regime, which is specifically effective in the regime with intensive particle collisions.
These methods have been further developed to consider real gas effects by introducing heat flux modification  \cite{xu_improved_2011}, rotational models  \cite{sha_liu_unified_2014}, and vibrational models  \cite{rui_zhang_unified_2023,zhao_wang_unified_2017}. They have been applied in various systems, including microflow  \cite{huang_unified_2013,liu_unified_2020}, compressible flow \cite{xu_parallel_2022,xu_discrete_2023}, binary gas mixture \cite{zhang_discrete_2019}, radiation transfer  \cite{wenjun_sun_unifed_2019,jiang_song__2021}, plasma  \cite{liu2021unified}, radiative plasma  \cite{quan2024radiative}, neutron transport  \cite{tan_time_2020}, phonon transport  \cite{guo_discrete_2016}, and electron-phonon coupling heat transfer  \cite{zhang2024electron}.
At the same time, the multiscale particle methods have been constructed as well  \cite{fei2020unified,fei2021efficient}.
All these multiscale methods have the unified preserving (UP) property in capturing the Navier-Stokes solution in the continuum regime  \cite{guo2023unified}.

Recently, numerous efforts have been made to enhance DVM-based algorithms, such as UGKS and DUGKS, for industrial applications by accelerating performance and reducing memory consumption. One notable memory reduction technique involves the introduction of an unstructured discrete velocity space (DVS) \cite{yuan2020conservative}, which significantly decreases the velocity space mesh in three-dimensional problems without compromising accuracy. The adaptive UGKS (AUGKS) \cite{xiao2020velocity} employs an adaptive velocity space decomposition to capture the non-equilibrium parts of the flow field, utilizing a discrete distribution function in corresponding areas while applying a continuous distribution function in others. This approach effectively conserves memory in near-continuous flow regimes. The performance of AUGKS in three-dimensional flows with rotational and vibrational non-equilibrium was further improved \cite{wei2024adaptive}.
Adaptive methods such as the adaptive DUGKS \cite{yang2023adaptive} and the adaptive UGKWP \cite{wei2023adaptive} have also been proposed. Regarding acceleration, several studies have integrated implicit methods into UGKS. The implicit UGKS has been developed for both steady \cite{zhu2016implicit} and unsteady \cite{zhu2019implicit} solutions, achieving convergence that is 1 to 2 orders of magnitude faster than explicit UGKS across all flow regimes by combining macroscopic predictions with microscopic implicit iterations. The IUGKS has been extended to three-dimensional thermal non-equilibrium flows, accounting for rotational and vibrational degrees of freedom \cite{zhang2024conservative}. Recently, the IAUGKS \cite{long2024implicit} was introduced, combining these advanced techniques.
In addition, other numerical techniques, such as reduce dimension by axisymmetric \cite{li_unified_2018}, high-order/low-order (HOLO) methods \cite{taitano2014moment,chacon2017multiscale}, memory reduction techniques \cite{chen2017unified,yang2018implicit}, fast evaluation of the full Boltzmann collision term \cite{mouhot2006fast,wu2013deterministic}, and adaptive refinement method \cite{chen2012unified}, can be also used to accelerate the deterministic method.

In the present work, we aim to improve the performance of UGKS by designing a new programming paradigm rather than focusing solely on algorithmic enhancements.
Hybrid parallel algorithms based on MPI/OpenMP have been designed \cite{ho2019multi,jiang2019implicit,baranger2012locally}, where MPI is used for parallelism in physical space and OpenMP in velocity space. However, due to OpenMP's relatively low parallel efficiency and the significant memory consumption in velocity space, it is challenging to apply these methods in large-scale computations. Subsequently, researchers \cite{li2016high,shuang2019parallel,zhang2022unified,wang_parallel_2022,zhang2024efficient} implemented MPI parallelism in both physical and velocity spaces, which reduces the memory requirements for individual cores, allowing the UGKS to tackle larger-scale problems.  Nonetheless, MPI parallelism in velocity space often involves reduction operations, leading to lower parallel efficiency. The main goal of this paper is to design a program framework for the UGKS that requires less memory and achieves higher parallel efficiency. First, the paper analyzes the UGKS to minimize the storage requirements in velocity space. For parallelism, the approach focuses solely on physical space but communicates point by point in velocity space to reduce the amount of data exchanged in each communication step. The framework maintains high parallel efficiency by combining this with non-blocking communication.

The organization of this paper is as follows. The basic model and formulation of the UGKS  are presented in section 2. Section 3 introduces the designation of the code, including memory usage and parallelization of the method. Several test cases, including a three-dimensional cavity and hypersonic flow around a cylinder, a sphere, and an X38-like vehicle, are presented in section 4. In addition, the computation performance is also discussed in this section. Finally, section 5 concludes the paper and outlines potential future work.

\section{Unified Gas-Kinetic Scheme}
\subsection{The BGK-Shakhov model}
The basic model is the BGK-Shakhov equation,
\begin{equation}
 \label{BGK}
 \frac{\partial f}{\partial t} + \boldsymbol{u}\cdot\nabla f =\frac{f^+-f}{\tau},
\end{equation}
 where $f=f(\boldsymbol{x},t,\boldsymbol{u},\boldsymbol{\xi})$ is the distribution function for gas molecules at physical space location $\boldsymbol{x}$ with microscopic translation velocity $\boldsymbol{u}$ and internal velocity $\boldsymbol{\xi}$, $\tau$ is particle collision time, and $f^+$ is the modified equilibrium distribution function.
The modified equilibrium distribution function is given by
\begin{equation*}
 f^+=g\left[ 1+(1-\text{Pr})\boldsymbol{c}\cdot\boldsymbol{q}\left(\frac{c^2}{\text{R}T}-5\right)/(5pRT)\right]=g+g^+,
\end{equation*}
where $g$ is the Maxwellian distribution, $\text{Pr}$ is the Prandtl number, $\boldsymbol{c}=\boldsymbol{u}-\boldsymbol{U}$ is the random velocity, $\boldsymbol{U}$ is the macroscopic velocity, $\boldsymbol{q}$ is the heat flux, $\text{R}$ is gas constant and $T$ is the temperature. The Maxwellian distribution is
\begin{equation*}
 g=\rho {\left(\frac{\lambda}{\pi}\right)}^\frac{K+D}{2}e^{-\lambda ((\boldsymbol{u}-\boldsymbol{U})^2+\boldsymbol{\xi}^2)},
\end{equation*}
where $\rho$ is density, $\lambda=m/(2k_B T)$, $m$ is molecule mass, $k_B$ is Boltzmann constant, $D$ is the spatial dimension, $K$ is the number of internal degrees of freedom and $\boldsymbol{\xi}^2=\xi_1^2+\xi_2^2+\cdots+\xi_K^2$.

The collision terms meet the requirement of conservative constraint or compatibility condition
\begin{equation*}
 \int (f^+-f)\Psi\td \boldsymbol{u}\td\boldsymbol{\xi},
\end{equation*}
where $\Psi=(1,\boldsymbol{u},\frac{1}{2}(\boldsymbol{u}^2+\boldsymbol{\xi}^2))^T$ is the collision invariants and $\td\boldsymbol{\xi}=\td\xi_1\td\xi_2\cdots\td\xi_K$.
The macroscopic variables can be calculated via
\begin{equation*}
 \begin{aligned}
  \boldsymbol{W}&=\left(\begin{matrix}
  \rho \\ \rho \boldsymbol{U} \\\rho E
  \end{matrix}\right)=\int \Psi f\td\boldsymbol{u}\td\boldsymbol{\xi}, \\
  \boldsymbol{q}& = \frac{1}{2}\int(\boldsymbol{u}-\boldsymbol{U}) (|\boldsymbol{u}-\boldsymbol{U}|^2+\boldsymbol{\xi}^2)f\td\boldsymbol{u} \td \boldsymbol{\xi}.
 \end{aligned}
\end{equation*}
\subsection{Reduced distribution function}
To reduce memory usage, we introduce two reduce distribution functions $h$ and $b$, defined by
\begin{equation*}
  \begin{aligned}
 h(\boldsymbol{x},t,\boldsymbol{u})&=\int f\td \boldsymbol{\xi}, \\
 b(\boldsymbol{x},t,\boldsymbol{u})&=\int \boldsymbol{\xi}^2f\td\boldsymbol{\xi},
  \end{aligned}
\end{equation*}
By multiplying Eq.(\ref{BGK}) by $1$ and $\boldsymbol{\xi}^2$ and integrating over the inner degrees, we can obtain
\begin{equation}\label{BGKReduce}
  \begin{aligned}
  \frac{\partial h}{\partial t} + \boldsymbol{u}\cdot\nabla h =\frac{h^+-h}{\tau},\\
  \frac{\partial b}{\partial t} + \boldsymbol{u}\cdot\nabla b =\frac{b^+-b}{\tau},
  \end{aligned}
\end{equation}
where the reduced equilibrium distributions are
\begin{equation*}
  \begin{aligned}
 h^+&=H+H^+,\\
 b^+&=B+B^+.
  \end{aligned}
\end{equation*}
The corresponding reduced Maxwellian distribution $g$ becomes,
\begin{equation*}
  \begin{aligned}
H&=\int g\td\boldsymbol{\xi}=\rho\left(\frac{\lambda}{\pi}\right)^{D/2}e^{-\lambda(\boldsymbol{u}-\boldsymbol{U})^2},\\
B&=\int\boldsymbol{\xi}^2 g\td\boldsymbol{\xi}=\frac{3-D+K}{2\lambda}H,
  \end{aligned}
\end{equation*}
and the corresponding terms related to $g^+$ becomes
\begin{equation*}
  \begin{aligned}
 H^+&=\int g^+\td\boldsymbol{\xi} = \frac{4(1-Pr)\lambda^2}{5\rho}(\boldsymbol{u}-\boldsymbol{U})\cdot \boldsymbol{q}(2\lambda(\boldsymbol{u}-\boldsymbol{U})^2-2-D)H,\\
 B^+&=\int\boldsymbol{\xi}^2 g^+\td\boldsymbol{\xi} = \frac{4(1-Pr)\lambda^2}{5\rho}(\boldsymbol{u}-\boldsymbol{U})\cdot \boldsymbol{q}\{[ 2\lambda(\boldsymbol{u}-\boldsymbol{U})^2-D](3-D+K)-2K\}\frac{H}{2\lambda}.
  \end{aligned}
\end{equation*}
In the following introduction, we use $f$ instead of $h$ and $b$ for simplicity.

\subsection{Finite volume method}
We define a discretization of the whole physical domain $\Omega$ with small cells $\Omega_i$
\begin{equation*}
 \Omega=\cup\Omega_i,\Omega_i\cap\Omega_j=\phi (i\neq j).
\end{equation*}
The whole velocity space is discretized by velocity space cell $\delta \boldsymbol{u}_k$. The integration of a value $Q$ over the whole velocity space can be expressed as
\begin{equation*}
 \int Q \td \boldsymbol{u} = \sum_{k}Q_k\mathcal{V}_k,
\end{equation*}
where $Q_k$ is the average of $Q$ over velocity space cell  $\delta \boldsymbol{u}_k$ and $\mathcal{V}_k=\int_{\delta \boldsymbol{u}_k}\td \boldsymbol{u}$ is the volume of velocity space cell $\delta \boldsymbol{u}_k$.

By taking the integration of Eq.(\ref{BGK}) or Eq.(\ref{BGKReduce}) over velocity space cell $\delta \boldsymbol{u}_k$, we can get the governing equation of distribution function at discretized velocity points
\begin{equation}\label{BGKdis}
    \frac{\partial f_{k}}{\partial t} + \boldsymbol{u}_k\cdot\nabla f_{k} =\frac{f^+_{k}-f_{k}}{\tau},
\end{equation}
where $f_{k}=f_{k}(\boldsymbol{x},t)=f(\boldsymbol{x},t,\boldsymbol{u}_k)$.
By take integration of above equation over physical cell $\Omega_i$ and from time $t^n$ to $t^{n+1}$, we can get
\begin{equation*}
 f_{i,k}^{n+1}=f_{i,k}^{n} -\frac{1}{|\Omega_i|}\sum_{j\in N(i)} S_{ij}\mathcal{F}_{ij,k} + \int_{t^{n}}^{t^{n+1}} \frac{f^+_{i,k}-f_{i,k}}{\tau_i}\td t,
\end{equation*}
where $f_{i,k}^n$ and $f_{i,k}^{n+1}$ are the average of distribution function over cell $\Omega_i$ and $\delta \boldsymbol{u}_k$ at time $t^n$ and $t^{n+1}$, $|\Omega_i|$ denotes the volume of cell $i$, $N(i)$ is the set of all interface-adjcent neighboring cells of cell $i$ and $j$ is one of the neighboring cells of $i$.  The interface between cell $i$ and $j$ is labeled as $ij$, having an area of $S_{ij}$. And the microscopic flux $\mathcal{F}_{ij,k}$ is
\begin{equation}\label{microscopicFlux}
  \mathcal{F}_{ij,k} = u_{k,n}\int_{t^n}^{t^{n+1}} f_{ij,k}\td t,
\end{equation}
where $u_{k,n}=\boldsymbol{u}_k\cdot\boldsymbol{n}_{ij}$ is the normal velocity via surface $ij$ with normal direction $\boldsymbol{n}_{ij}$, and $f_{ij,k}$ is the average distribution fuction over cell $\delta \boldsymbol{u}_k$ at the center of cell interface $ij$.

By taking moments of the above equation and considering the compatibility condition, we can get the governing equation of macroscopic conservative values
\begin{equation}\label{macroscopicUpdata}
  \boldsymbol{W}_i^{n+1}=\boldsymbol{W}_i^n-\frac{1}{\Omega_i}\sum_{j\in N(i)} S_{ij} \boldsymbol{F}_{ij}  ,
\end{equation}
where $\boldsymbol{W}_i^n$ and $\boldsymbol{W}_{i}^{n+1}$ are cell-averaged conservative value at time $t^n$ and $t^{n+1}$, and the macroscopic flux is defined by
\begin{equation}\label{macroscopicFlux}
  \boldsymbol{F}_{ij}=\sum_{k} \mathcal{V}_k\left(\begin{matrix}
    \mathcal{H}_{ij,k} \\ \boldsymbol{u}_k \mathcal{H}_{ij,k} \\\frac{1}{2}( \boldsymbol{u}_k^2\mathcal{H}_{ij,k}+\mathcal{B}_{ij,k})
  \end{matrix}\right),
\end{equation}
where the $ \mathcal{H}_{ij,k}$ and $ \mathcal{B}_{ij,k}$ is defined by
\begin{equation*}
  \begin{aligned}
  \mathcal{H}_{ij,k} &= u_{k,n}\int_{t^n}^{t^{n+1}} h_{ij,k}\td t,\\
  \mathcal{B}_{ij,k} &= u_{k,n}\int_{t^n}^{t^{n+1}} b_{ij,k}\td t.
\end{aligned}
\end{equation*}

\subsection{Gas evolution model}
Along the characteristic line, the integral solution of BGK equation (\ref{BGKdis}) gives
\begin{equation*}
 f_{k}(\boldsymbol{r},t,\boldsymbol{\xi})=\frac{1}{\tau}\int_{t_0}^t f^+_{k}(\boldsymbol{r}^\prime,t^\prime)e^{-(t-t^\prime)/\tau}\td t^\prime + e^{-t/\tau}f_{0,k}(\boldsymbol{r}-\boldsymbol{u}_k t),
\end{equation*}
where $f_{0,k}(\boldsymbol{r})$ is the initial distribution function at the beginning of each step $t_n$, and $f^+(\boldsymbol{r},t)$ is the effective equilibrium state distributed in space and time around $\boldsymbol{r}$ and $t$. The integral solution gives a multiscale modeling of evolution processes from an initial non-equilibrium distribution $f$ to an equilibrium distribution $f^+$ by collision.

To achieve second-order accuracy, the initial distribution function $f_{0,k}(\boldsymbol{r})$ is approximated as
\begin{equation*}
 f_{0,k}(\boldsymbol{r}) = \begin{cases}
 f_{k}^l+\boldsymbol{r}\cdot \nabla f_{k}^l,&u_{k,n}>0,\\
 f_{k}^r+\boldsymbol{r}\cdot \nabla  f_{k}^r,&u_{k,n}<0,\\
 \end{cases}
\end{equation*}
where $f_{k}^l$ and $f_{k}^r$ are the reconstructed initial distribution functions at the left and right sides of the interface. The equilibrium state is approximated as
\begin{equation*}
 f^+_{k}(\boldsymbol{r},t) \approx g_{0,k}+ g^+_{0,k} + \boldsymbol{r}\cdot\nabla g_{0,k}+\frac{\partial g_{0,k}}{\partial t}t.
\end{equation*}
$g_0$ is the Maxwellian distribution function derived from the conservative variables
contributed by all particles transported from both sides of the interface. Details of the distribution functions and derivatives are demonstrated in earlier work  \cite{xu2015direct,xu2010unified}.

In summary, the distribution function at the cell interface is
\begin{equation}\label{fullSolution}
 f_{k}(\boldsymbol{0},t,\boldsymbol{\xi})=\begin{cases}
 c_1f_{k}^l +c_2\boldsymbol{u}_k\cdot\nabla f_{k}^l + c_3(g_{0,k}+g^+_{0,k})+c_4\boldsymbol{u}_k\cdot\nabla g_{0,k}+c_5\partial_t \partial g_{0,k},& u_{k,n}>0, \\
 c_1f_{k}^r +c_2\boldsymbol{u}_k\cdot\nabla f_{k}^r + c_3(g_{0,k}+g^+_{0,k})+c_4\boldsymbol{u}_k\cdot\nabla g_{0,k}+c_5\partial_t \partial g_{0,k},& u_{k,n}<0,
  \end{cases}
\end{equation}
and
\begin{equation*}
  \begin{aligned}
 c_1 &= e^{-t/\tau}, \\
 c_2 &=-t e^{-t/\tau}, \\
 c_3 &=1- e^{-t/\tau}, \\
 c_4 &= te^{-t/\tau}-\tau(1-e^{-t/\tau}), \\
 c_5 &= t-\tau(1-e^{-t/\tau}). \\
  \end{aligned}
\end{equation*}

\subsection{Boundary condition}
This paper adopts ghost cells to construct the boundary fluxes. Two types of boundary conditions are considered. This first type is artificially defined boundaries, such as inlets, outlets, and far fields. The conservation variables at the ghost cells are treated with the same boundary conditions as Euler equations, where the distribution function is set to Maxwellian distribution corresponding to the conservative variables. When calculating the fluxes, the distribution functions at cell interfaces are selected based on the velocity direction.

The second type is real wall boundaries, which are treated with the Maxwellian isothermal boundary conditions. The microscopic distribution function at the wall is defined as
\begin{equation*}
f_{w,k}=\begin{cases}
 f_{in,k}+\boldsymbol{u}_k\cdot\nabla f_{in,k}t,& u_{k,n}>0 \\
 g_w, & u_{k,n}<0
  \end{cases},
\end{equation*}
where $u_{k,n}=\boldsymbol{u}_k\cdot \boldsymbol{n}_w$ is the projection of micro velocity on wall-normal direction $\boldsymbol{n}_w$, the $f_{in,k}$ and $\nabla f_{in,k}$ is obtained by one-sided interpolation from the interior region, and $g_w$ is defined as
\begin{equation*}
 g_w=\rho_w\left(\frac{m}{2\pi kT_w}\right)e^{-\frac{m((\boldsymbol{u}-\boldsymbol{U}_w)^2+\boldsymbol{\xi}^2)}{2kT_w}},
\end{equation*}
where $T_w$ and $\boldsymbol{U}_w$ is given wall temperature and velocity, and $\rho_w$ is calculated by
\begin{equation}
 \label{maxWalldes}
 \sum_{k}\mathcal{V}_k u_{k,n}\int_{t^n}^{t^{n+1}} h_{w,k}\td t=0,
\end{equation}
which means no particle penetrating the wall.

Then, we can obtain the flux of the microscopic distribution function
\begin{equation*}
\mathcal{F}_{w,k}=\begin{cases}
u_{k,n} (f_{in,k}+\boldsymbol{u}_k\cdot\nabla f_{in,k}t),& u_{k,n}>0 \\
 u_{k,n}g_w, & u_{k,n}<0
 \end{cases},
\end{equation*}
and the macroscopic flux $\boldsymbol{F}_w={(F_{w0},F_{w1},F_{w2},F_{w3},F_{w4})}^T$ can be obtained through Eq.(\ref{macroscopicFlux}). And by denoting the surface flux of momentum equation as $\boldsymbol{F}_{mw}=(F_{w1},F_{w2},F_{w3})$, we can obtain surface pressure
\begin{equation*}
 p_w = \boldsymbol{F}_{mw}\cdot \boldsymbol{n}_w,
\end{equation*}
the shear stress
\begin{equation*}
 \sigma_w=|\boldsymbol{F}_{mw}-p_w\boldsymbol{n}_w|,
\end{equation*}
and the heat flux
\begin{equation*}
 q_w=F_{w4}.
\end{equation*}
\subsection{Source term}
The source term can be discretized by the trapezoidal rule,
\begin{equation*}
\int_{t^{n}}^{t^{n+1}} \frac{f^+_{i,k}-f_{i,k}}{\tau_i}\td t = \frac{\Delta t}{2}\left(\frac{f^{+(n+1)}_{i,k}-f^{n+1}_{i,k}}{\tau^{n+1}_i} + \frac{f^{+(n)}_{i,k}-f^n_{i,k}}{\tau^n_i}\right),
\end{equation*}
where the $f^{+(n+1)}$ and $\tau^{n+1}$ can be obtained through the macroscopic conservative values $\boldsymbol{W}^{n+1}$. Considering the microscopic govern equation, the distribution function can be updated by
\begin{equation}\label{microscopicUpdata}
 f_{i,k}^{n+1}= \left(1+\frac{\Delta t}{2\tau^{n+1}_i}\right)^{-1}\left[  f_{i,k}^{n} -\frac{1}{\Omega_i}\sum_{j\in N(i)} S_{ij}\mathcal{F}_{ij,k} + \frac{\Delta t}{2}\left(\frac{f^{+(n+1)}_{i,k}}{\tau^{n+1}_i} + \frac{f^{+(n)}_{i,k}-f^n_{i,k}}{\tau^n_i}\right)\right].
\end{equation}

\section{Algorithm}

\subsection{Basic UGKS algorithm}
The algorithm of the UGKS is summarized as follows:
\begin{enumerate}[label=\textbf{Step \arabic*}]
  \item Calculate time step and heat flux.
  \item Reconstruction. The constraint least-square reconstruction and gradients compressor factors \cite{zhang2023slidingmesh} are used to get the gradients of macroscopic values, and the constraint least-square reconstruction and Venkatakrishnan limiter are used to get the gradients of microscopic values.
 \item Evaluate flux. Calculate the microscopic flux by Eq.(\ref{microscopicFlux}) and Eq.(\ref{fullSolution}), and get the corresponding macroscopic flux through Eq.(\ref{macroscopicFlux}).
 \item Update flow fields. Apply Eq.(\ref{macroscopicUpdata}) to obtain the macroscopic quantities of the full field at the $n+1$ step. Then calculate Eq.(\ref{microscopicUpdata}) to update the microscopic distribution function.
\end{enumerate}
\subsection{Two stage update of microscopic distribution function}
Due to the original algorithm, we need to store all the gradients and spatial residuals of the distribution functions, which requires a significant amount of memory because of the large number of velocity space meshes. To address this, we can consider updating the distribution functions one by one, thus eliminating the need to store the gradients of all distribution functions.

However, according to Eq.(\ref{microscopicUpdata}), the update of the microscopic distribution functions requires the macroscopic values at time $t^{n+1}$ through $f^{+(n+1)}$. Therefore, we must store the spatial residuals and distribution functions at time $t^n$ to perform the updates. We propose splitting the update process into two steps. The first step is given by

\begin{equation}\label{microS1}
  \tilde{f}_{i,k}^{n+1} = f_{i,k}^{n} - \frac{1}{\Omega_i}\sum_{j\in N(i)} S_{ij}\mathcal{F}_{ij,k} + \frac{\Delta t}{2} \frac{f^{+(n)}_{i,k} - f^n_{i,k}}{\tau^n_i}.
\end{equation}
The second step is defined as
\begin{equation}\label{microS2}
 f_{i,k}^{n+1} = \left(1 + \frac{\Delta t}{2\tau^{n+1}_i}\right)^{-1}\left(\tilde{f}_{i,k}^{n+1} + \frac{\Delta t}{2}\frac{f^{+(n+1)}_{i,k}}{\tau^{n+1}_i}\right).
\end{equation}
In this approach, we can update the distribution function one by one in velocity space using Eq.(\ref{microS1}), which only requires storing the gradients and residuals for one distribution function. Furthermore, after obtaining $\tilde{f}_{i,k}^{n+1}$, the variable $f_{i,k}^n$ is no longer needed, allowing us to reuse the same block of memory to store $\tilde{f}_{i,k}^{n+1}$.

Additionally, several terms need further clarification.
\begin{itemize}
 \item In the calculation of the Shakhov model, the heat flux is needed, which needs the momentum of distribution at the cell interface. However, we use the averaged heat flux of left and right cells for simplicity.
 \item In the calculation of microscopic flux, the equilibrium part and time coefficients only depend on macroscopic values, so we can calculate it first and store it at the cell interface.
 \item As shown in Eq.(\ref{maxWalldes}), the Maxwellian wall boundary condition needs the whole velocity space values to get density, so we need to store the distribution functions at the wall and evolute flux individually.
\end{itemize}
As a result, the UGKS algorithm has been shown in Algorithm \ref{alg1}.

\begin{algorithm}
 \caption{Algorithm for Time Step Calculation and Flux Evolution}
 \label{alg1}
 \begin{algorithmic}[1]
   \STATE Calculate the time step $\Delta t$ based on the CFL condition and compute heat flux for each cell.
   \STATE Reconstruct the macroscopic values.
   \STATE Calculate the equilibrium state and time coefficients for cell interfaces.
   \FOR{$k=1$ to $k_{\text{max}}$}
     \STATE Reconstruct the microscopic values for $f_{k}$.
     \STATE Evolve the flux $\mathcal{F}_{k}$ using Eq.(\ref{microscopicFlux}) and calculate their contribution to the macroscopic flux using Eq.(\ref{macroscopicFlux}).
     \STATE Interpolate $f_{k}$ and their gradients at the boundary.
     \STATE Update $f_{k}^n$ to $\tilde{f}_{k}^{n+1}$ using Eq.(\ref{microS1}).
   \ENDFOR
   \STATE Evolve the flux at the boundary.
   \STATE Update macroscopic values using Eq.(\ref{macroscopicUpdata}).
   \STATE Update microscopic values using Eq.(\ref{microS2}).
 \end{algorithmic}
\end{algorithm}
\subsection{Parallelization of the UGKS}

In this section, we present the parallelization of our code based on the Message Passing Interface (MPI). First, we need to identify the messages that must be exchanged:

 \begin{itemize}
  \item The macroscopic reconstruction requires the macroscopic values of adjacent cells.
     \item The calculation of the equilibrium state and time coefficients necessitates the macroscopic values, their gradients, and the heat flux of adjacent cells.
     \item The microscopic reconstruction requires the microscopic distribution functions of adjacent cells.
     \item The flux evolution requires the microscopic distribution functions and their gradients of adjacent cells.
 \end{itemize}

\begin{figure}[!h]
  \centering
  \includegraphics[width=6cm]{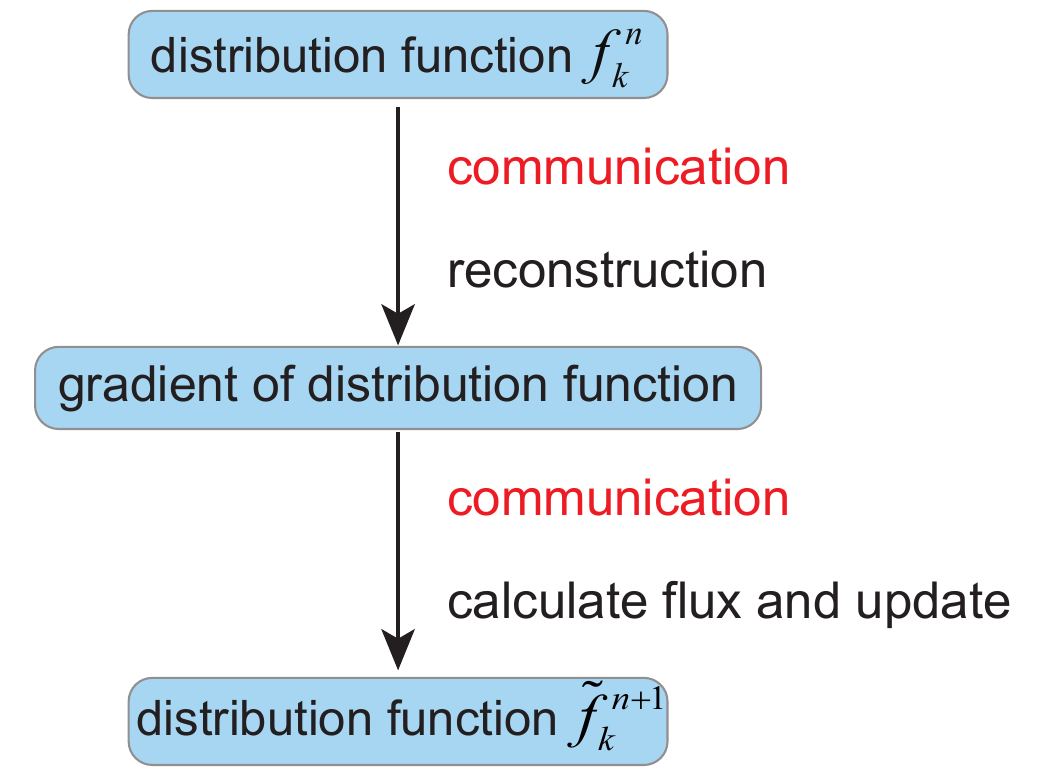}
  \caption{The update of the microscopic distribution function $f_k^n$ to $\tilde{f}_k^{n+1}$.}\label{updatef}
\end{figure}

Compared with the macroscopic values, the microscopic values need more times of exchanges because of the huge number of velocity space mesh cells.  The evolution of the microscopic distribution function from $f_k^n$ to $\tilde{f}_k^{n+1}$ is illustrated in Figure \ref{updatef}.
It is important to note that we need to perform two message exchanges at each microscopic speed point. We must first exchange the microscopic distribution functions and then exchange their gradients, as the gradients depend on the microscopic distribution functions. However, the microscopic distribution functions at different microscopic velocity points are independent. Therefore, we can combine the two communications by reconstructing and evolving different velocity points within the same loop.
\begin{figure}[!h]
  \centering
  \includegraphics[width=12cm]{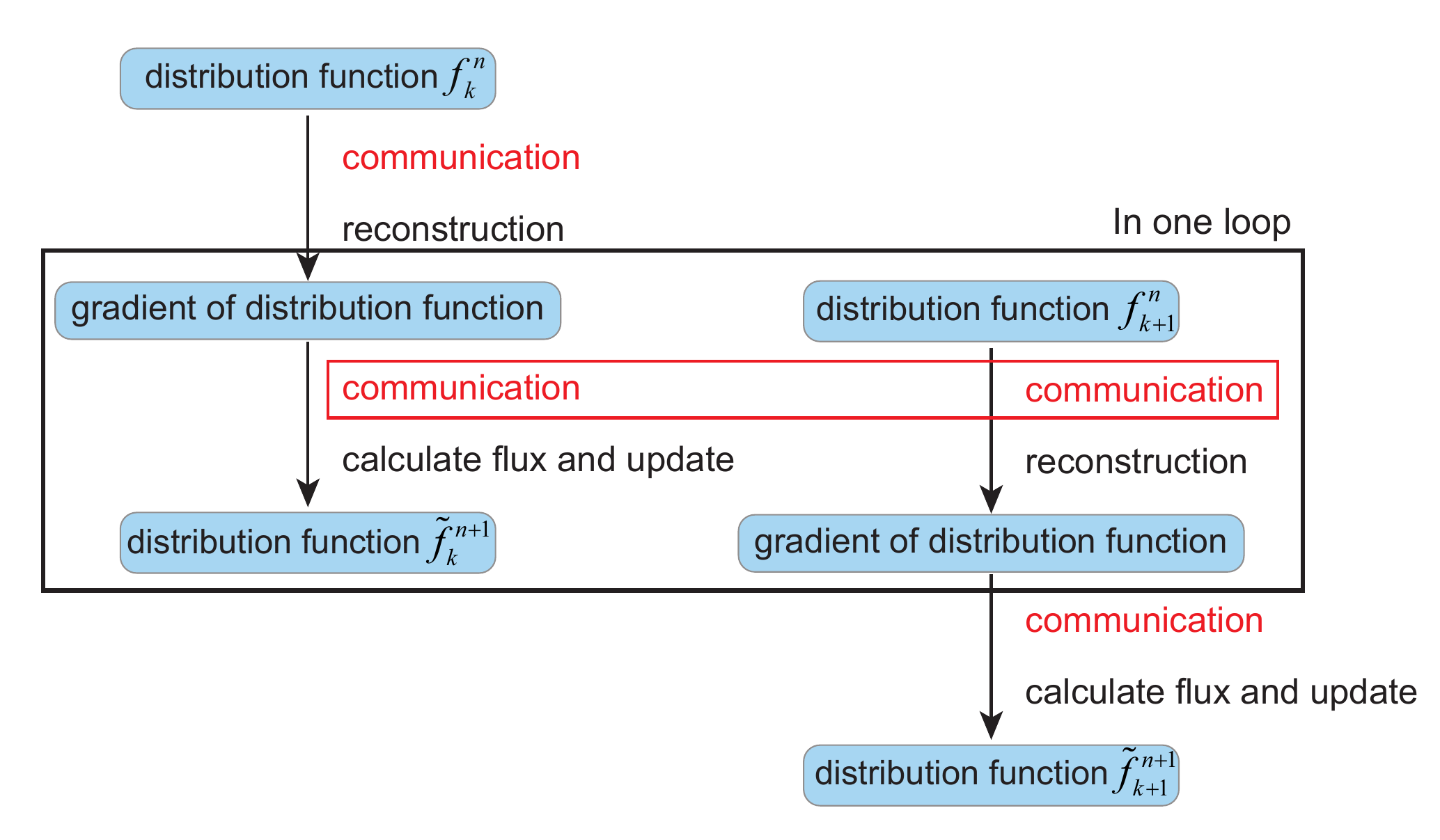}
  \caption{Combining communications by reconstructing and evolving different velocity points within the same loop.}\label{updatefmpi}
\end{figure}
As shown in Figure \ref{updatefmpi}, we perform the evolution of $f_k$ and the reconstruction of $f_{k+1}$ within a single loop.

\begin{figure}[!h]
 \centering
 \includegraphics[width=8cm]{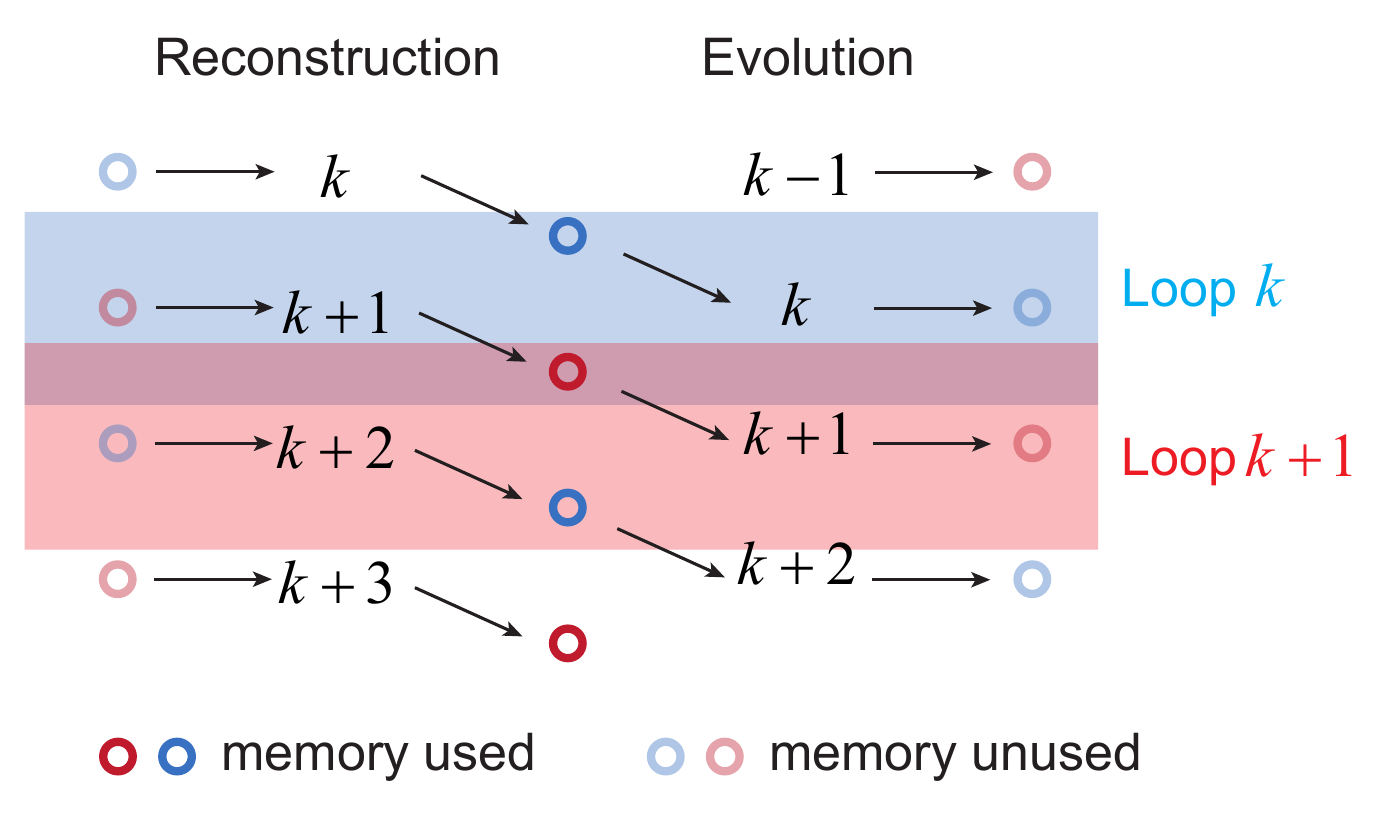}
 \caption{Memory usage for the distribution function gradient.}\label{gradientmemory}
\end{figure}

Next, we consider memory usage. The entire space of the distribution function must be stored. However, the gradient of the distribution function does not need to be stored once the spatial residuals are obtained. In the new algorithm designed for MPI communication, we require two blocks of memory to store the gradients. As illustrated in Figure \ref{gradientmemory}, we only need to store two gradients for each loop.
For example, in loop $k$, we only need to store the gradients $\nabla f_k^n$ and $\nabla f_{k+1}^n$. In the subsequent loop $k+1$, the gradient $\nabla f_k^n$ is no longer needed, allowing that block of memory to be repurposed for storing $\nabla f_{k+2}^n$.

\begin{figure}[!h]
 \centering
 \includegraphics[width=15cm]{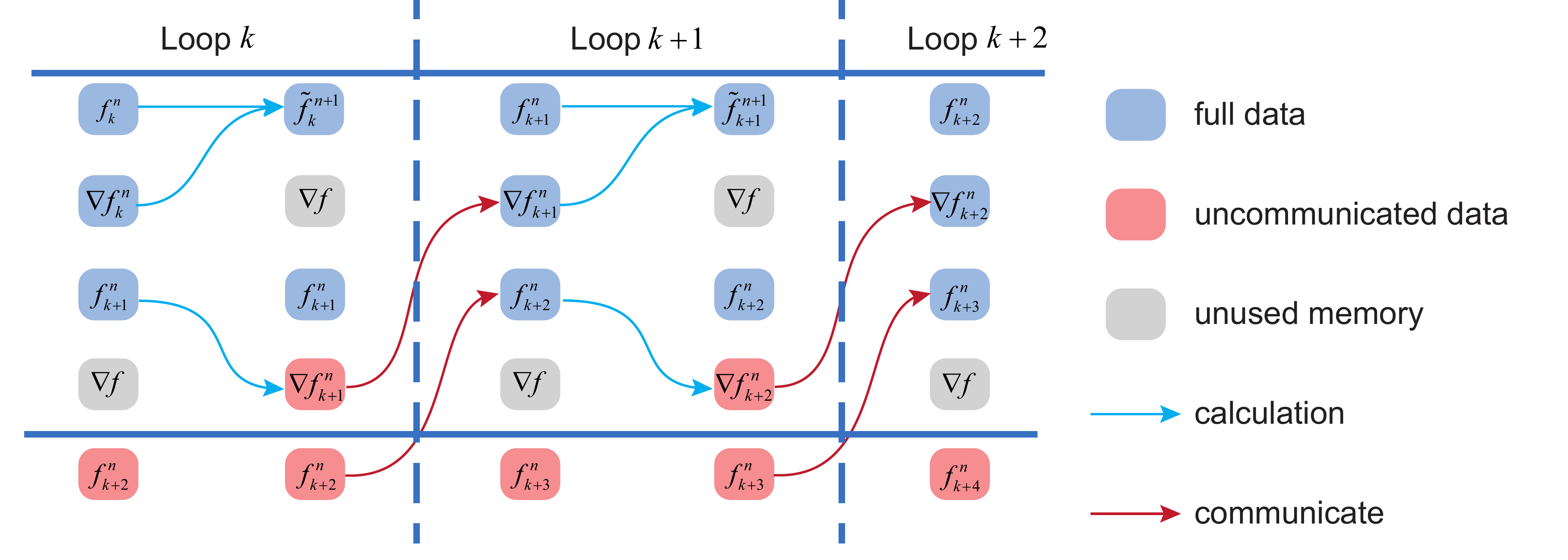}
 \caption[]{The summary of the calculation processes, message communication and memory usage of the upadtion from $f_k^n$ to $\tilde{f}_k^{n+1}$.}\label{updatesummary}
\end{figure}

In summary, we illustrate the calculation processes, message communication, and memory usage in Figure \ref{updatesummary}.
Non-blocking communication is also used in this process. Considering other parts of the algorithm, the whole parallelized algorithm is shown as as Algorithm \ref{alg2}.
\begin{algorithm}[!h]
 \caption{Parallel Algorithm for the UGKS}
 \label{alg2}
 \begin{algorithmic}[1]
  \STATE Calculate the time step $\Delta t$ based on the CFL condition and compute heat flux for each cell.
  \STATE \textbf{Send macroscopic values, heat flux, and microscopic distribution function $f_1$.}
  \STATE \textbf{Receive macroscopic values, heat flux, and microscopic distribution function $f_1$.}
  \STATE Reconstruct macroscopic values.
  \STATE Reconstruct microscopic values for $f_1$ and store their gradients as $\nabla f_1$.
  \STATE \textbf{Send gradients of macroscopic values, microscopic distribution function $f_2$, and gradients of microscopic values $\nabla f_{1}$.}

  \FOR{$k=1$ to $k_{\text{max}}$}
    \STATE Let $\& \nabla f_{f} \gets (k \mod 2 == 1) \, ? \, \nabla f_{1} \, : \, \nabla f_{2}$ and $\&\nabla f_{r} \gets (k \mod 2 == 1) \, ? \, \nabla f_{2} \, : \, \nabla f_{1}$.

    \IF {$k==1$}
      \STATE \textbf{Receive gradients of macroscopic values, distribution function $f_{2}$, and gradients of microscopic $\nabla f_{f}$.}
      \STATE Calculate equilibrium state and time coefficients for cell interfaces.
    \ELSE
      \STATE \textbf{Receive distribution function $f_{k+1}$ and gradients of microscopic $\nabla f_{f}$.}
    \ENDIF

    \STATE Reconstruct microscopic values for $f_{k+1}$ and store their gradients as $\nabla f_{r}$.
    \STATE \textbf{Send microscopic distribution function $f_{k+2}$ and gradients of microscopic values $\nabla f_{r}$.}

    \STATE Evolve flux $\mathcal{F}_{k}$ with $f_{k}$ and $\nabla f_{f}$ using Eq.(\ref{microscopicFlux}), and their contribution to macroscopic flux using Eq.(\ref{macroscopicFlux}).
    \STATE Interpolate $f_{k}$ and their gradients into the boundary.
    \STATE Update $f_{k}^n$ to $\tilde{f}_{k}^{n+1}$ using Eq.(\ref{microS1}).
  \ENDFOR

  \STATE Evolve flux at the boundary.
  \STATE Update macroscopic values using Eq.(\ref{macroscopicUpdata}).
  \STATE Update microscopic values using Eq.(\ref{microS2}).
 \end{algorithmic}
\end{algorithm}

\section{Numerical example}
\subsection{Hypersonic flow around a circular cylinder}
\begin{figure}[!h]
 \centering
 \includegraphics[width=8cm]{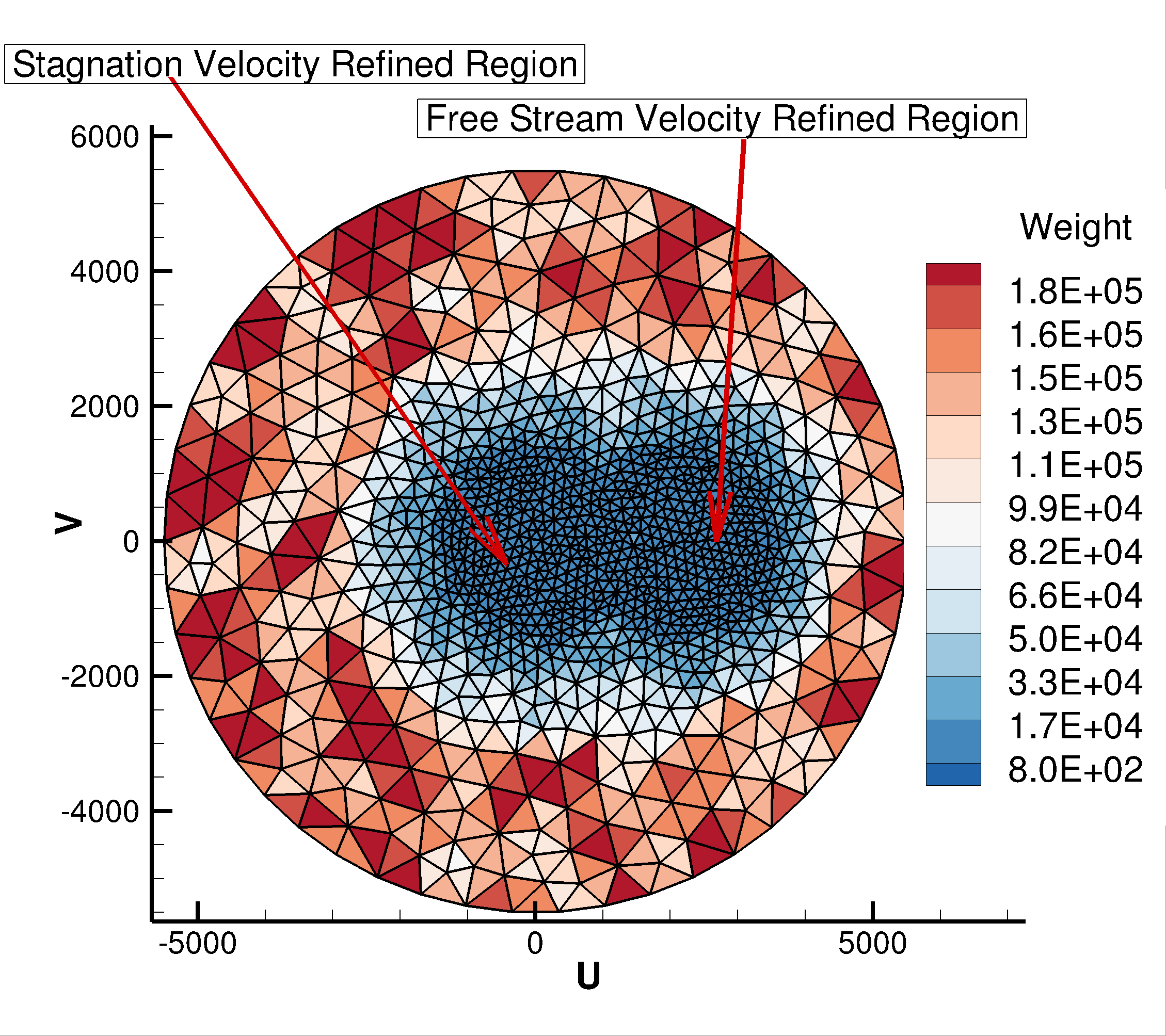}
 \caption[]{Unstructured DVS mesh with 2,060 cells used for hypersonic flow at $\text{Kn}_\infty=0.1$ and $\text{Ma}_\infty=5$ passing over a cylinder by the current UGKS program.}\label{cylinderMa5DVS}
\end{figure}

Hypersonic gas flow of argon with Prandtl number $\text{Pr}=2/3$ around a circular cylinder has been simulated at $\text{Ma}_\infty=5$ and $\text{Kn}_\infty=0.1$. The characteristic length used to define the Knudsen number is the cylinder diameter $D=2$ m. The temperature of free stream flow is $T_\infty=273$ K, and an isothermal wall with a fixed temperature of $T_w=273 \text{K}$ is applied for the cylinder surface. The physical domain is discretized by 12,600 ($180\times70$) quadrilateral cells with the height of first layer mesh $h=0.001$ m, while the unstructured discrete velocity space (DVS) mesh consists of 2,060 cells as shown in Figure \ref{cylinderMa5DVS}. The DVS is discretized in a circle region with the center $0.4\times(U_\infty,0)$ with a total radius of $6\sqrt{RT_s}$ where $T_s$ is the stagnation temperature of the free stream flow. The unstructured DVS mesh is refined at zero velocity point with a radius of $3\sqrt{RT_w}$ and the free stream point with a radius of $3\sqrt{RT_\infty}$.

\begin{figure}[!h]
 \centering
 \subfigure[Mach number]{\includegraphics[width=8cm]{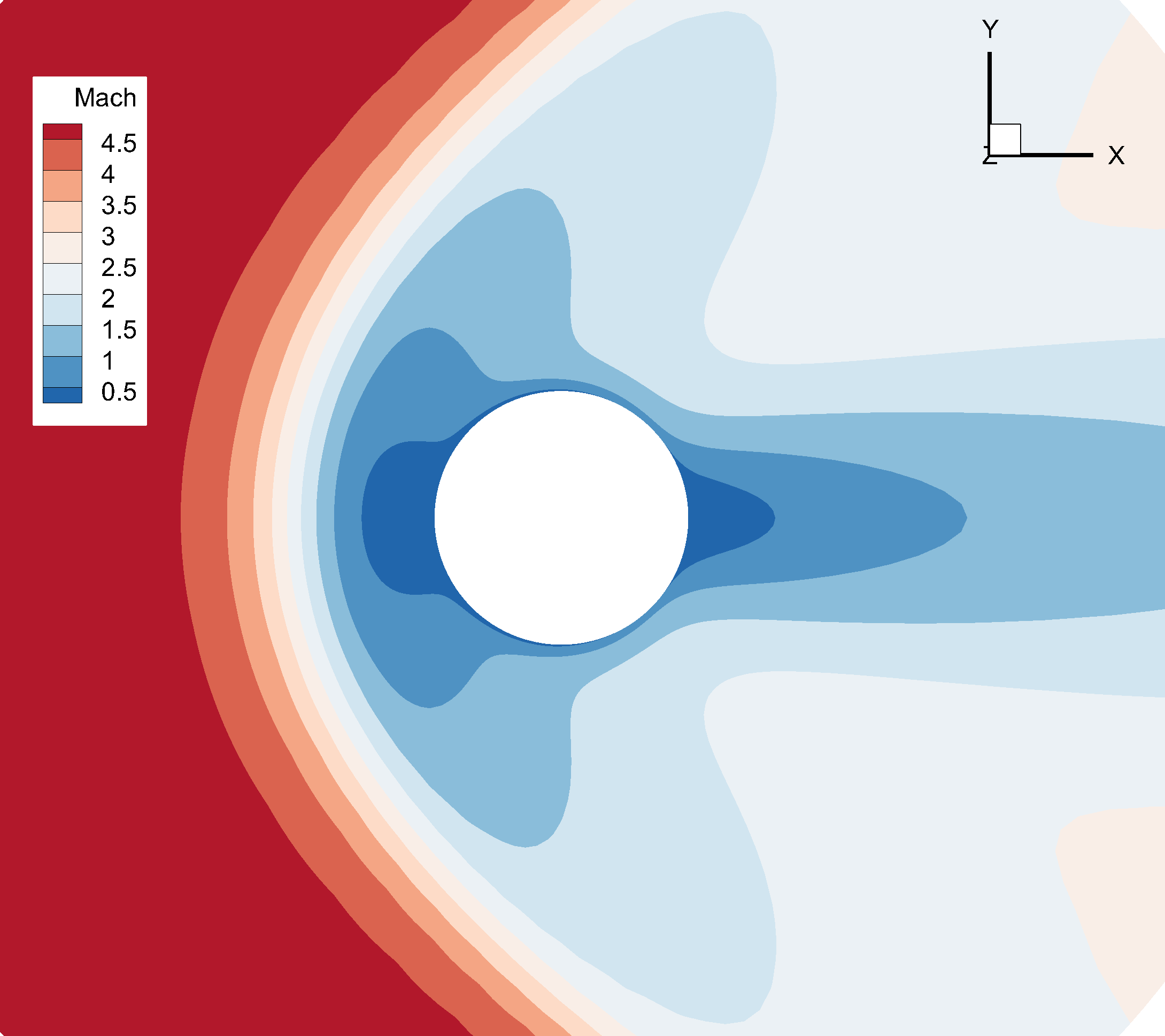}}
 \subfigure[Temperature]{\includegraphics[width=8cm]{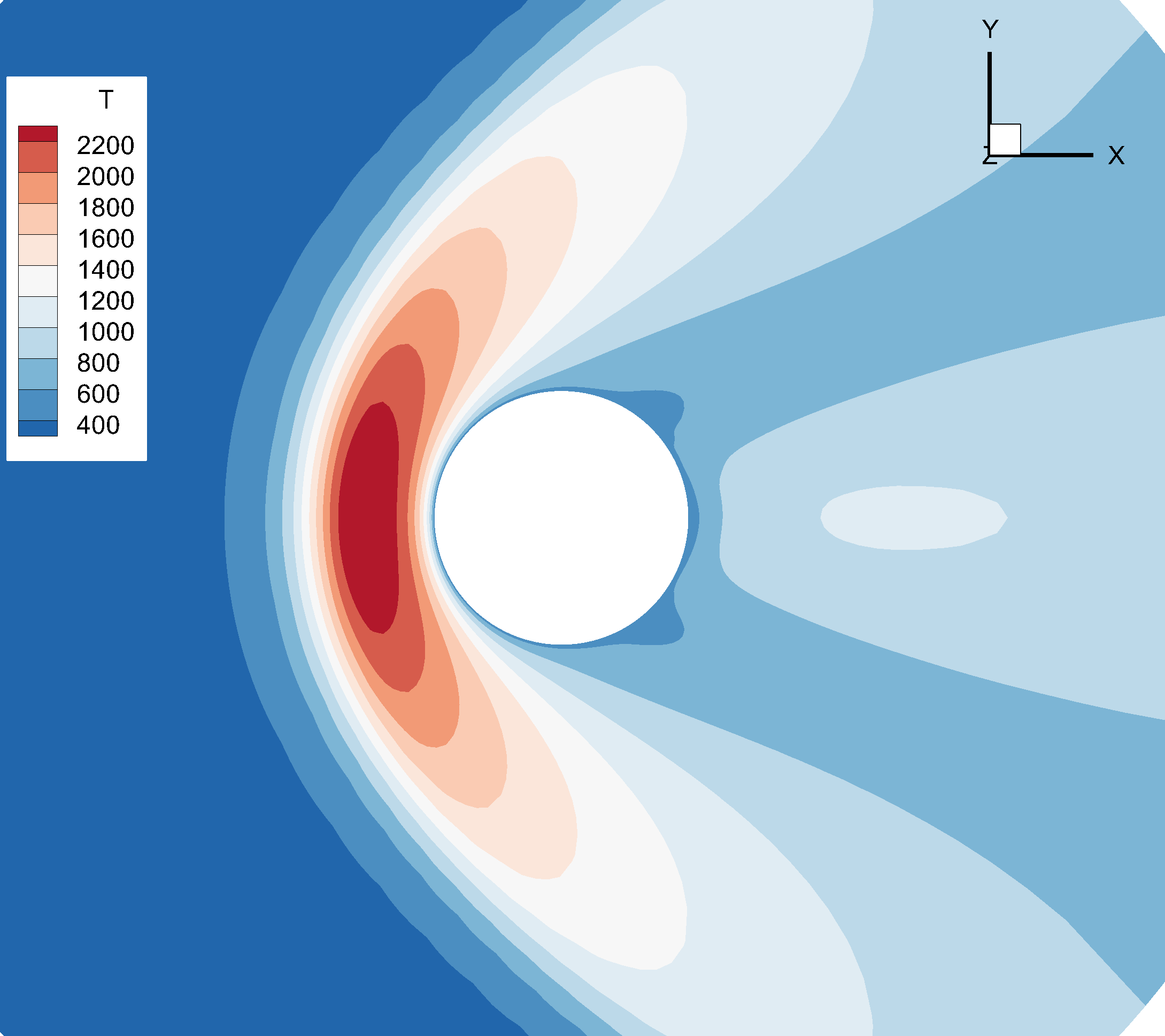}}
\caption[]{Contours of hypersonic flow at $\text{Kn}_\infty=0.1$ and $\text{Ma}_\infty=5$ passing over a circular cylinder by the current UGKS program.}\label{cylinderMa5contor}
\end{figure}

\begin{figure}[!h]
 \centering
 \subfigure[Pressure coefficient]{\includegraphics[width=5cm]{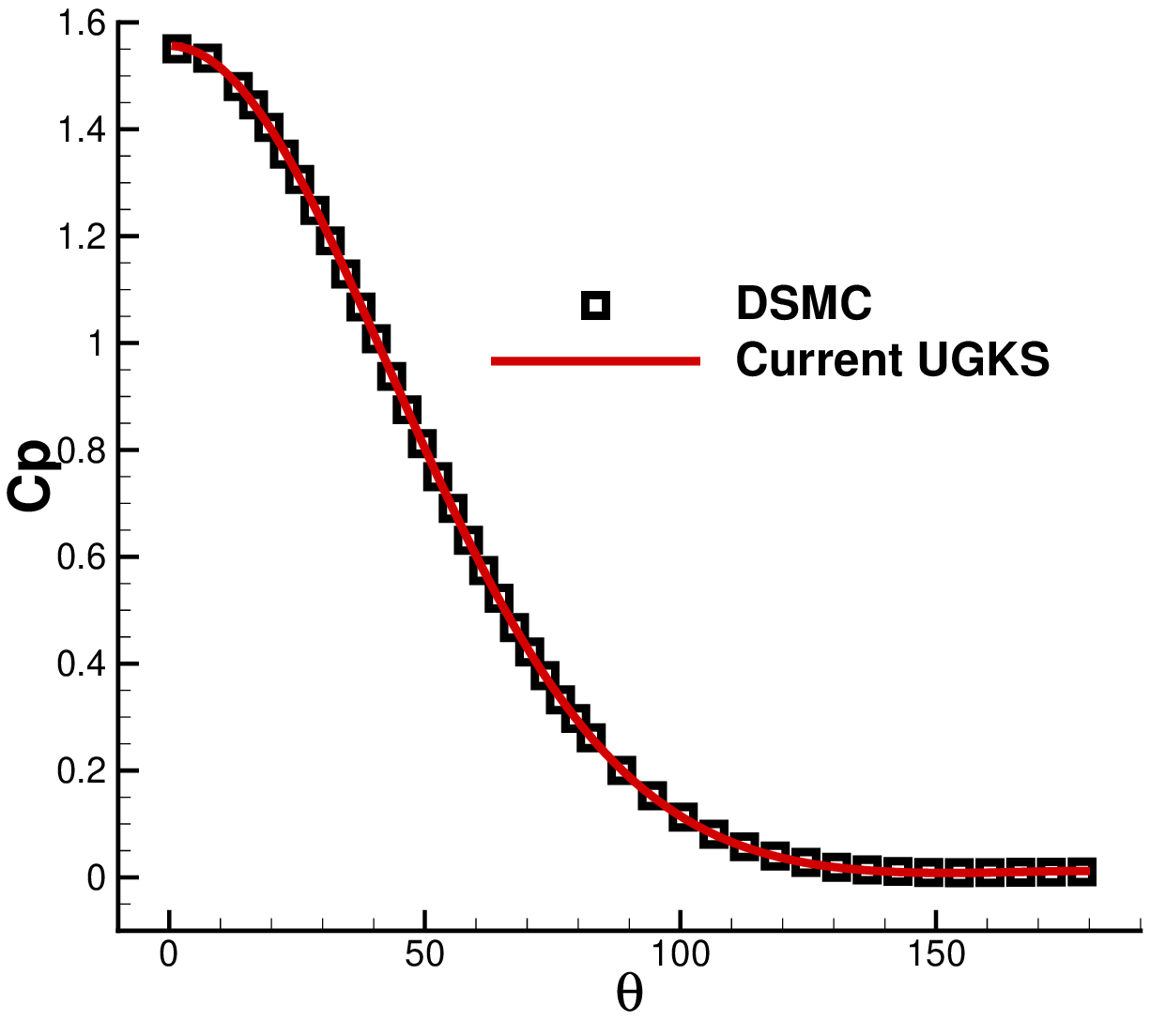}}
 \subfigure[Shear stress coefficient]{\includegraphics[width=5cm]{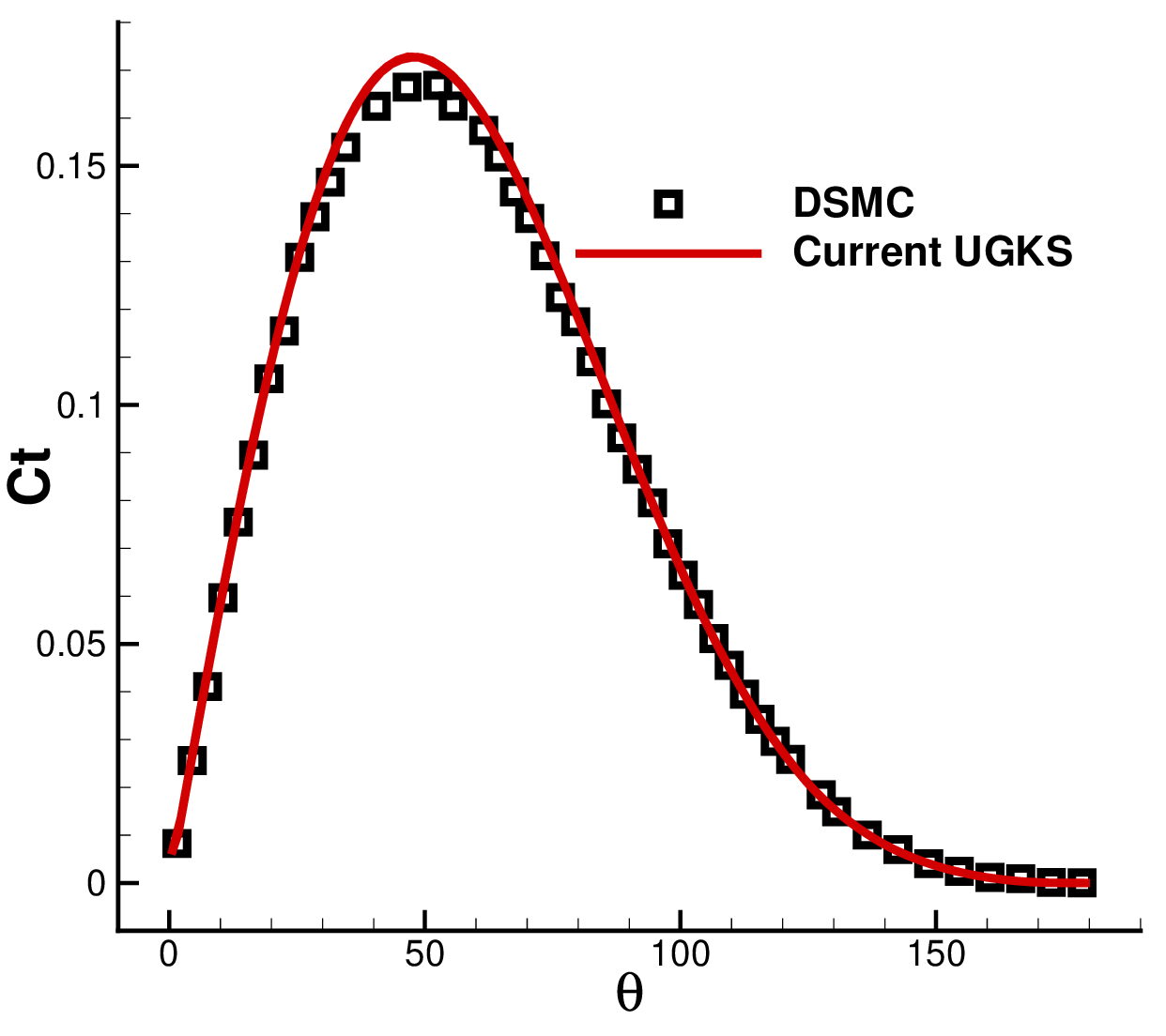}}
 \subfigure[Heat flux coefficient]{\includegraphics[width=5cm]{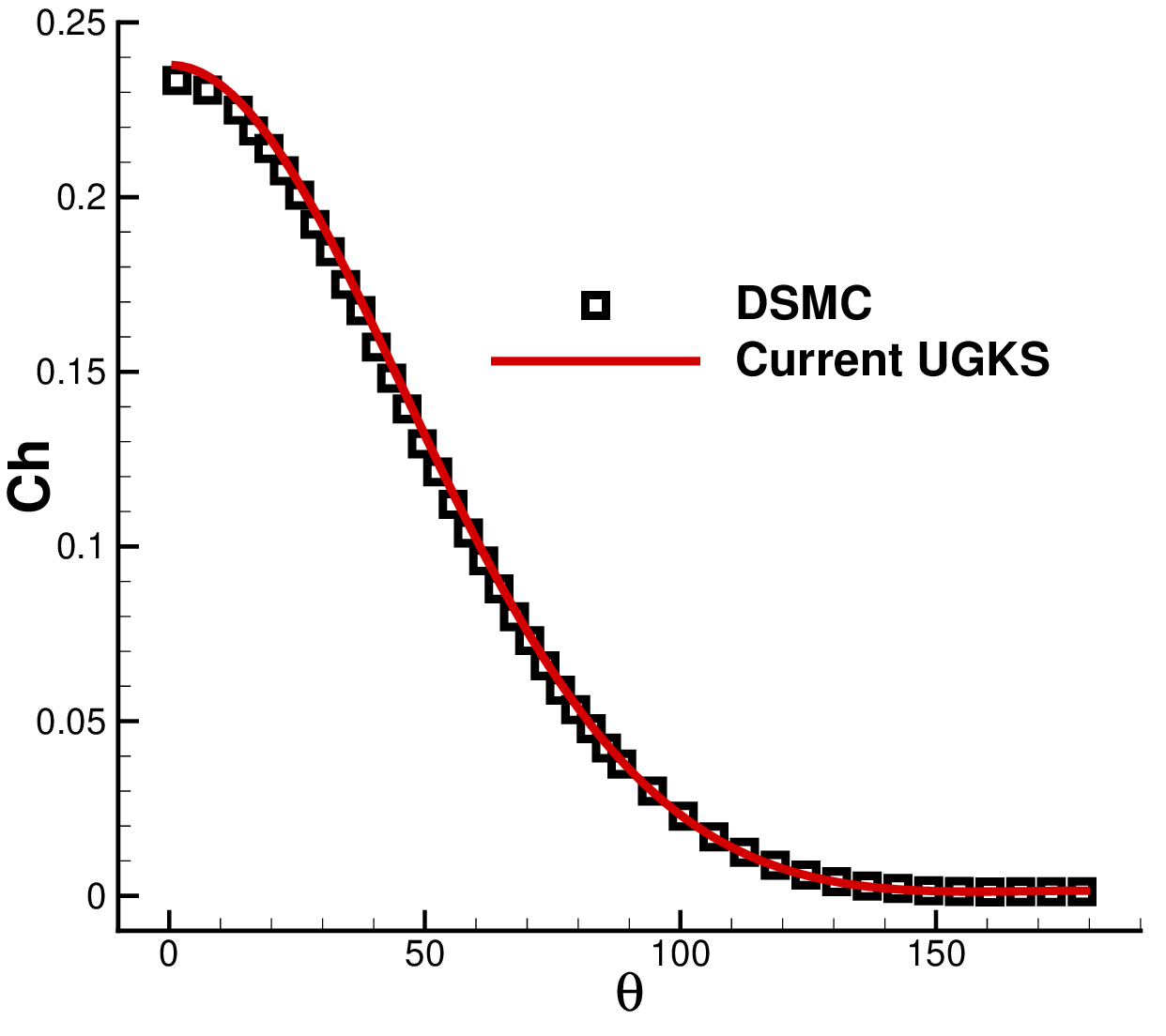}}
 \caption[]{Surface quantities of hypersonic flow at $\text{Kn}_\infty=0.1$ and $\text{Ma}_\infty=5$ passing over a circular cylinder by the current UGKS program compared with the DSMC method  \cite{zhang2024conservative}.}\label{cylinderMa5surface}
\end{figure}

The contours of Mach number and temperature are plotted in Figure \ref{cylinderMa5contor}.
To verify the result quantificationally, the non-dimensionalized surface coefficients (the pressure coefficient $C_p$, the shear stress coefficient $C_t$, and the heat flux coefficient $C_h$) are defined by
\begin{equation*}
 C_p=\frac{p_s}{\frac{1}{2}\rho_\infty U_\infty^2},
 C_t=\frac{f_s}{\frac{1}{2}\rho_\infty U_\infty^2},
 C_h=\frac{h_s}{\frac{1}{2}\rho_\infty U_\infty^3},
\end{equation*}
where $U_\infty$ is the incoming flow speed calculated by free-stream Mach number $\text{Ma}_\infty$, $p_s$ is the surface pressure, $f_s$ is the surface friction and $h_s$ is the surface heat flux. As shown in Figure \ref{cylinderMa5surface}, our results agree well with the results of the DSMC method  \cite{zhang2024conservative}.

\begin{figure}[!h]
 \centering
 \includegraphics[width=8cm]{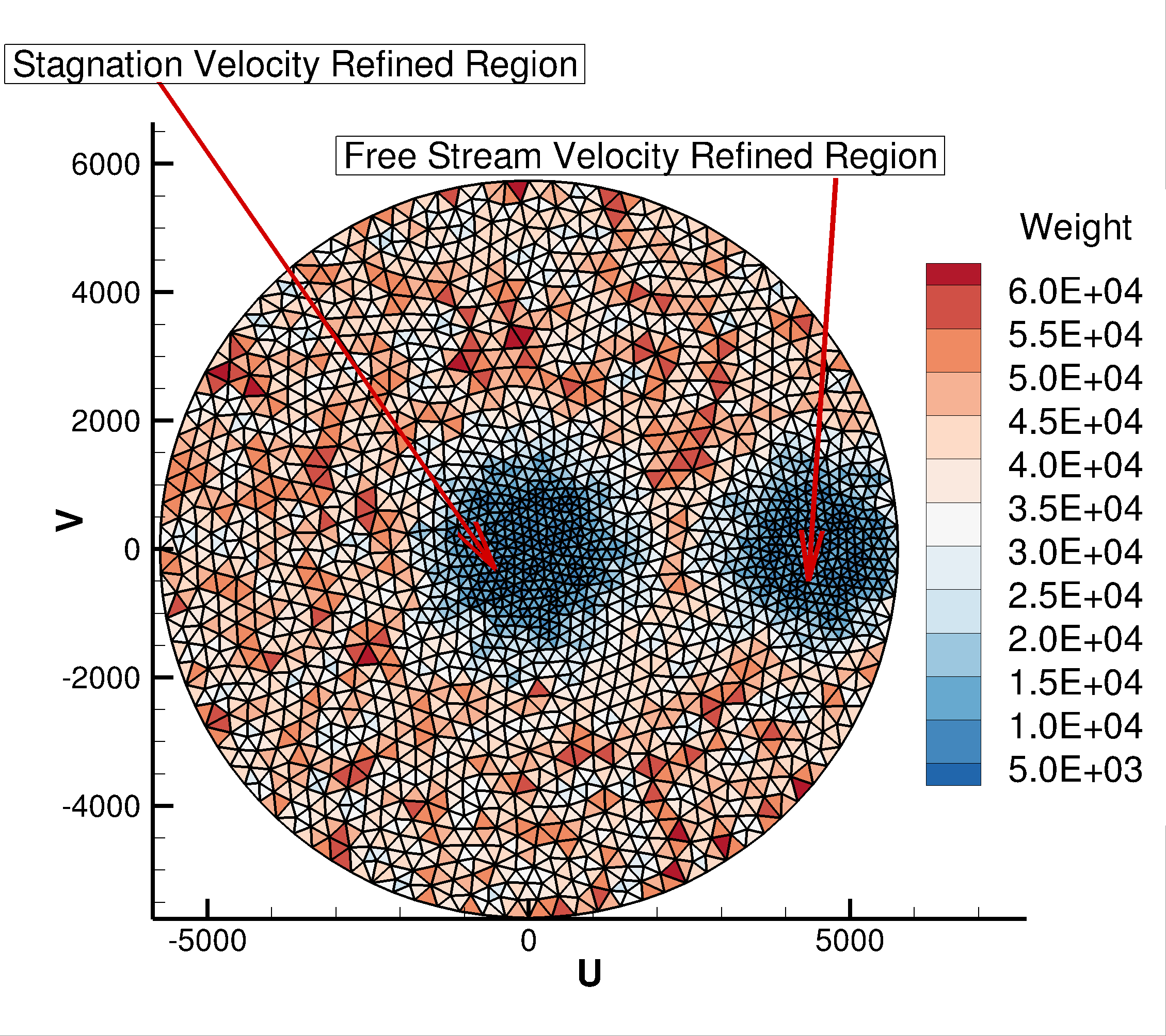}
 \caption[]{Unstructured DVS mesh with 3,420 cells used for hypersonic flow at $\text{Kn}_\infty=0.01$ and $\text{Ma}_\infty=15$ passing over a cylinder by the current UGKS program}\label{cylinderMa15DVS}
\end{figure}

To further verify the computational accuracy and robustness of the current scheme, hypersonic flow passing over a circular cylinder at a huge Mach number $\text{Ma}_\infty=15$ and $\text{Kn}_\infty=0.01$ is simulated. The characteristic length used to define the Knudsen number is the cylinder diameter $D=1$ m. The temperature of free stream flow is $T_\infty=273$ K, and an isothermal wall with a fixed temperature of $T_w=273$ K is applied for the cylinder surface. The physical domain is discretized by 14,000 ($200\times70$) quadrilateral cells with the height of first layer mesh $h=1\times10^{-4}$ m, while the unstructured discrete veslocity space (DVS) mesh consists of 3,420 cells, as shown in Figure \ref{cylinderMa15DVS}. The DVS is discretized in a circle region with the center $0.4\times(U_\infty,0)$ with a total radius of $5\sqrt{RT_s}$ where $T_s$ is the stagnation temperature of the free stream flow. The unstructured DVS mesh is refined at zero velocity point with a radius of $5\sqrt{RT_w}$ and the free stream point with a radius of $5\sqrt{RT_\infty}$.

\begin{figure}[!h]
 \centering
 \subfigure[Mach number]{\includegraphics[width=8cm]{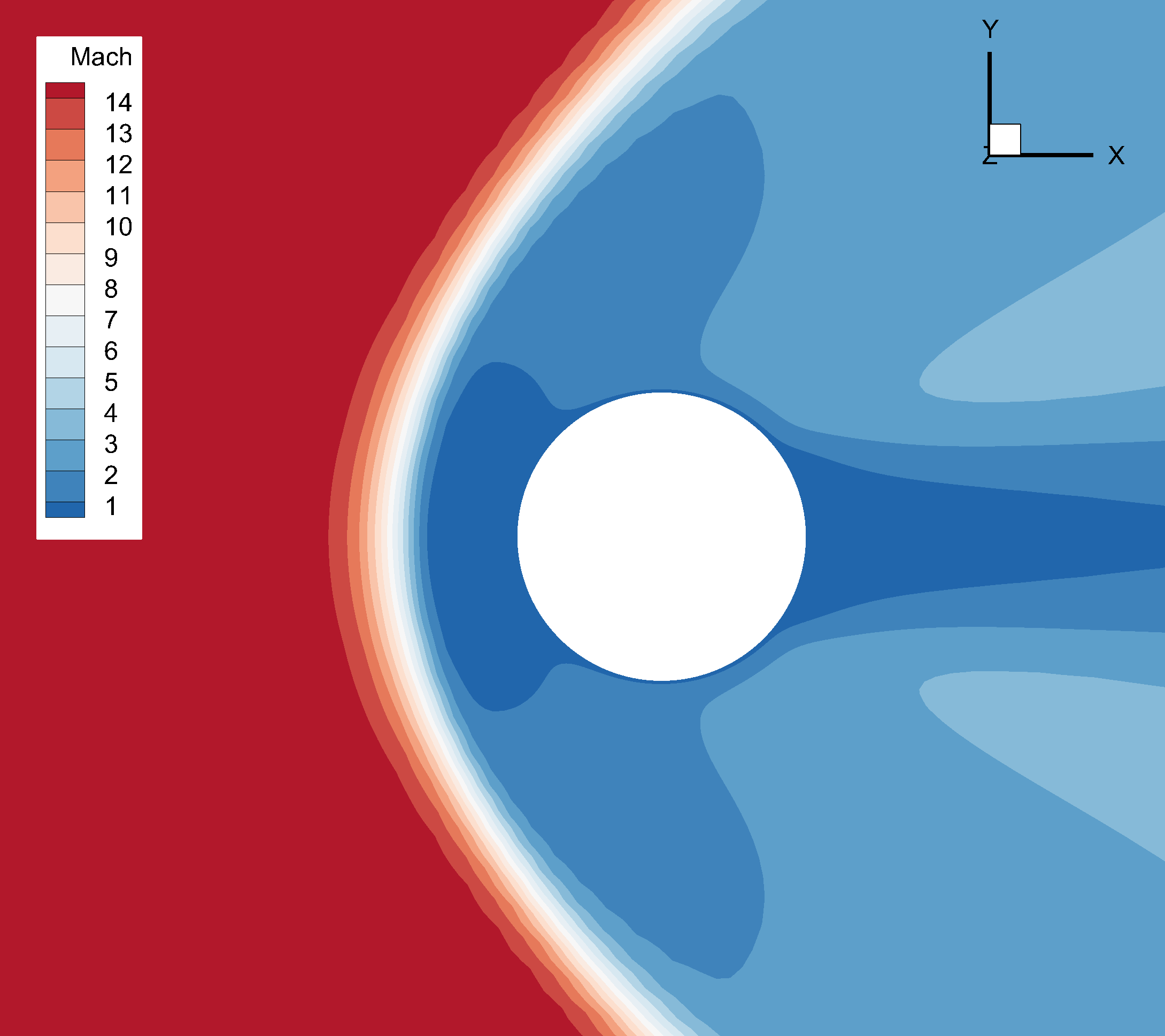}}
 \subfigure[Temperature]{\includegraphics[width=8cm]{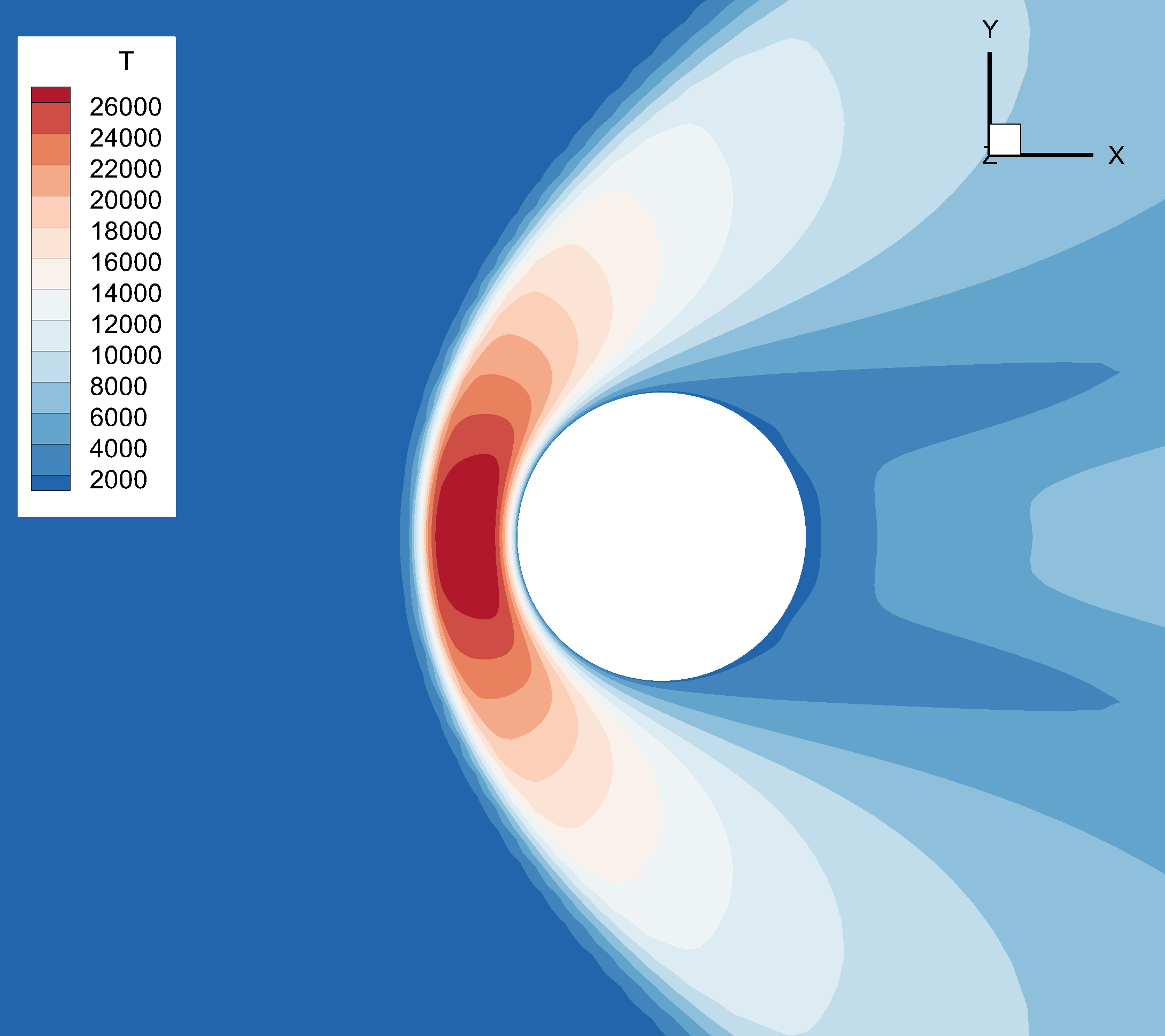}}
\caption[]{Contours of hypersonic flow at $\text{Kn}_\infty=0.01$ and $\text{Ma}_\infty=15$ passing over a circular cylinder by the current UGKS program.}\label{cylinderMa15contor}
\end{figure}

\begin{figure}[!h]
 \centering
 \subfigure[Pressure coefficient]{\includegraphics[width=5cm]{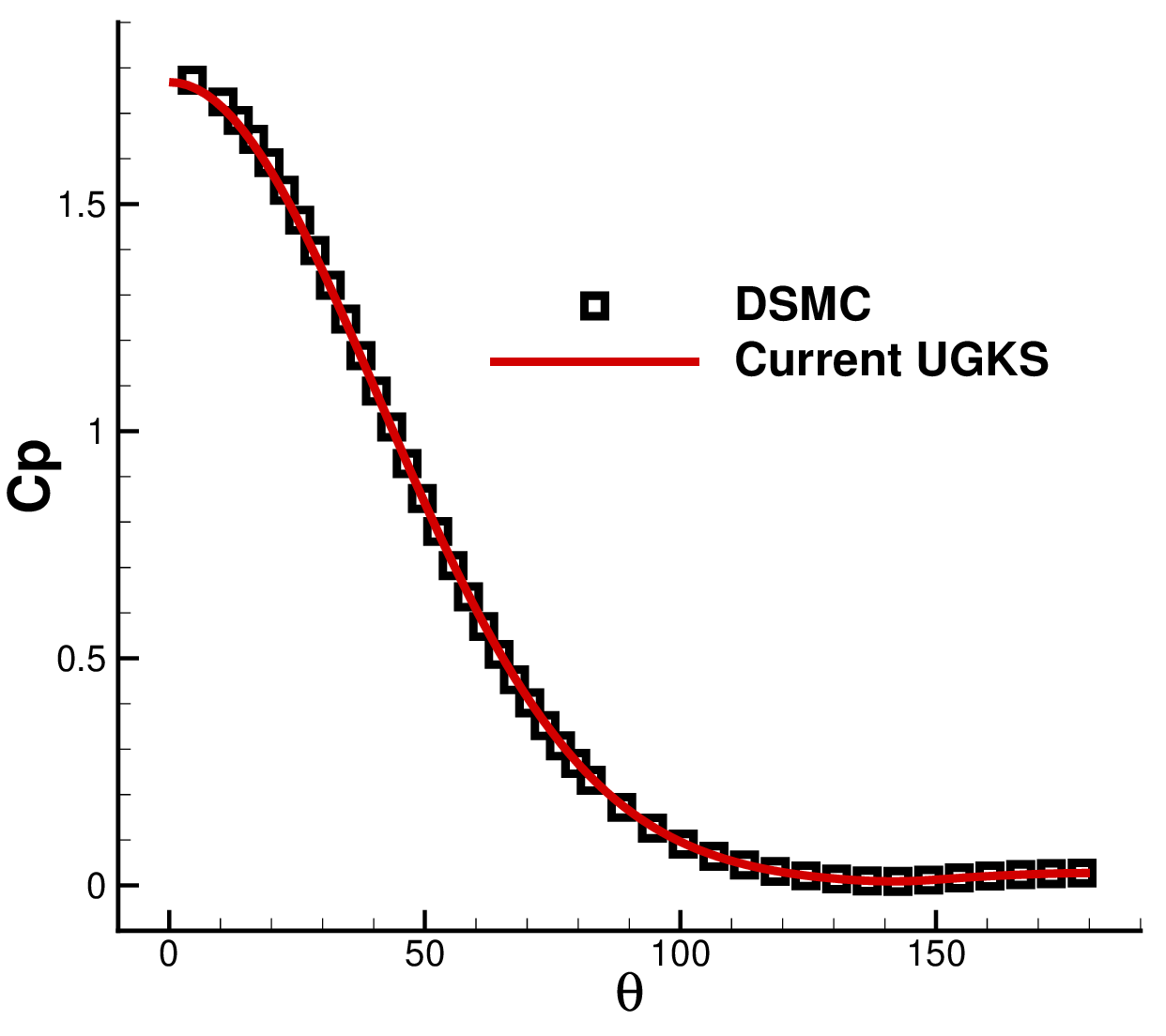}}
 \subfigure[Shear stress coefficient]{\includegraphics[width=5cm]{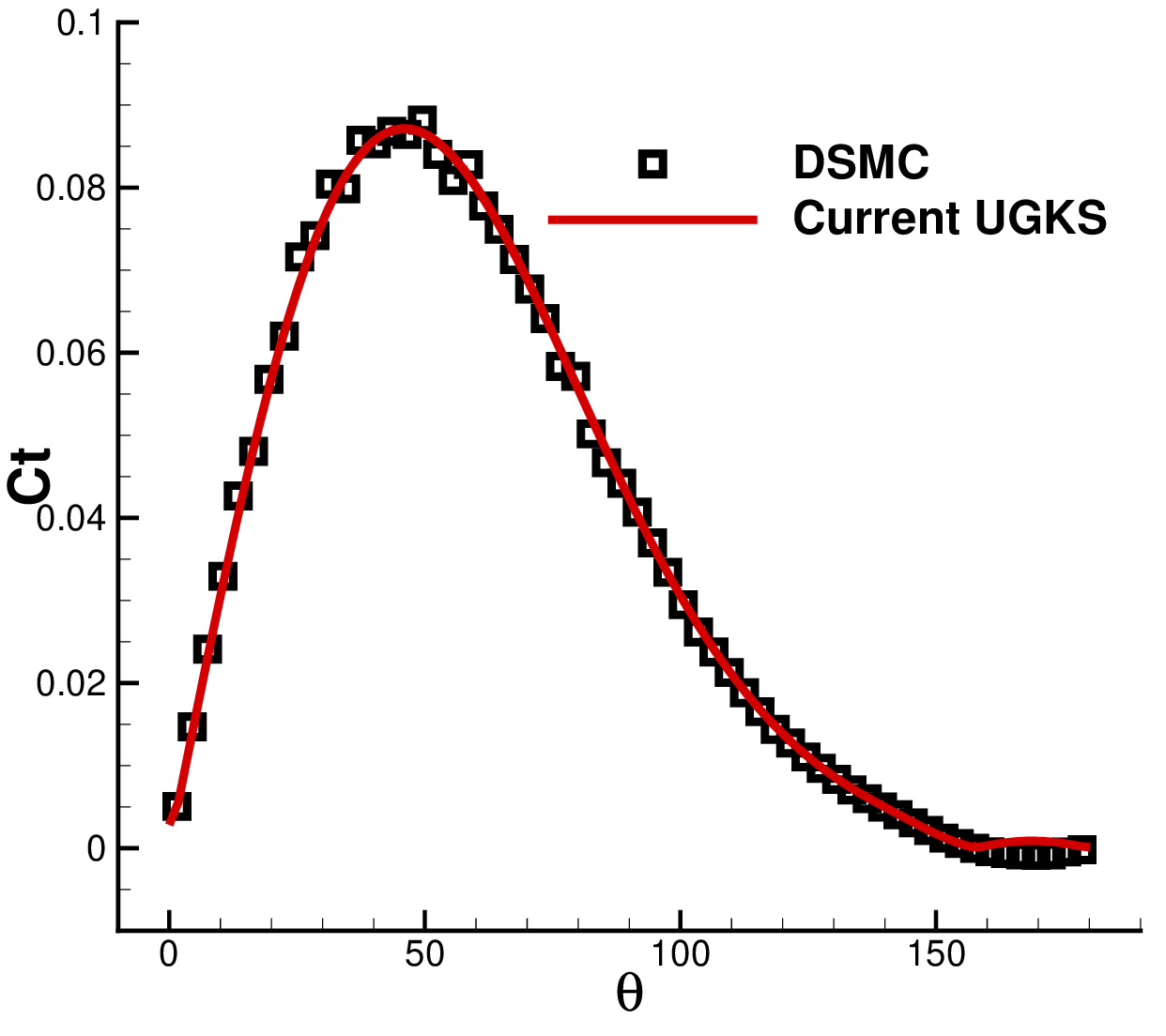}}
 \subfigure[Heat flux coefficient]{\includegraphics[width=5cm]{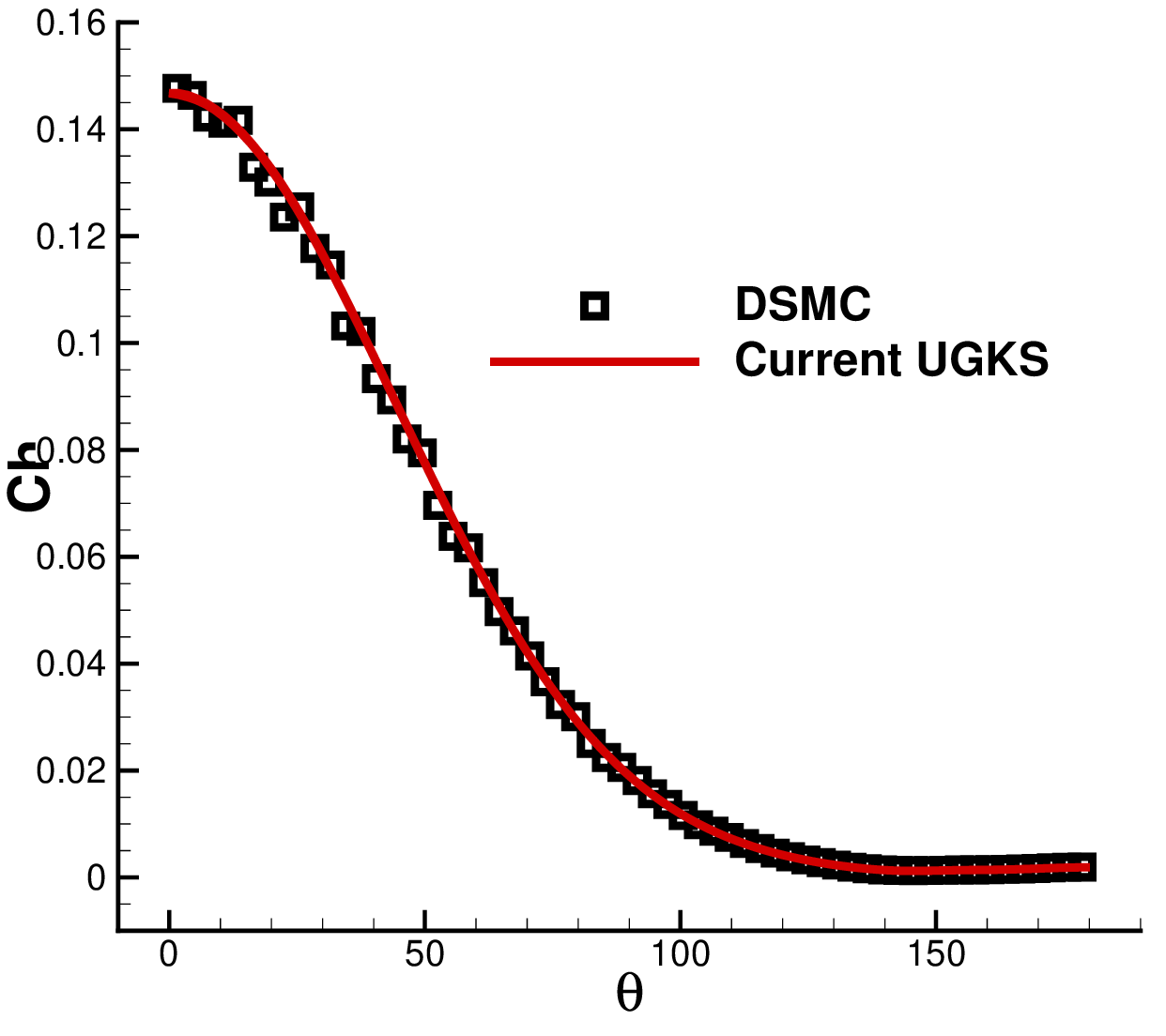}}
 \caption[]{Surface quantities of hypersonic flow at $\text{Kn}_\infty=0.01$ and $\text{Ma}_\infty=15$ passing over a circular cylinder by the current UGKS program compared with the DSMC method \cite{zhang2024conservative}.}\label{cylinderMa15surface}
\end{figure}

The contours of Mach number and temperature of hypersonic flow at $\text{Kn}=0.01$ and $\text{Ma}=15$ passing over a circular cylinder are shown in Figure \ref{cylinderMa15contor}. To further verify our code, the surface quantities compared with the DSMC method \cite{zhang2024conservative} are shown in Figure \ref{cylinderMa15surface}.

\subsection{Supersonic flow around a sphere}
\begin{figure}[!h]
 \centering
 \includegraphics[width=8cm]{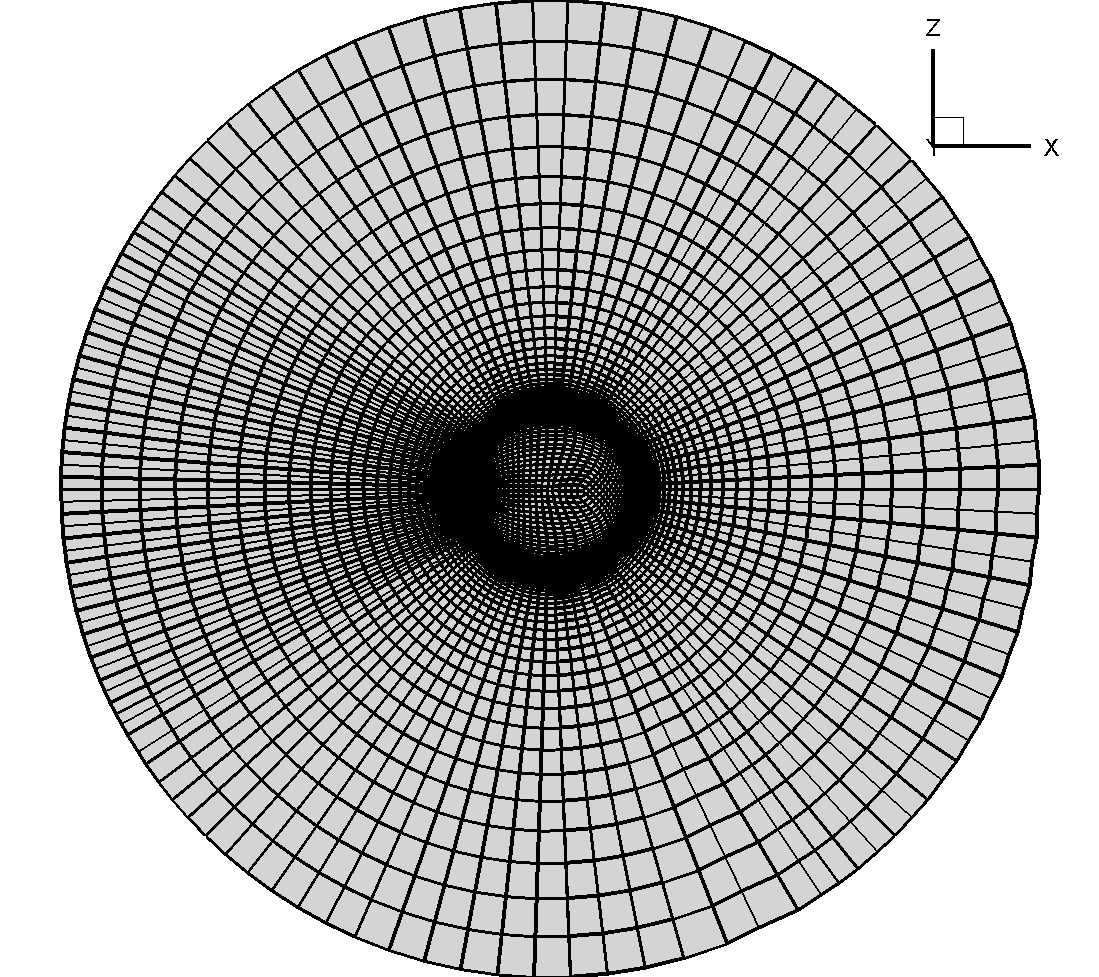}
 \includegraphics[width=8cm]{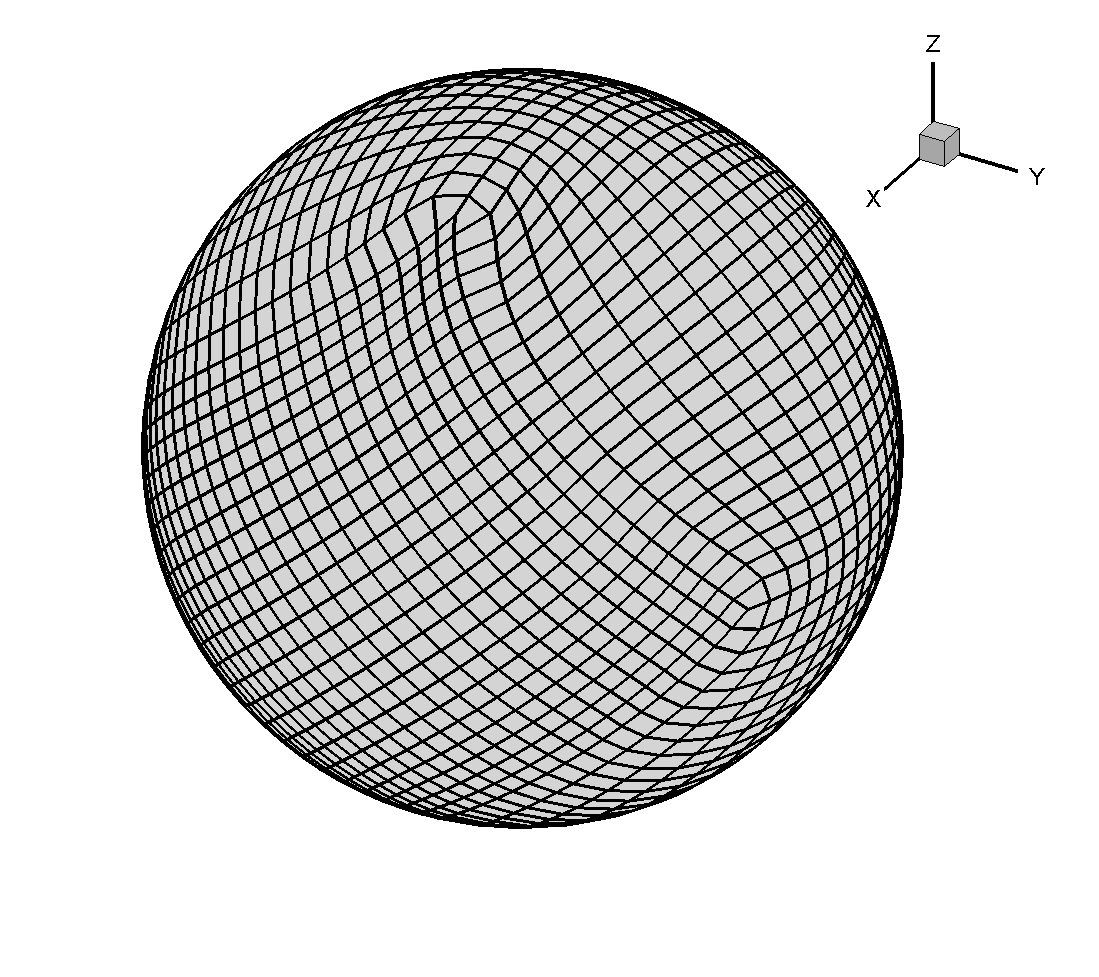}
 \caption[]{Physical domain mesh used for supersonic flow at $\text{Ma}_\infty=4.25$ passing over spheresphere by the current UGKS program.}\label{spheremesh}
\end{figure}

\begin{figure}[!h]
 \centering
 \includegraphics[width=8cm]{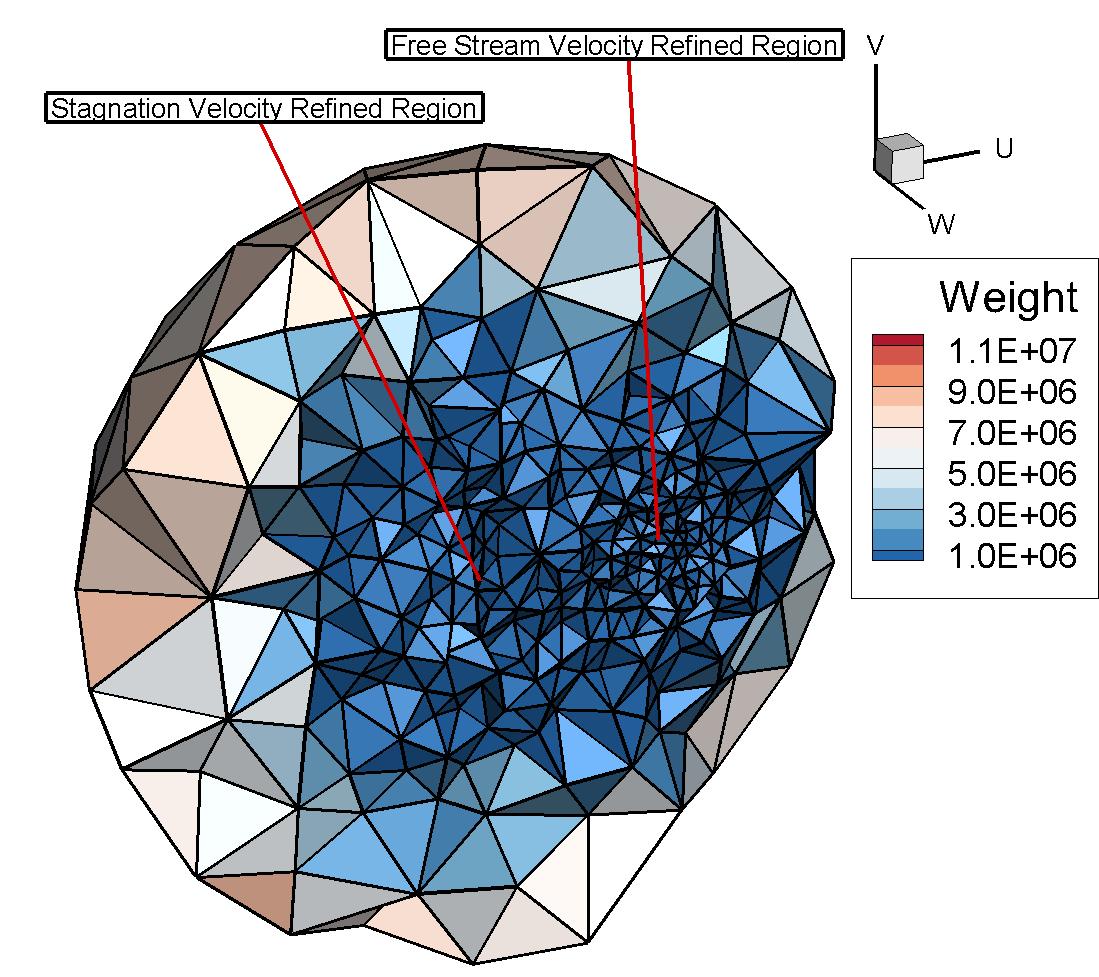}
 \caption[]{Unstructured DVS mesh consisting of 11,740 cells for supersonic flow at $\text{Ma}_\infty=4.25$ passing over sphere by the current UGKS program.}\label{spheredvs}
\end{figure}

To verify further our code in three-dimensional supersonic flow, the supersonic flow passing over a sphere at Mach number $4.25$ for $\text{Kn}_\infty = 0.031$ and $\text{Kn}_\infty = 0.121$ are simulated for argon gas with Prandtl number $\text{Pr}=0.72$. The characteristic length is the sphere diameter $D = 2$ m to define the Knudsen number. The physical domain contains $3,082\times 37$ hexahedron cells. The physical domain mesh and sphere surface mesh are shown in Figure \ref{spheremesh}. The height of the first layer mesh is set as $5\times 10^{-3}$ m, and the computational domain is set as a spherical domain with the radius $R<6$ m.
Figure \ref{spheredvs} illustrates the section view of unstructured DVS mesh with 11,740 cells. The DVS is discretized into a sphere mesh with a radius of $5\sqrt{RT_{\max}}$, where $T_{\max}$ is the maximum temperature estimated by GKS result as the initial field of the UGKS. The sphere center is located at the zero point. The velocity space near the zero and free stream velocity points are refined.

\begin{figure}[!h]
 \centering
 \subfigure[Mach number]{\includegraphics[width=8cm]{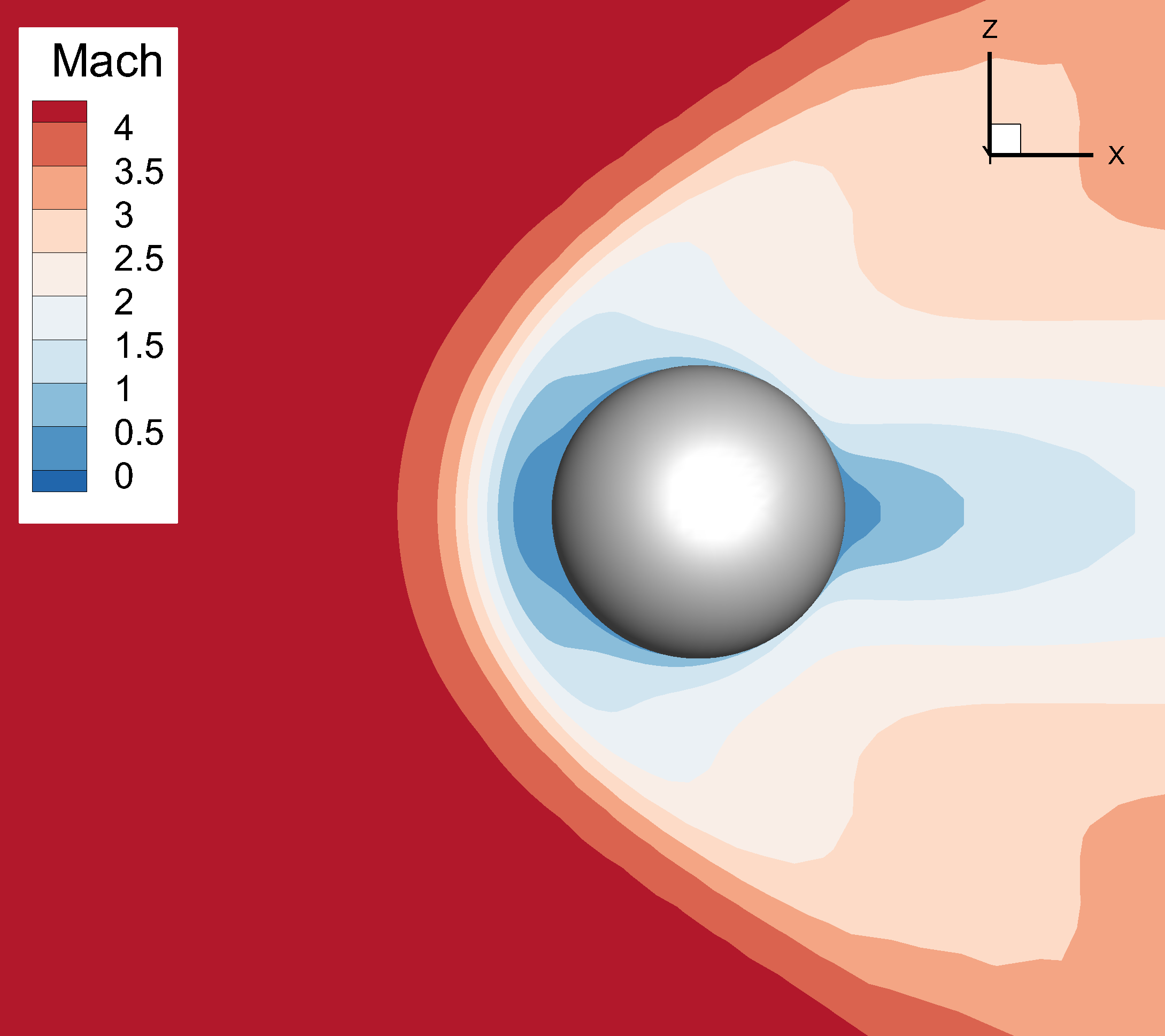}}
 \subfigure[Temperature]{\includegraphics[width=8cm]{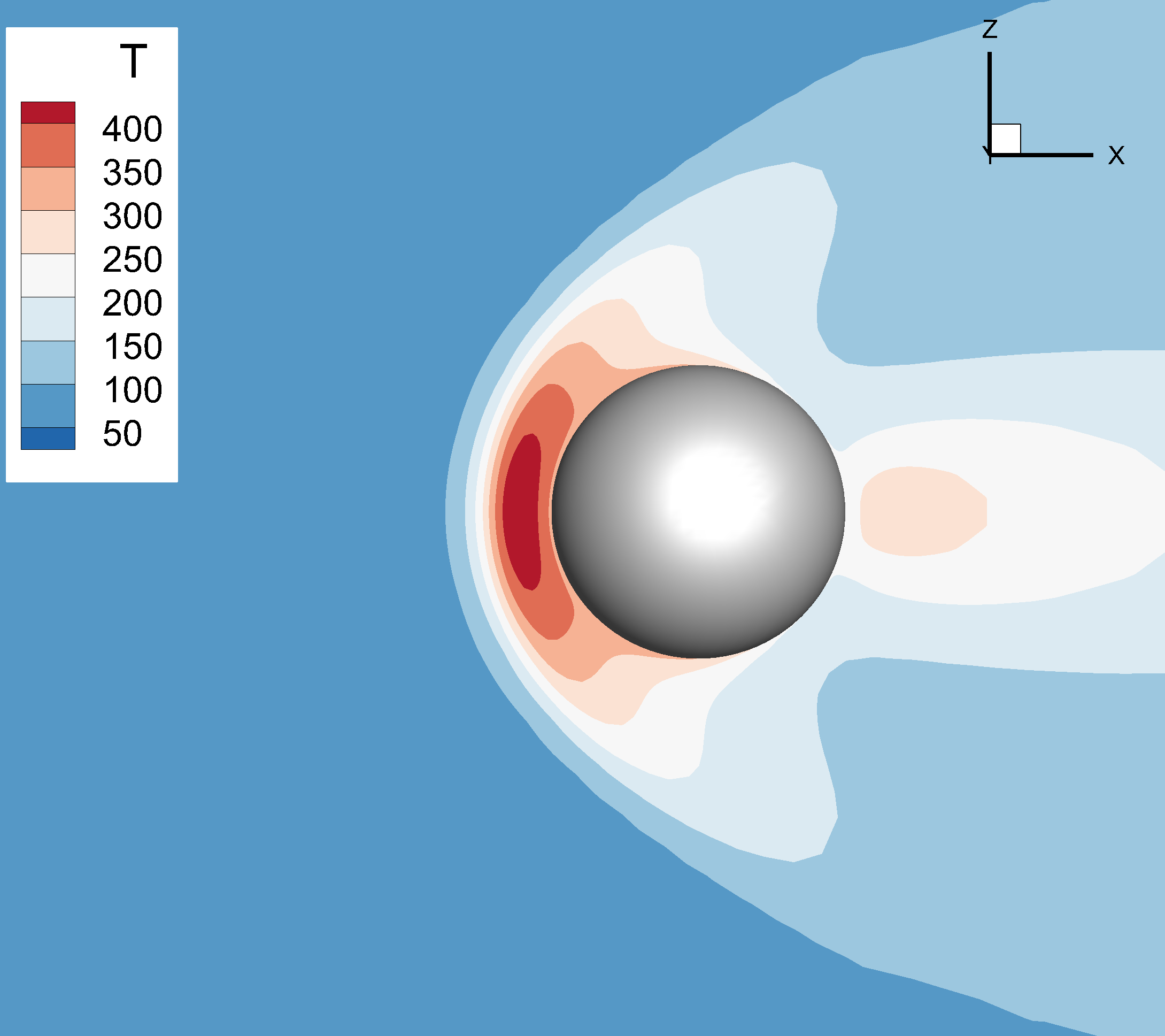}}
 \caption[]{Contours of supersonic flow around a sphere at $\text{Ma}_\infty=4.25$ for $\text{Kn}_\infty=0.031$ by the current UGKS program.}\label{sphere031contor}
\end{figure}

\begin{figure}[!h]
 \centering
 \subfigure[Mach number]{\includegraphics[width=8cm]{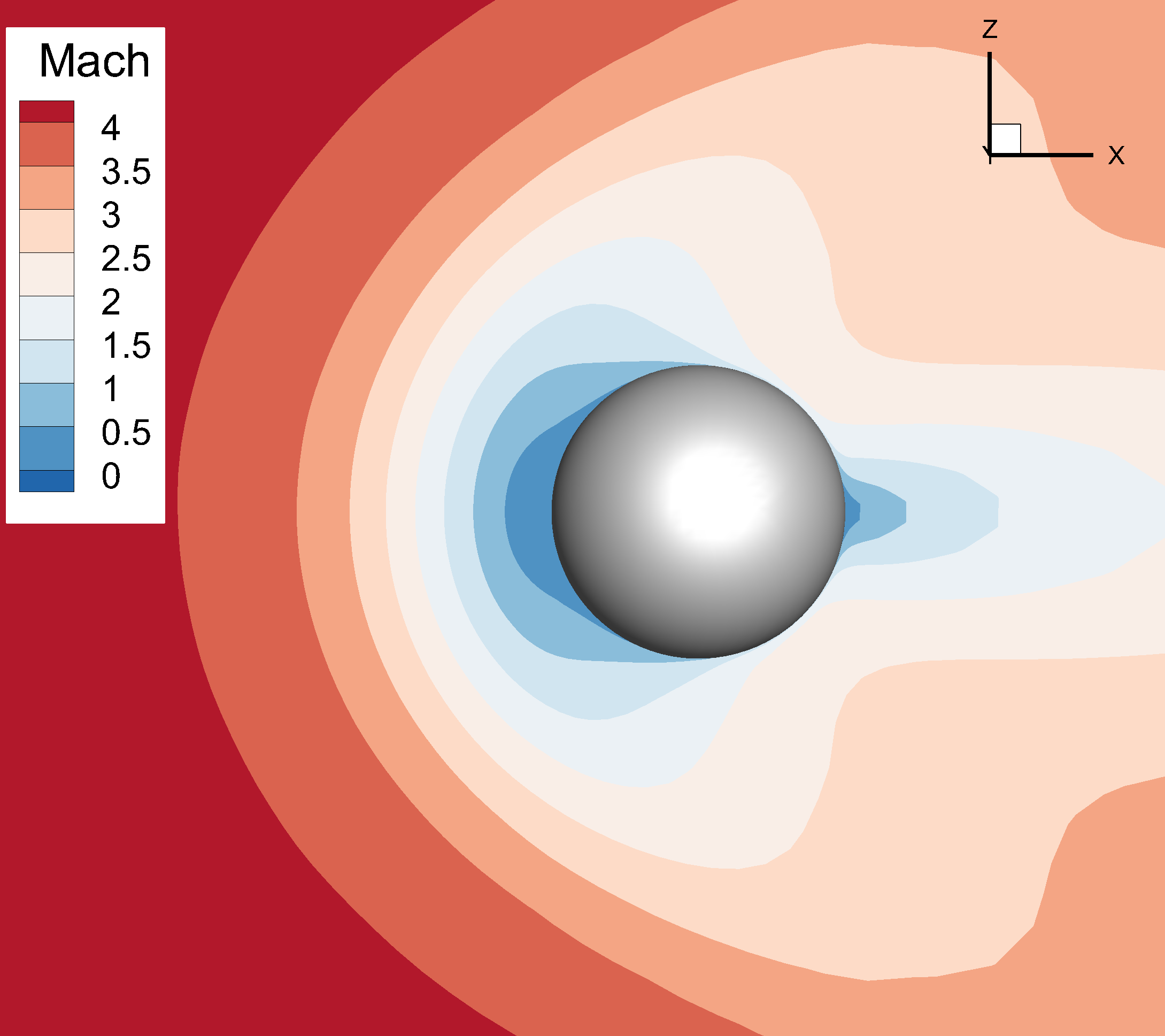}}
 \subfigure[Temperature]{\includegraphics[width=8cm]{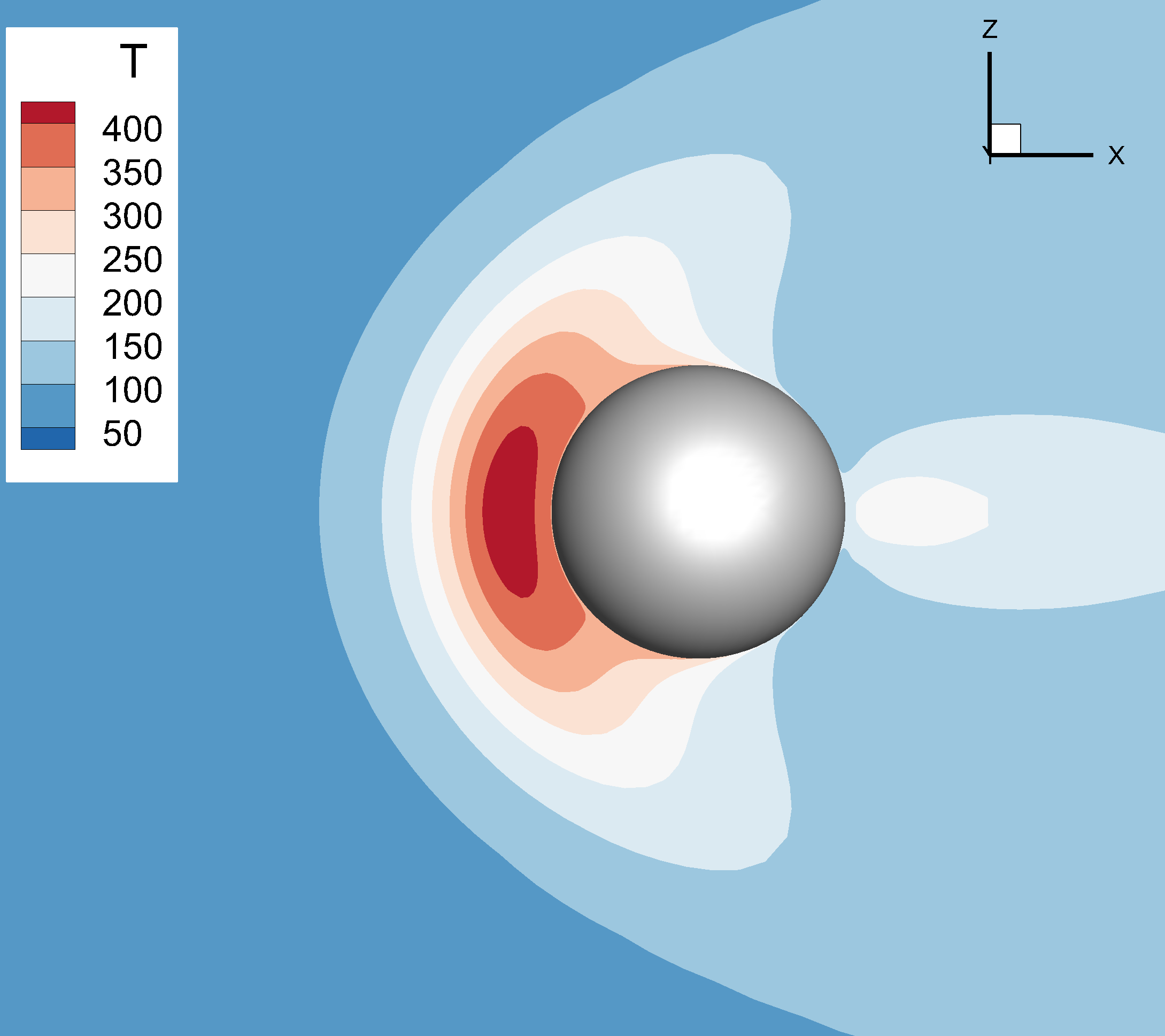}}
 \caption[]{Contours of supersonic flow around a sphere at $\text{Ma}_\infty=4.25$ for $\text{Kn}_\infty=0.121$ by the current UGKS program.}\label{sphere121contor}
 \end{figure}

\begin{figure}[!h]
 \centering
 \subfigure[Pressure coefficient]{\includegraphics[width=5cm]{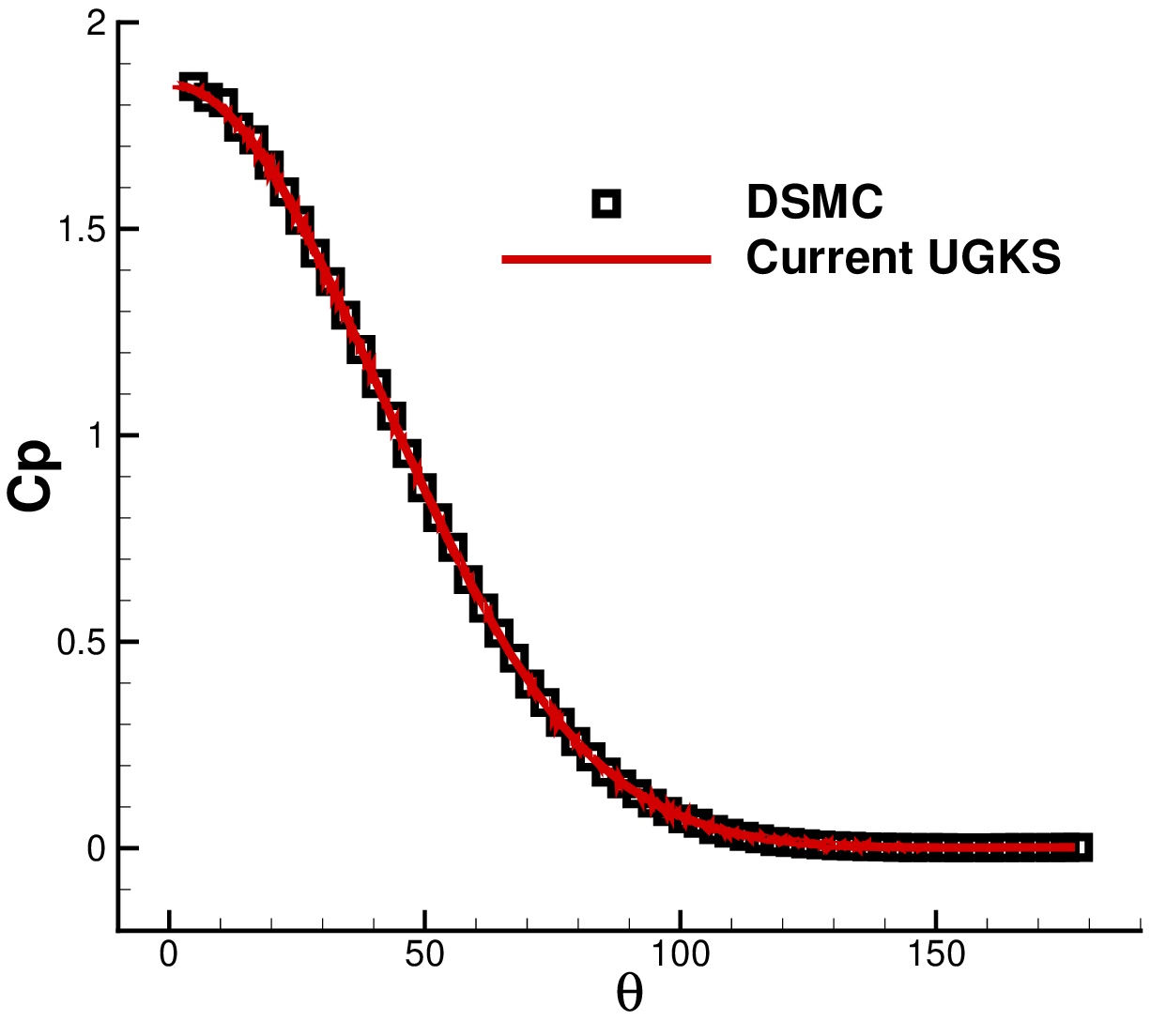}}
 \subfigure[Shear stress coefficient]{\includegraphics[width=5cm]{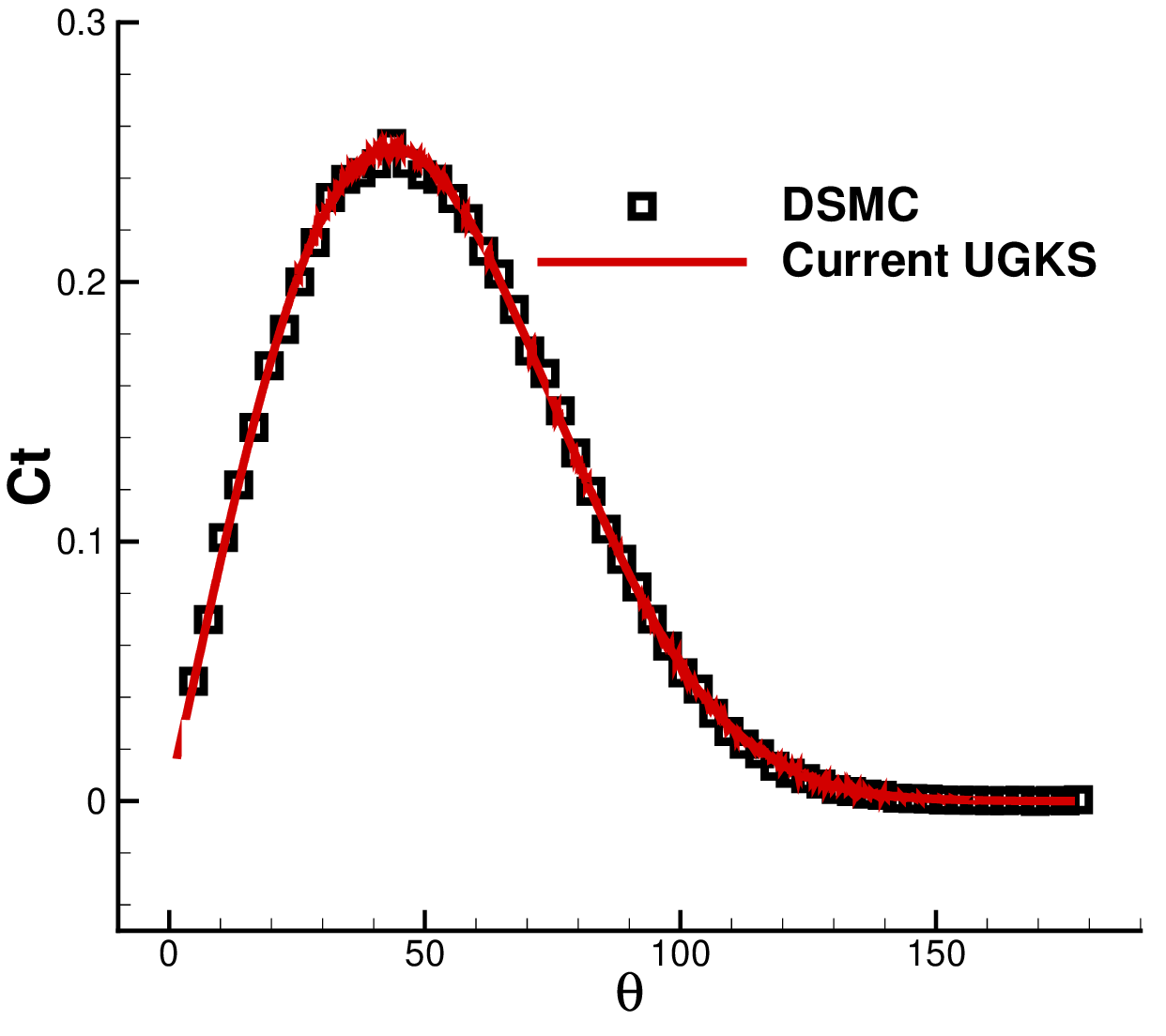}}
 \subfigure[Heat flux coefficient]{\includegraphics[width=5cm]{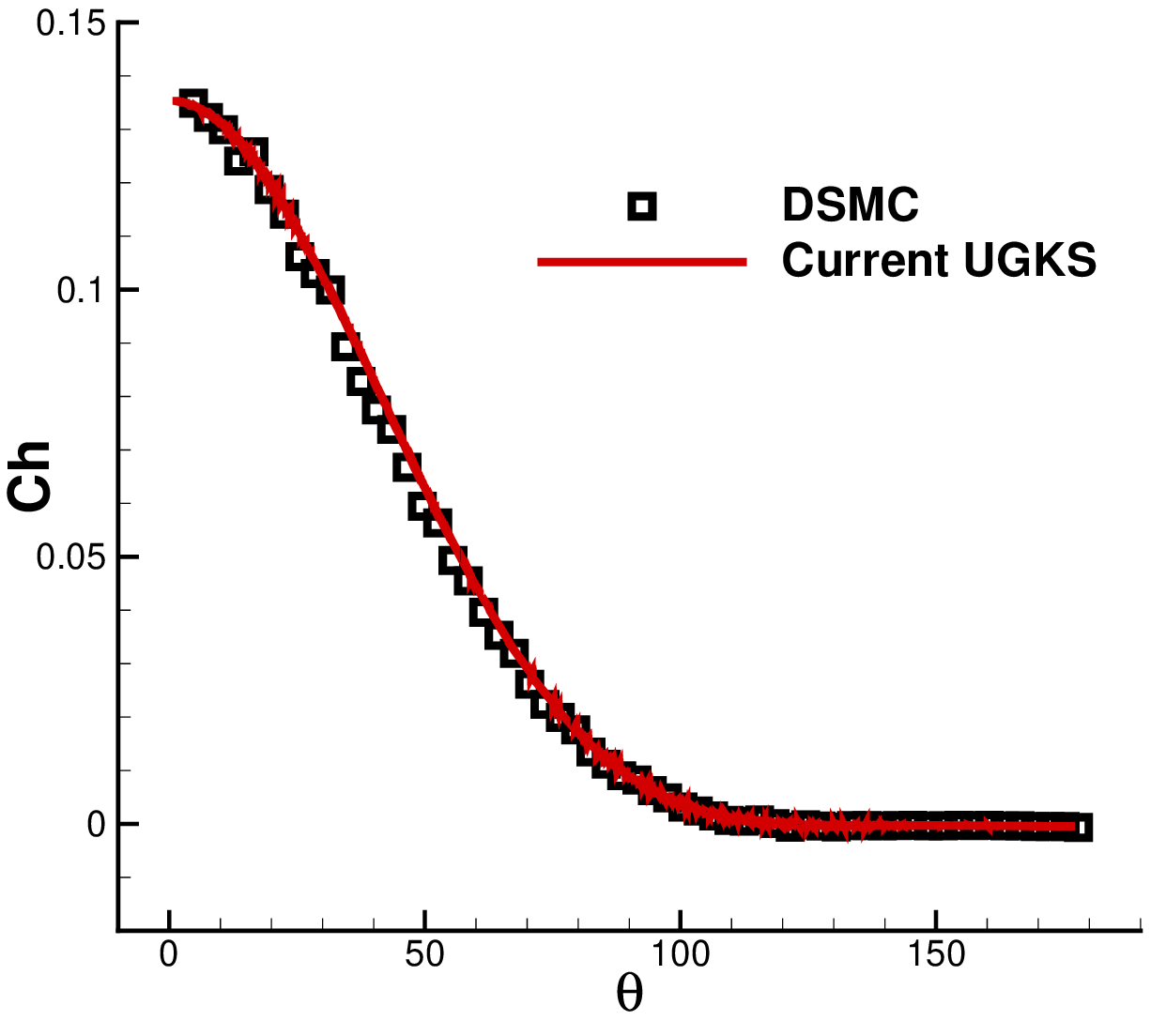}}
 \caption[]{Surface quantities of supersonic flow around a sphere at $\text{Ma}_\infty=4.25$ for $\text{Kn}_\infty=0.031$ by the current UGKS program compared with the DSMC method \cite{zhang2024conservative}.}\label{sphere121surface}
\end{figure}

\begin{figure}[!h]
 \centering
 \subfigure[Pressure coefficient]{\includegraphics[width=5cm]{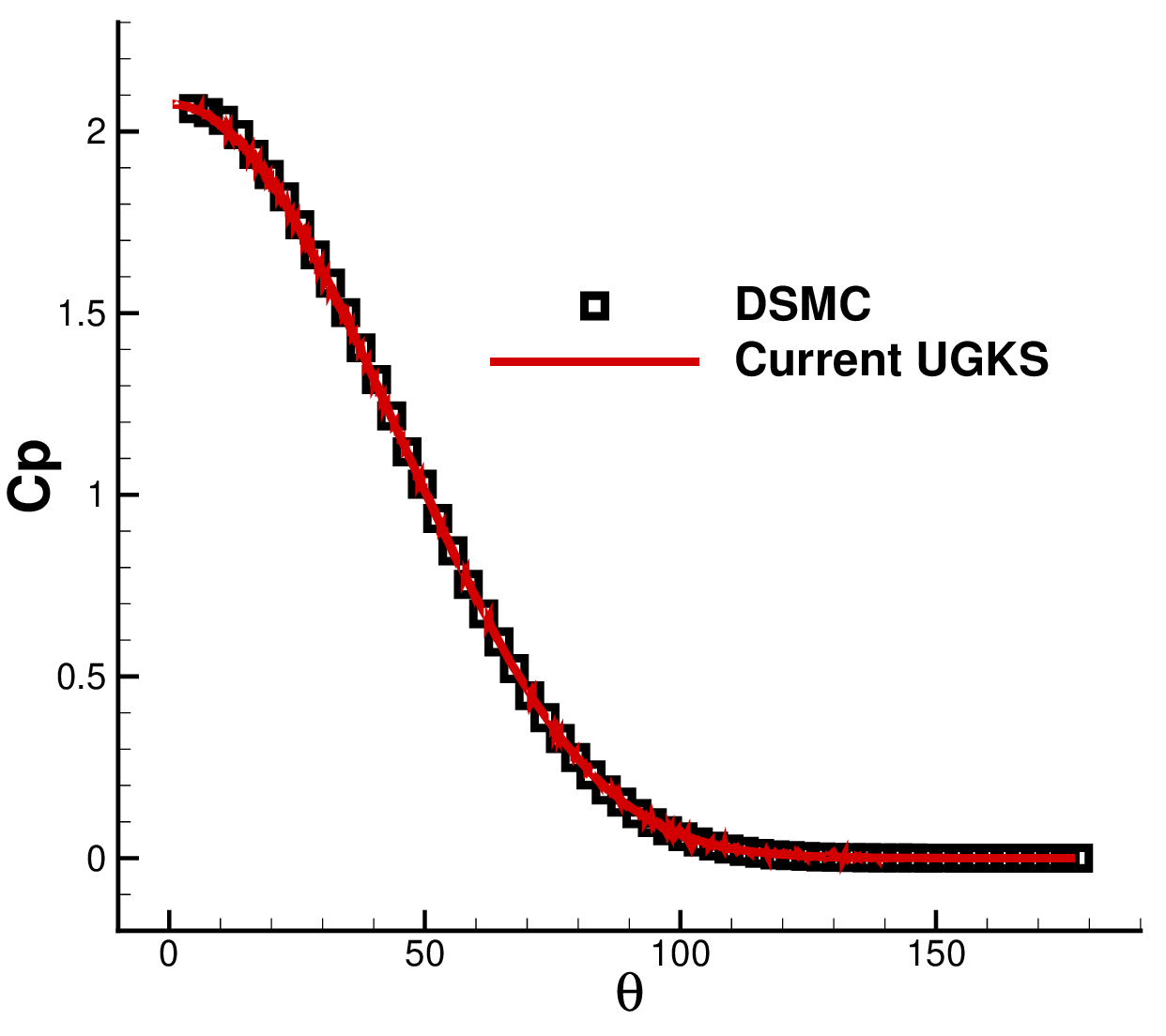}}
 \subfigure[Shear stress coefficient]{\includegraphics[width=5cm]{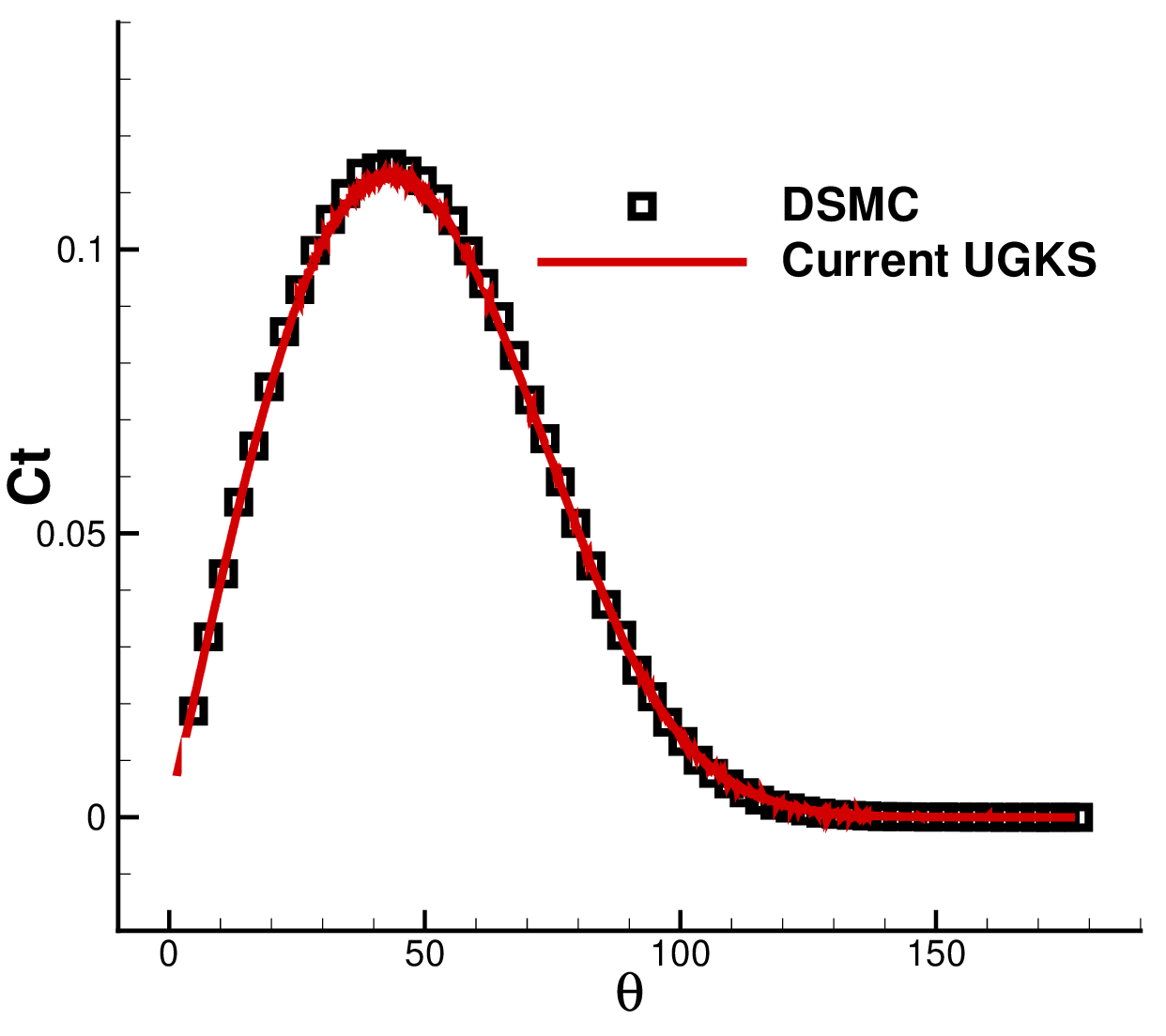}}
 \subfigure[Heat flux coefficient]{\includegraphics[width=5cm]{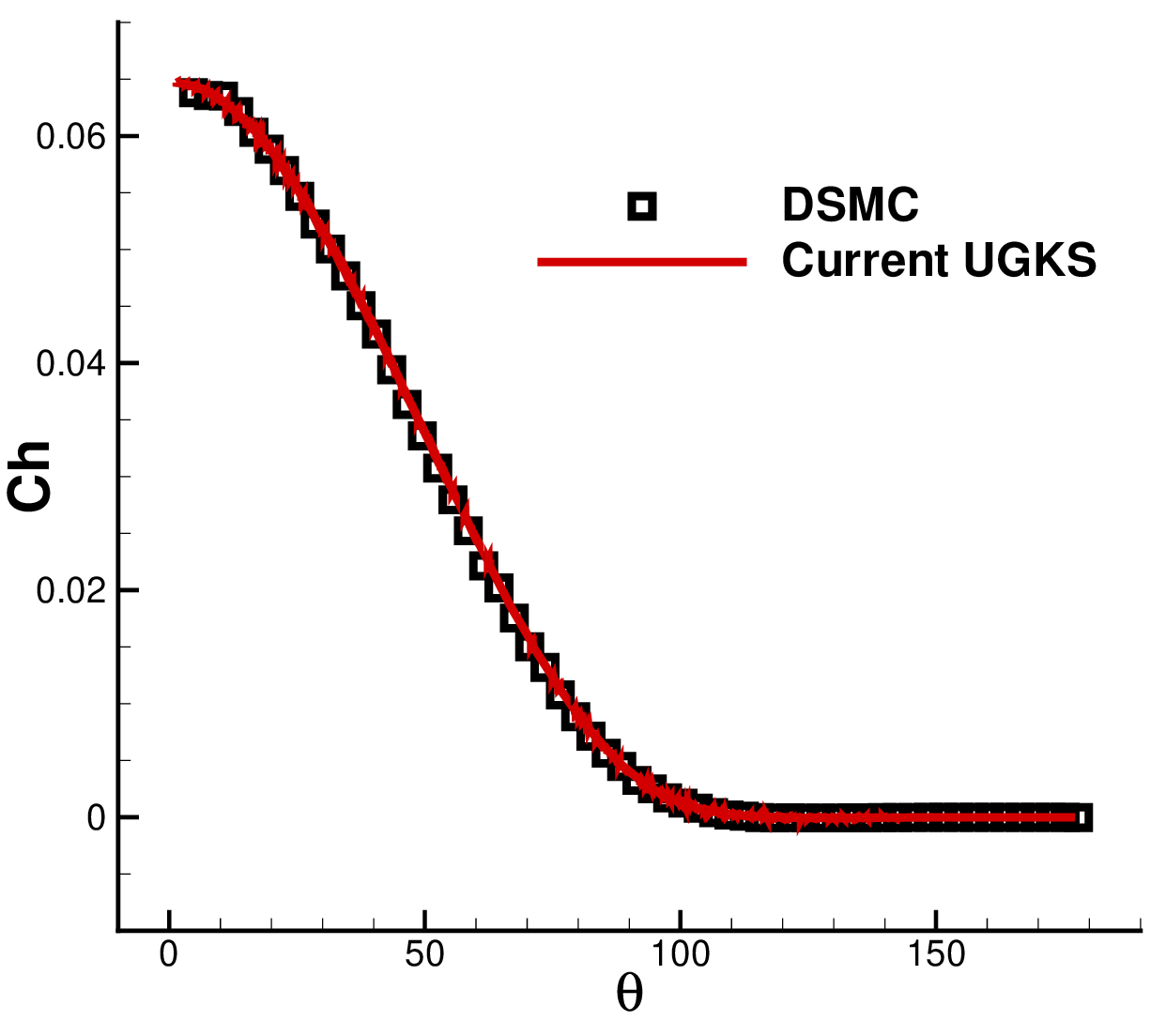}}
 \caption[]{Surface quantities of supersonic flow around a sphere at $\text{Ma}_\infty=4.25$ for $\text{Kn}_\infty=0.121$ by the current UGKS program compared with the DSMC method \cite{zhang2024conservative}.}\label{sphere031surface}
\end{figure}

Figures \ref{sphere031contor} and \ref{sphere121contor} depict the contour of Mach number and temperature by our code for different $\text{Kn}_\infty$. To further verify the ability to accurately predict surface quantities, the pressure coefficient, shear stress coefficient, and heat flux coefficient are compared with DSMC results \cite{zhang2024conservative}, shown in Figure \ref{sphere031surface} and Figure \ref{sphere121surface}. The result of the pressure coefficient, shear stress coefficient, and heat flux coefficient agree well with the DSMC result.

\subsection{Lid-driven cavity flow}
\begin{figure}[!h]
 \centering
 \includegraphics[width=8cm]{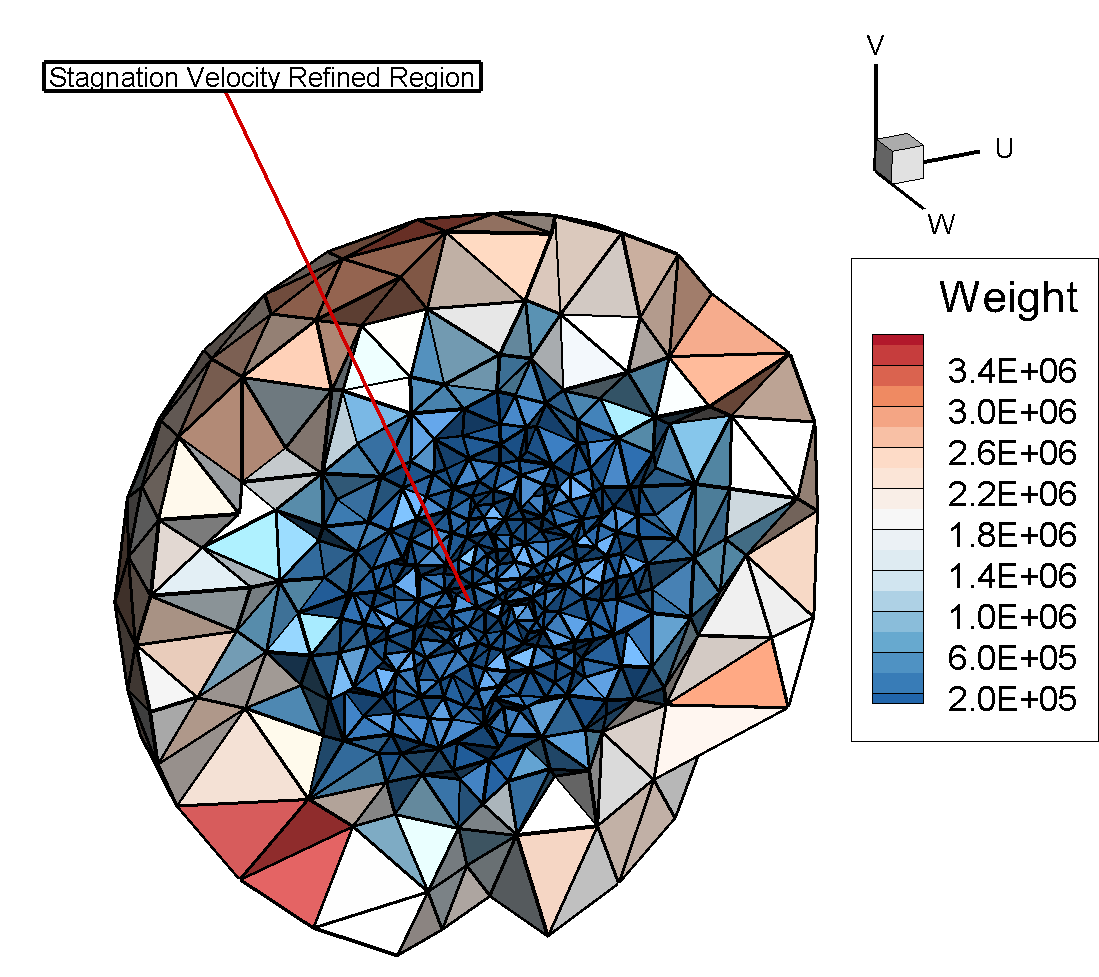}
 \caption[]{Unstructured DVS mesh used for cavity flow by the current UGKS program.}\label{cavitydvs}
\end{figure}

The three-dimensional lid-driven cavity flows at different Knudsen numbers are used to verify our code. The monatomic gas with the Prandtl number $2/3$ is used in this case. The gas constant is set as $R=208.14$ $\text{J}/\text{kg}\cdot\text{K} $ and $\omega=0.74$. Three Knudsen numbers, $1,0.1$ and $0.01$, are tested in this section. The physical domain is $[0,1]^3$ and discretized with $20^3$ hexahedron cells. The wall temperature is 273 K, and the upper wall moves with $x$-direction velocity of $50.567$ m/s. The initial condition is set as a zero-velocity field with a temperature of 273 K. Figure \ref{cavitydvs} illustrates the section view of unstructured DVS mesh with 17,182 cells. The DVS is discretized into a sphere mesh with a radius of $5\sqrt{RT_w}$, whose center is located at the zero point. The velocity space near the zero point is refined.

\begin{figure}[!h]
 \centering
 \subfigure[Velocity profiles along the central lines compared with the DVM \cite{yang2019improved}]{\includegraphics[width=8cm]{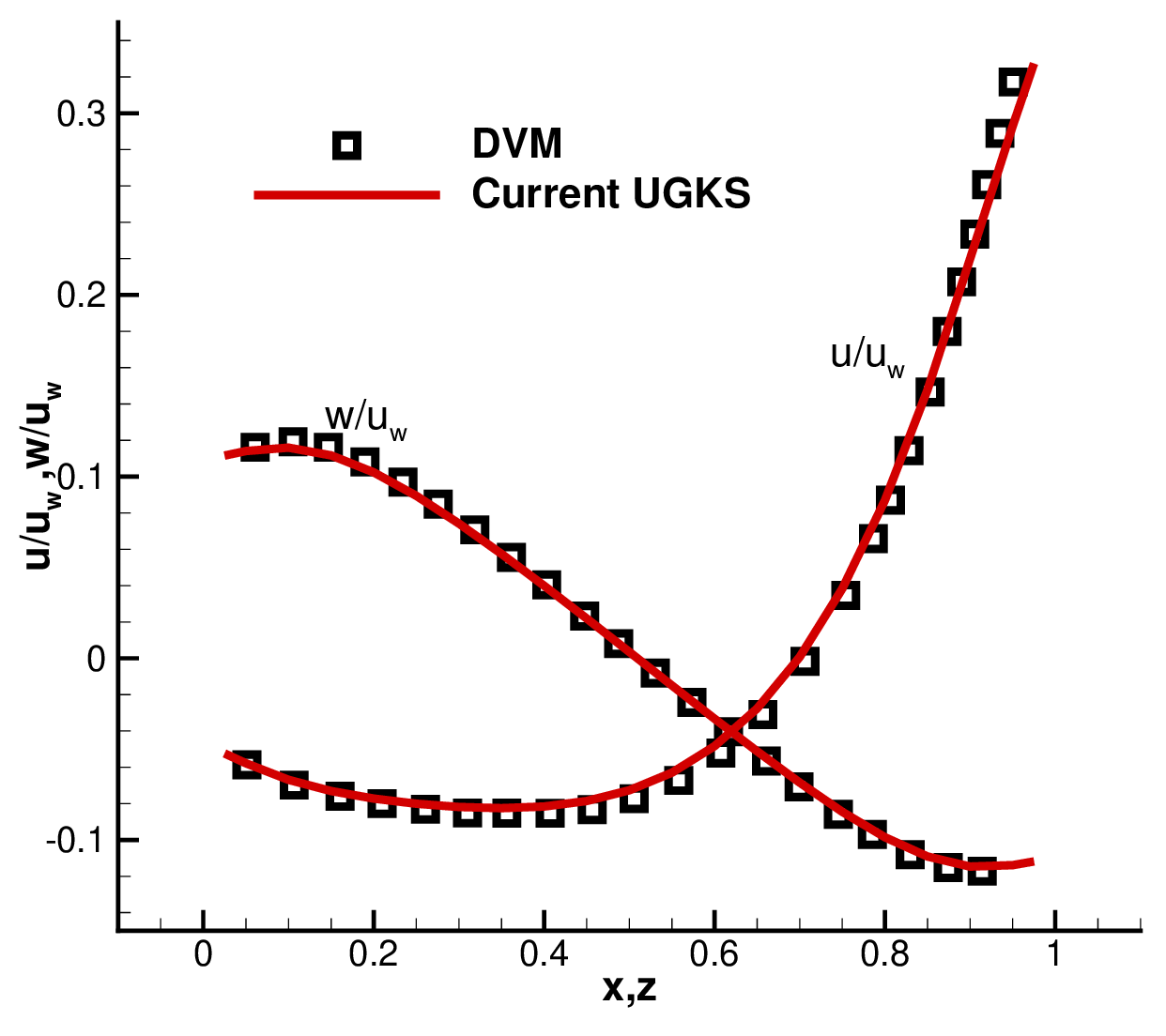}}
 \subfigure[Temperature contour]{\includegraphics[width=8cm]{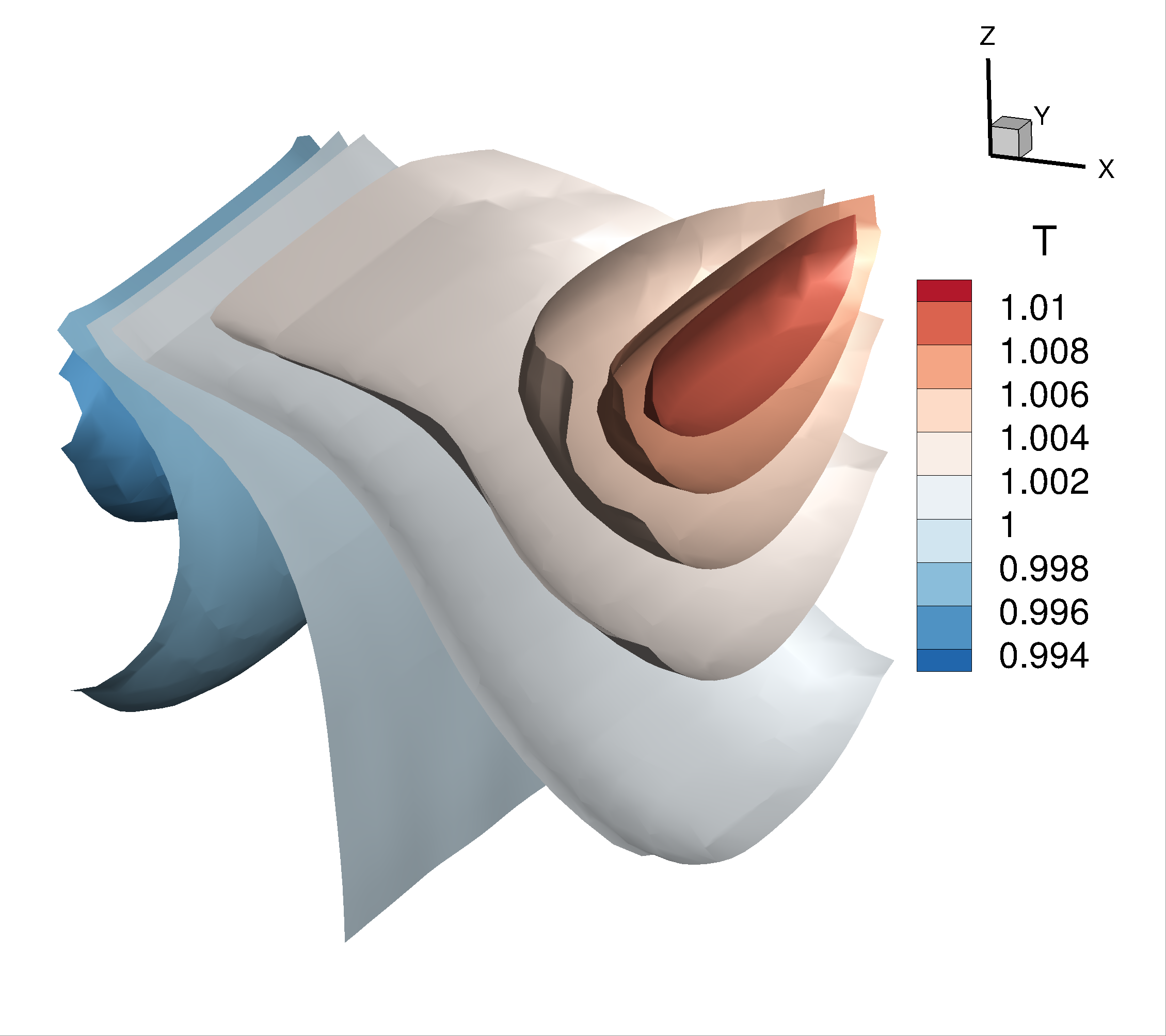}}
 \caption[]{3D lid-driven cavity flow at $\text{Kn} = 1$ by the current UGKS program.}\label{cavitykn1}
\end{figure}

\begin{figure}[!h]
 \centering
 \subfigure[Velocity profiles along the central lines compared with the DVM \cite{yang2019improved}]{\includegraphics[width=8cm]{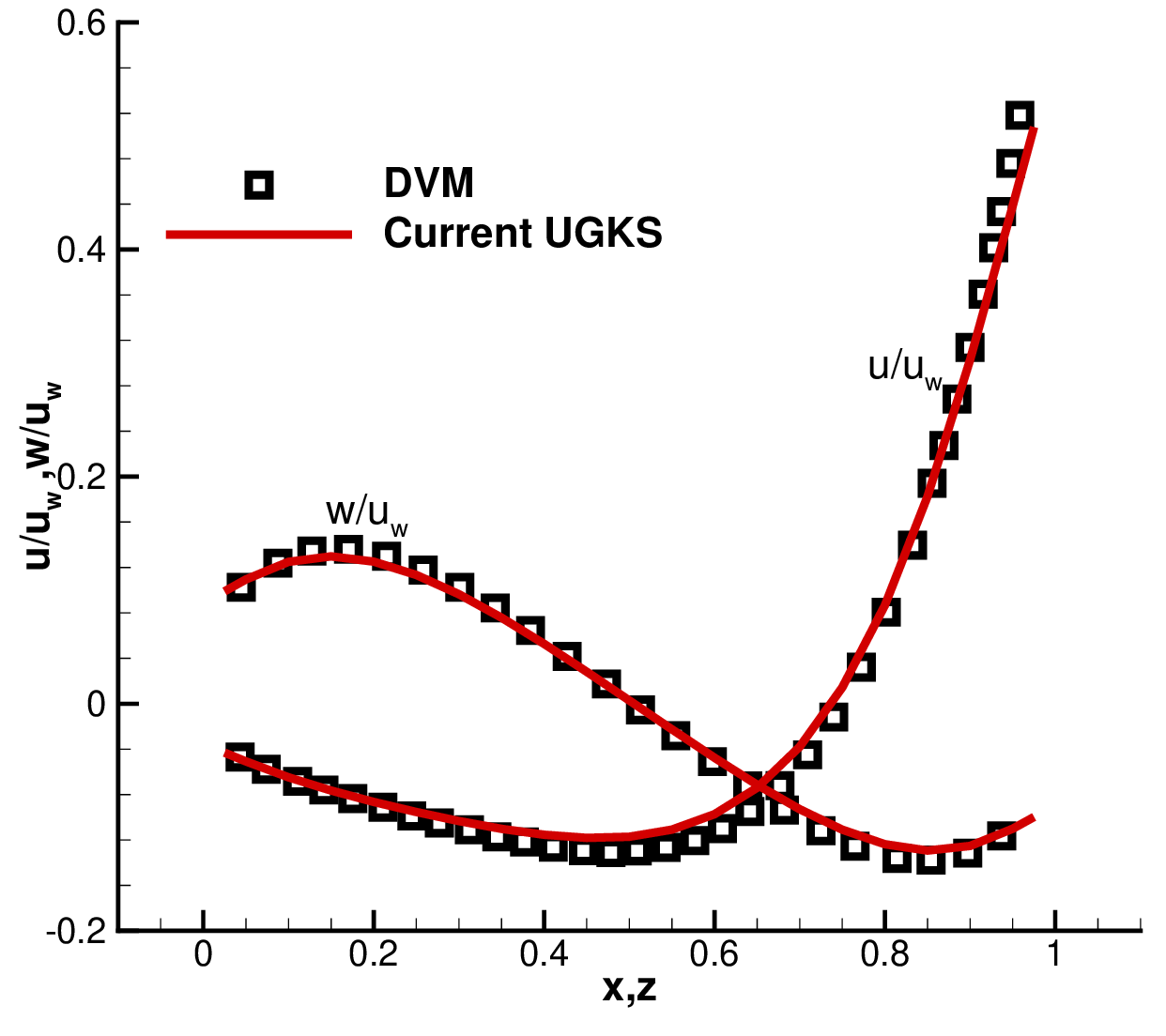}}
 \subfigure[Temperature contour]{\includegraphics[width=8cm]{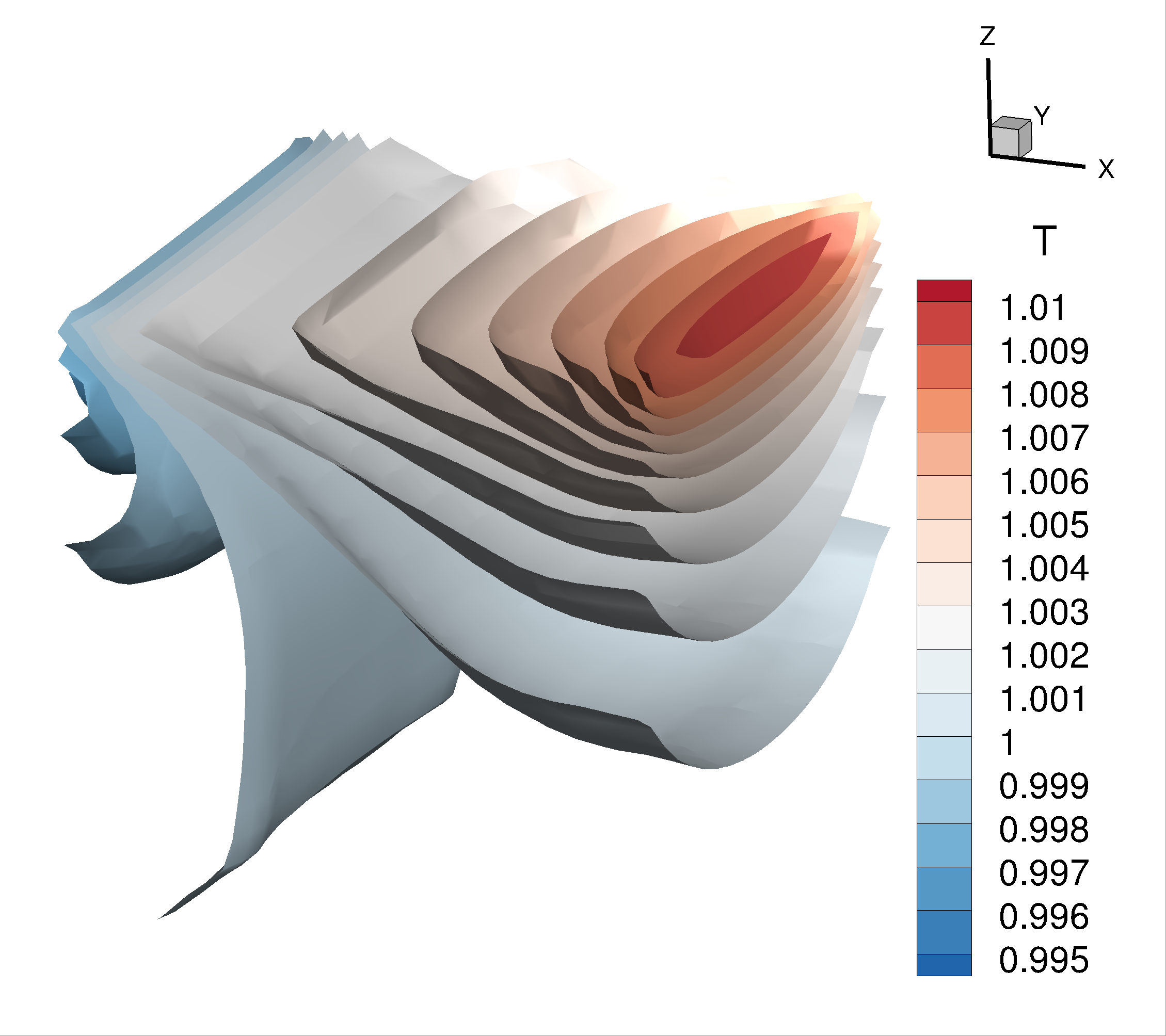}}
 \caption[]{3D lid-driven cavity flow at $\text{Kn} = 0.1$ by the current UGKS program.}\label{cavitykn0.1}
\end{figure}

\begin{figure}[!h]
 \centering
 \subfigure[Velocity profiles along the central lines compared with the DVM \cite{yang2019improved}]{\includegraphics[width=8cm]{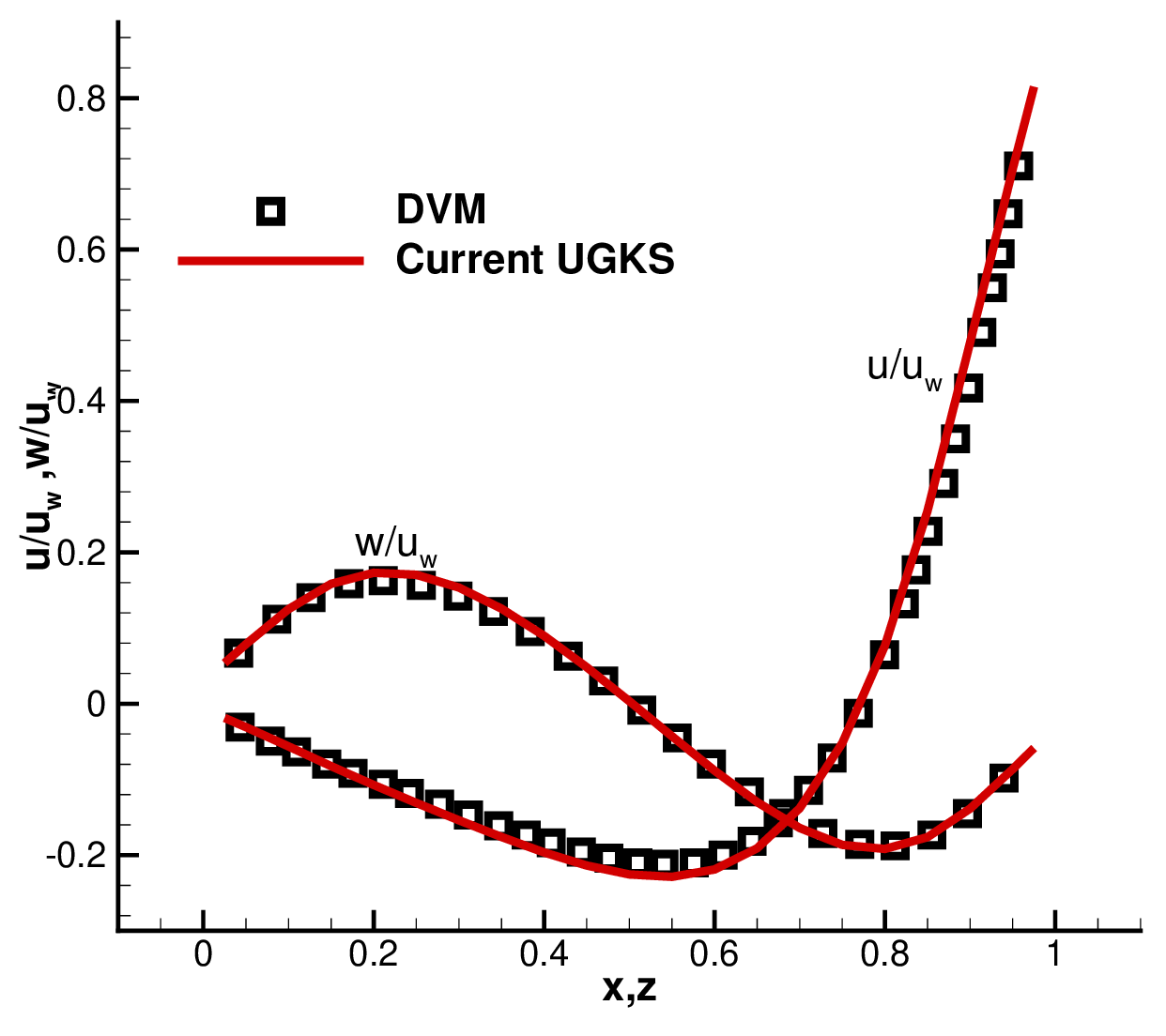}}
 \subfigure[Temperature contour]{\includegraphics[width=8cm]{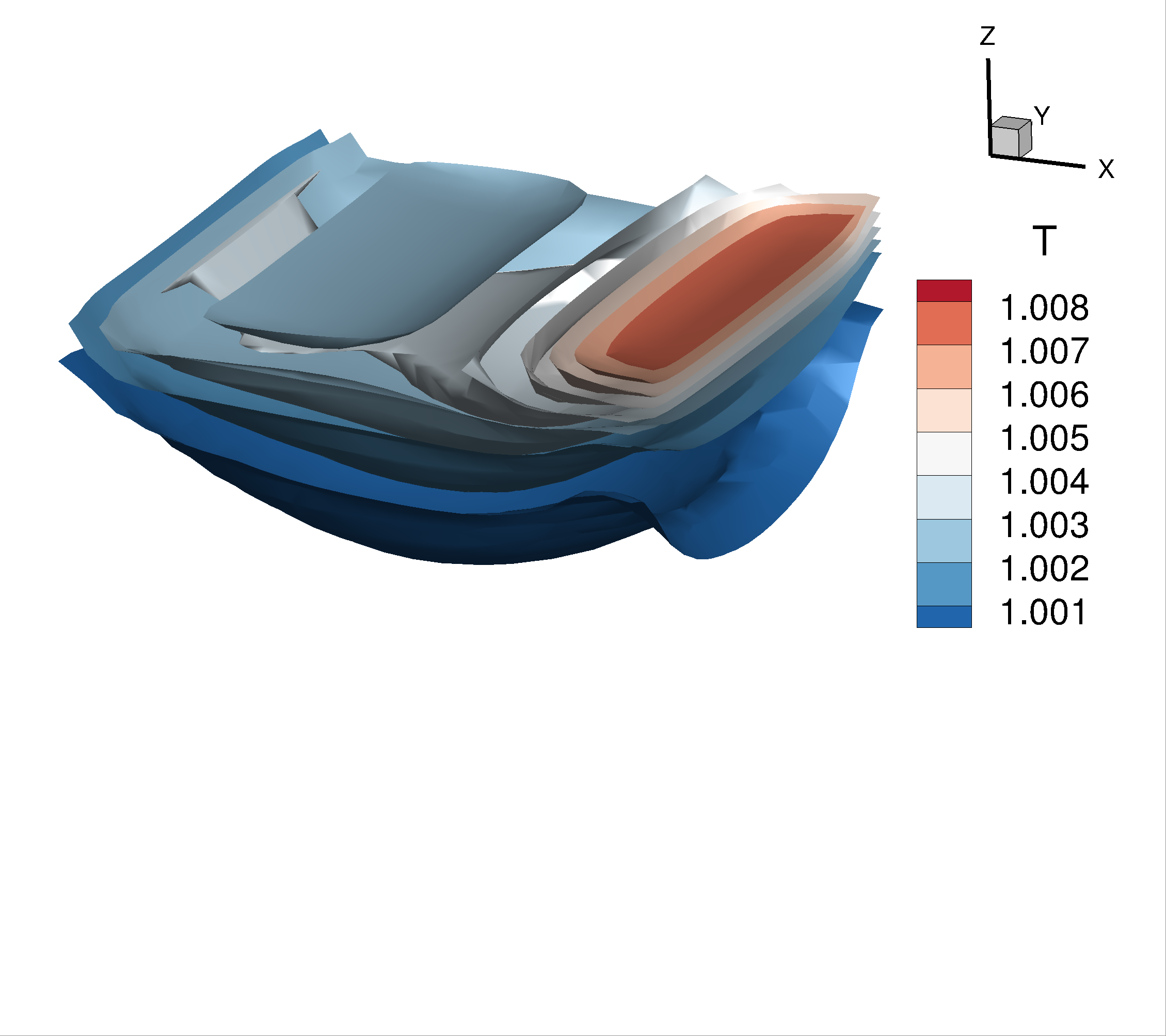}}
 \caption[]{3D lid-driven cavity flow at $\text{Kn} = 0.01$ by the current UGKS program.}\label{cavitykn0.01}
\end{figure}

The computational results are presented in Figures \ref{cavitykn1}, \ref{cavitykn0.1}, and \ref{cavitykn0.01}. These figures include the velocity profiles along the central lines, compared with reference data obtained by the discrete velocity method(DVM) \cite{yang2019improved} and the temperature contours. These results indicate that our scheme can effectively address problems ranging from rarefied flow to near-continuous flow.


\begin{figure}[!h]
 \centering
 \subfigure[Computational times and parallel efficiency]{\includegraphics[width=8cm]{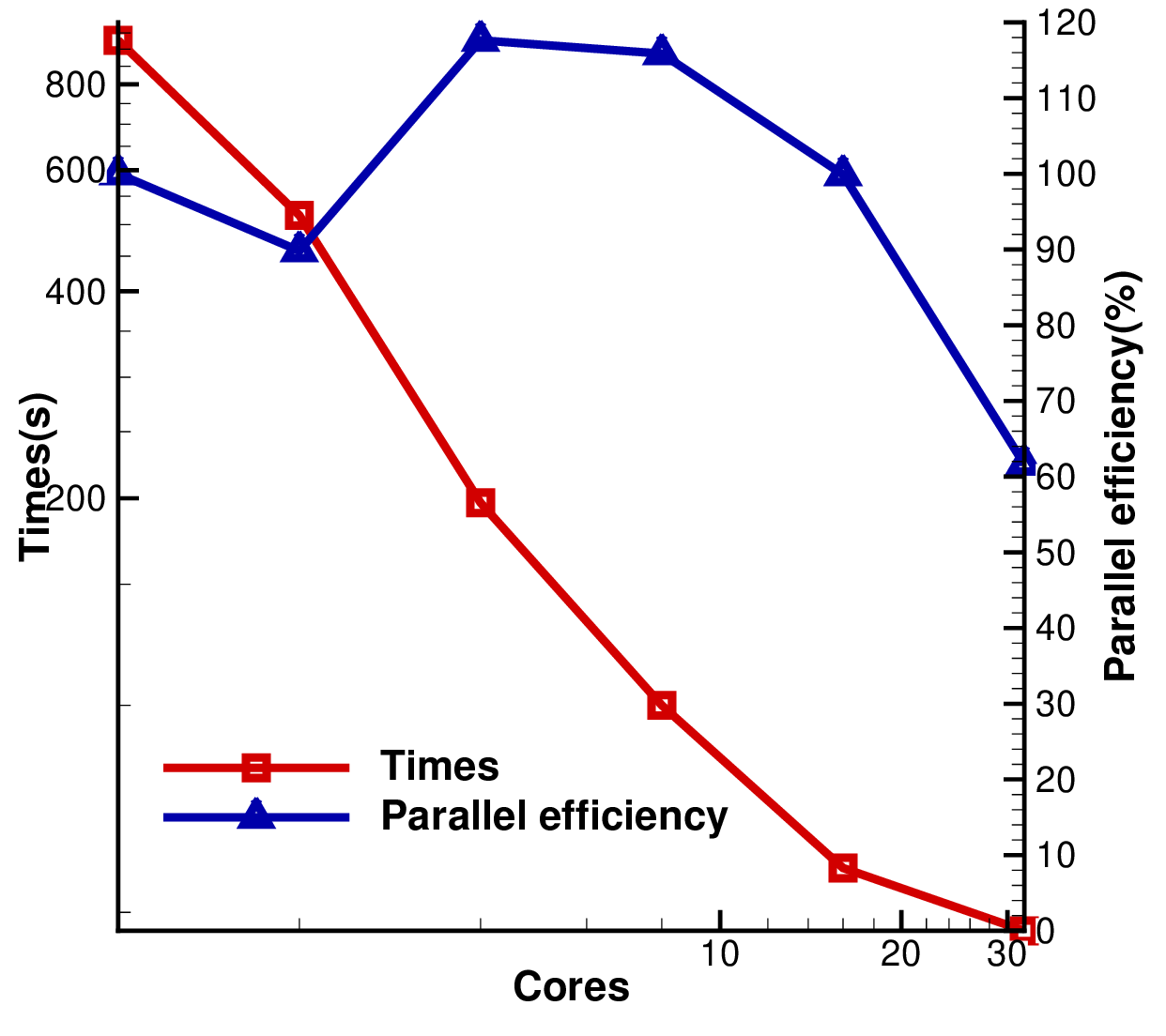}\label{cavityTime}}
 \subfigure[Cache and L1 misses ratio]{\includegraphics[width=8cm]{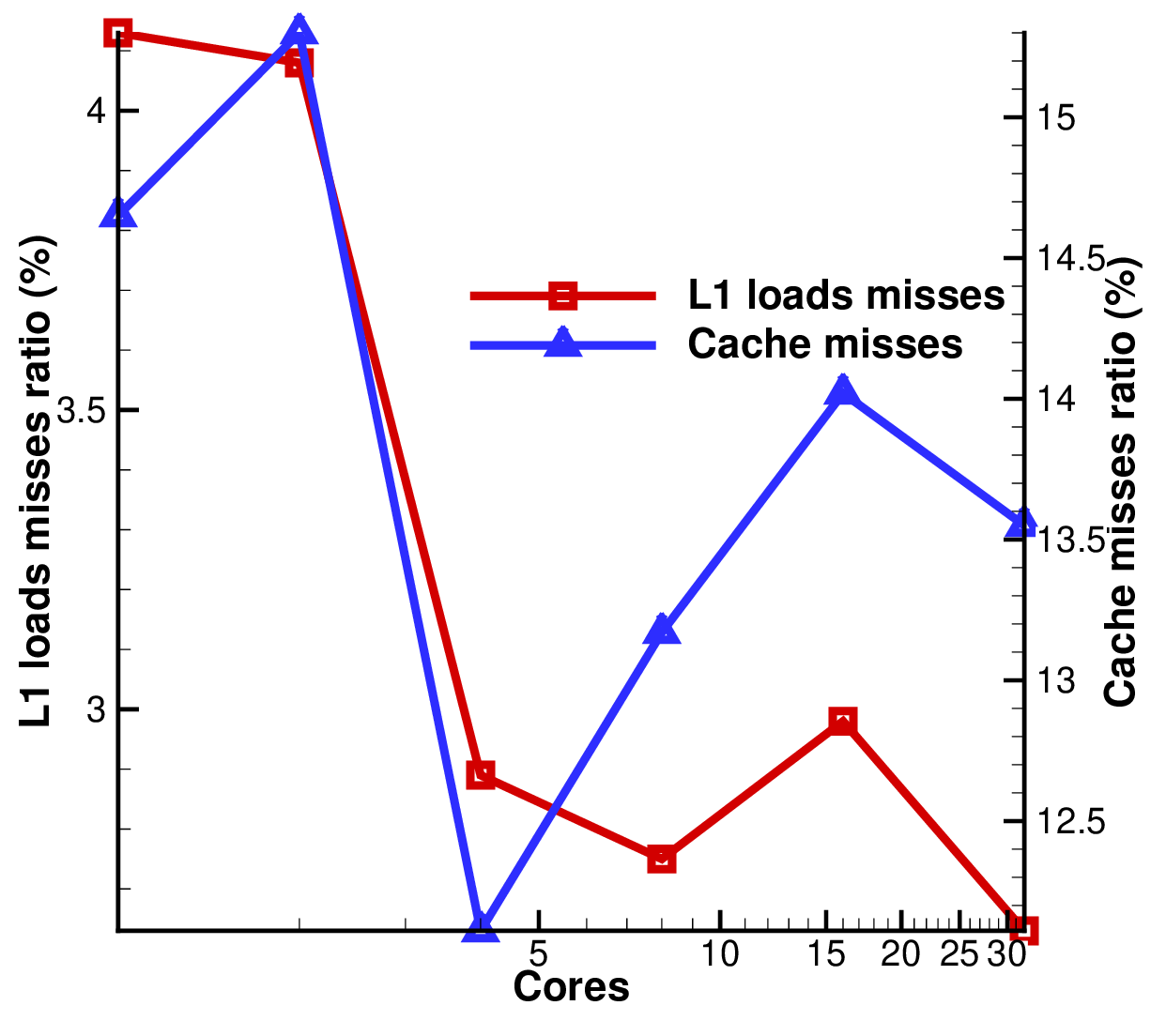}\label{cavityCache}}
 \caption[]{Parallel computational performance of cavity by the current UGKS program.}
\end{figure}

We now discuss the parallel efficiency of our code when applied to small-scale problems. The testing platform consists of an AMD Ryzen Threadripper 2990WX CPU operating at 4.2 GHz, featuring 32 cores and 128 GB of memory. The code is compiled using GCC version 11.4.0 with the \texttt{-O2} optimization flag and is linked to Open MPI version 4.1.5. We conducted tests using varying cores, ranging from 1 to 32. The computational times for running ten steps and parallel efficiency are presented in Figure \ref{cavityTime}. The results indicate that parallel efficiency approaches 100\% when utilizing fewer than 16 cores. Additionally, as illustrated in Figure \ref{cavityCache}, both the cache miss ratio and L1 miss ratio are lower when employing 4 and 8 cores, which explains the observed parallel efficiency exceeding 100\%.

Next, we discuss memory usage. The memory consumption is approximately 5.3 GB for computations using a single core. We consider a configuration with 8,000 cells and ($6 \times 20 \times 20 = 2,400$) boundary faces to analyze theoretical memory usage. Each cell at one velocity space point requires two doubles to store the reduced distribution function. We need to store both the distribution functions and their gradients for wall surfaces, necessitating eight doubles for each face at one velocity space point.
In total, the memory required to store all distribution functions is given by
$ (8,000 \times 2 + 2,400 \times 8) \times 17,182 \text{ doubles} $,
which amounts to approximately 4.5 GB. In this case, only 0.8 GB (less than 20\% of the theoretical memory) is needed to store other data.

\subsection{Hypersonic flow around an X38-like space vehicle}
This case is used to study our program's capability to handle a larger number of grids. Hypersonic flow at $\text{Ma}_\infty=8.0$ passing over an X38-like vehicle for $\text{Kn}_\infty =0.00275$ at angles of attack of $\text{AoA}=20^{\circ}$ is simulated. The reference length to define the Knudsen number is $L_{ref}= 0.28 $ m. In this case, all flow regimes are involved due to the hypersonic free stream flow in the transition regime and the complex geometric shape, posing a great challenge to the numerical solver. The free stream temperature is $T_\infty=56$ K, and an isothermal wall is applied to the vehicle surface with $T_w=302$ K.

\begin{figure}[!h]
 \centering
 \subfigure[Half of the domain]{\includegraphics[width=8cm]{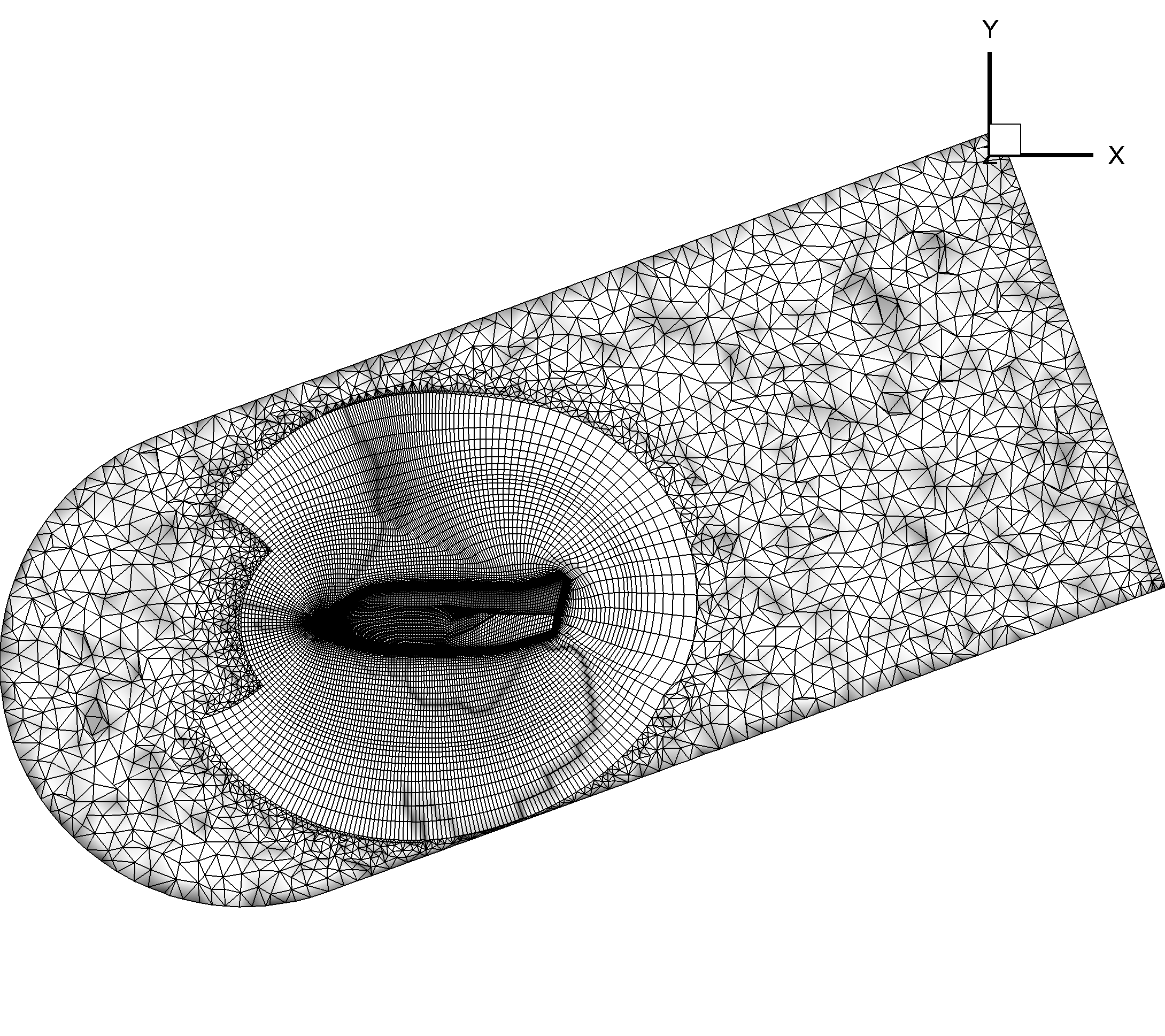}}
 \subfigure[Close view of the vehicle surface]{\includegraphics[width=8cm]{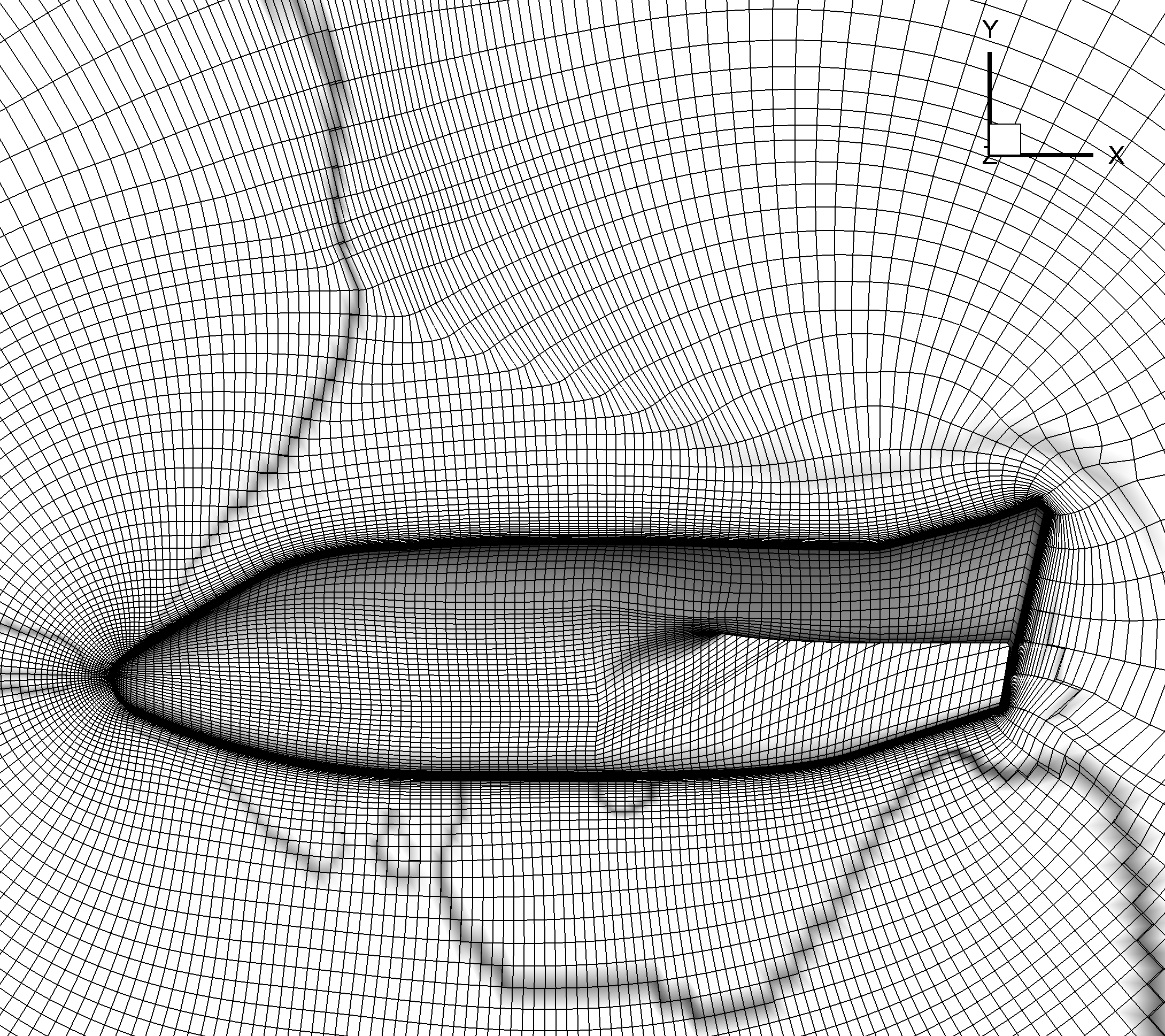}}
 \caption[]{Section view of the physical mesh of X38-like space vehicle with 600,078 cells by the current UGKS program.}\label{x38phyMesh}
\end{figure}

\begin{figure}[!h]
 \centering
 \includegraphics[width=8cm]{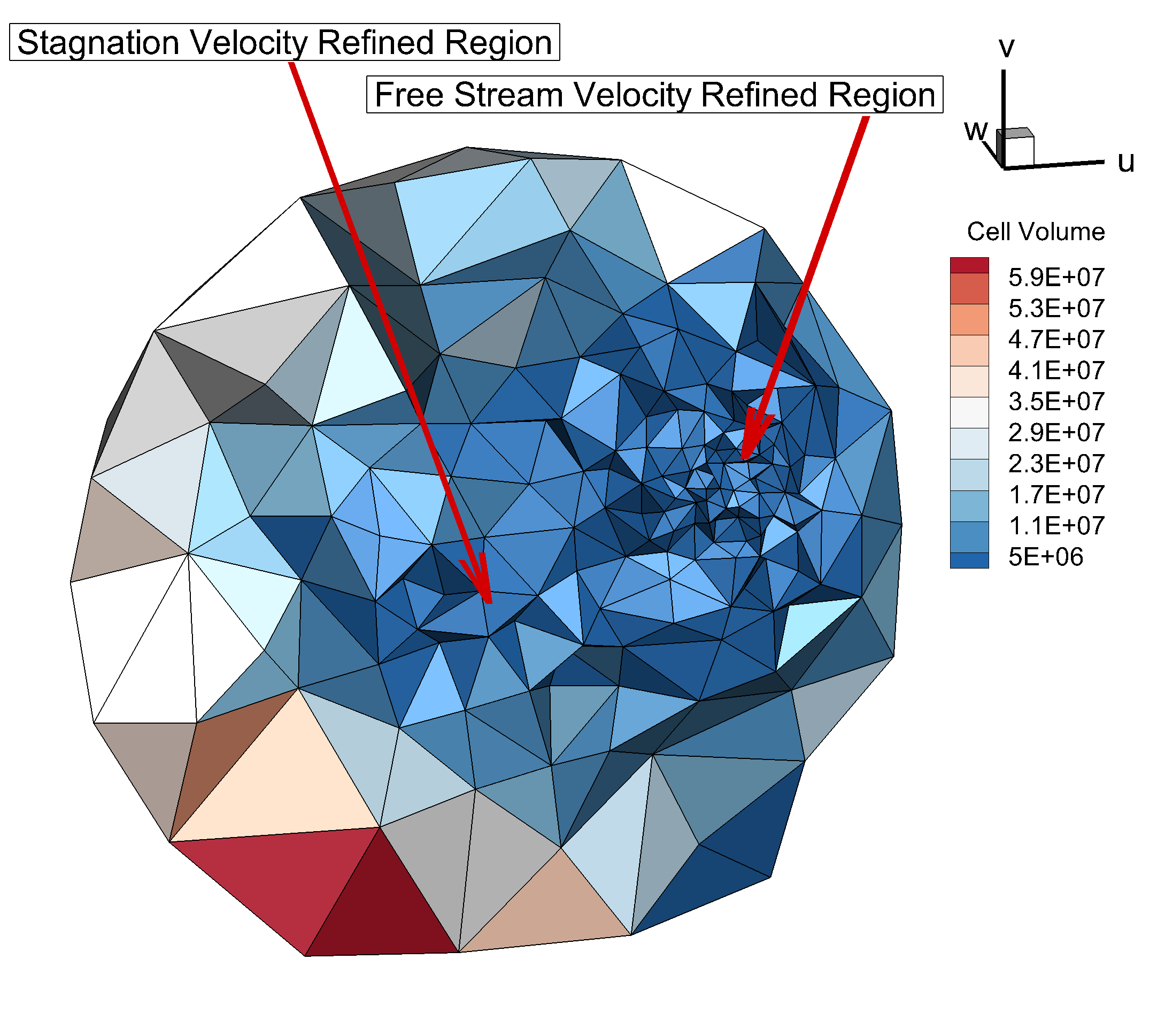}
 \caption[]{Unstructured DVS mesh with 4,548 cells used for hypersonic flow at $\text{Mach}_\infty=8.0$ and $\text{Kn}_\infty=0.00275$ passing over an X38-like space vehicle by the current UGKS program.}\label{x38DVSmesh}
\end{figure}

The physical mesh consists of 1,058,685 cells with the cell height of the first layer mesh $h=1.5\times10^{-4} $ m, as shown in Figure \ref{x38phyMesh}.
The unstructured DVS mesh is depicted in Figure \ref{x38DVSmesh}.
The DVS is discretized into 4,548 cells in a sphere mesh with a radius of $4\sqrt{RT_s}$. The velocity space near the zero and free stream velocity points are refined within a spherical region of radius $r=\sqrt{RT_w}$ and $r=\sqrt{RT_\infty}$ respectively.

\begin{figure}[!h]
 \centering
 \subfigure[Mach number]{\includegraphics[width=8cm]{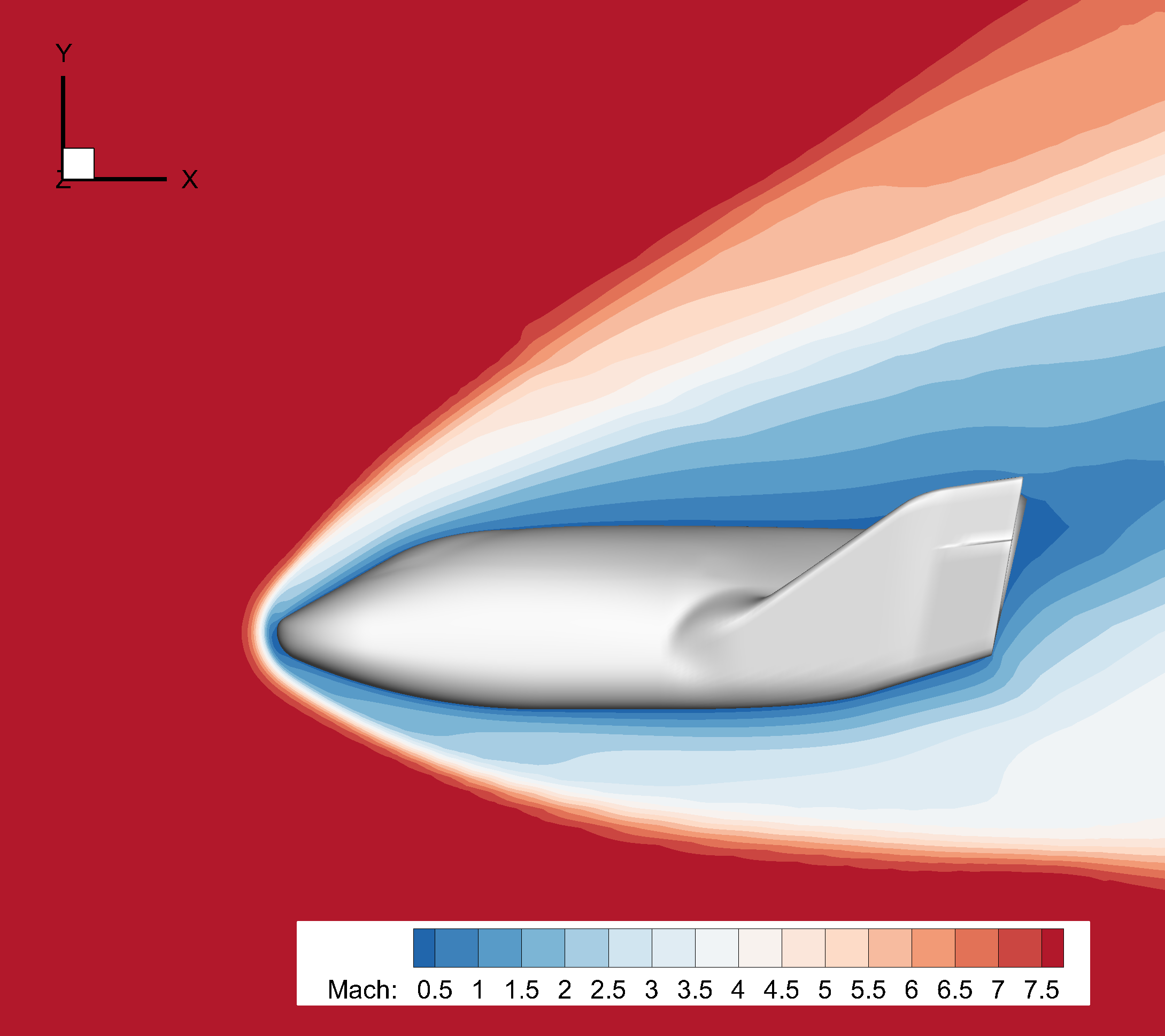}}
 \subfigure[Temperature]{\includegraphics[width=8cm]{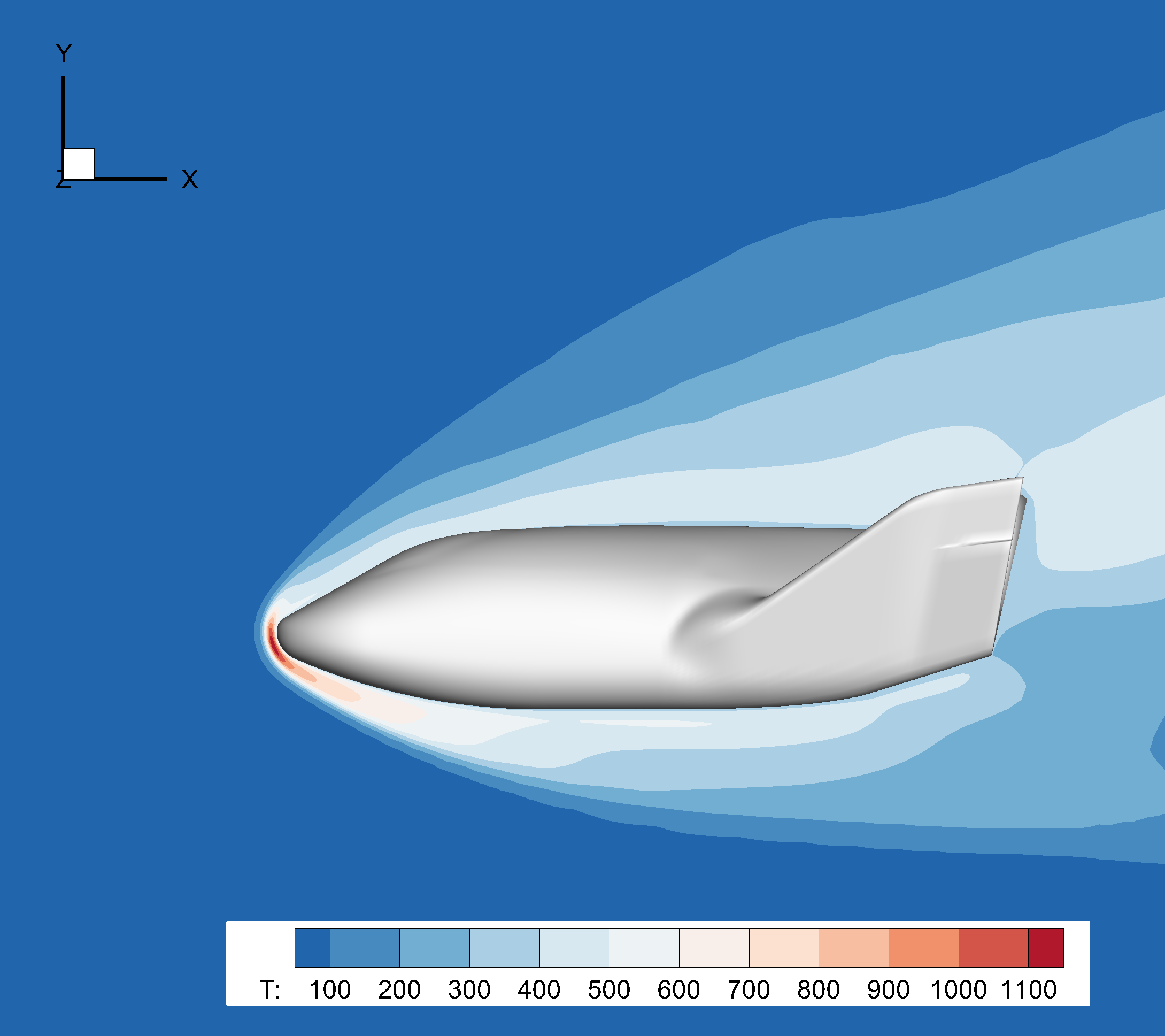}}
 \caption[]{The contour of hypersonic flow around an X38-like space vehicle at $\text{Mach}_\infty=8.0$ and $\text{Kn}_\infty=0.00275$ by the current UGKS program.}\label{x38contour}
\end{figure}

\begin{figure}[!h]
 \centering
 \subfigure[Pressure coefficient]{\includegraphics[width=5cm]{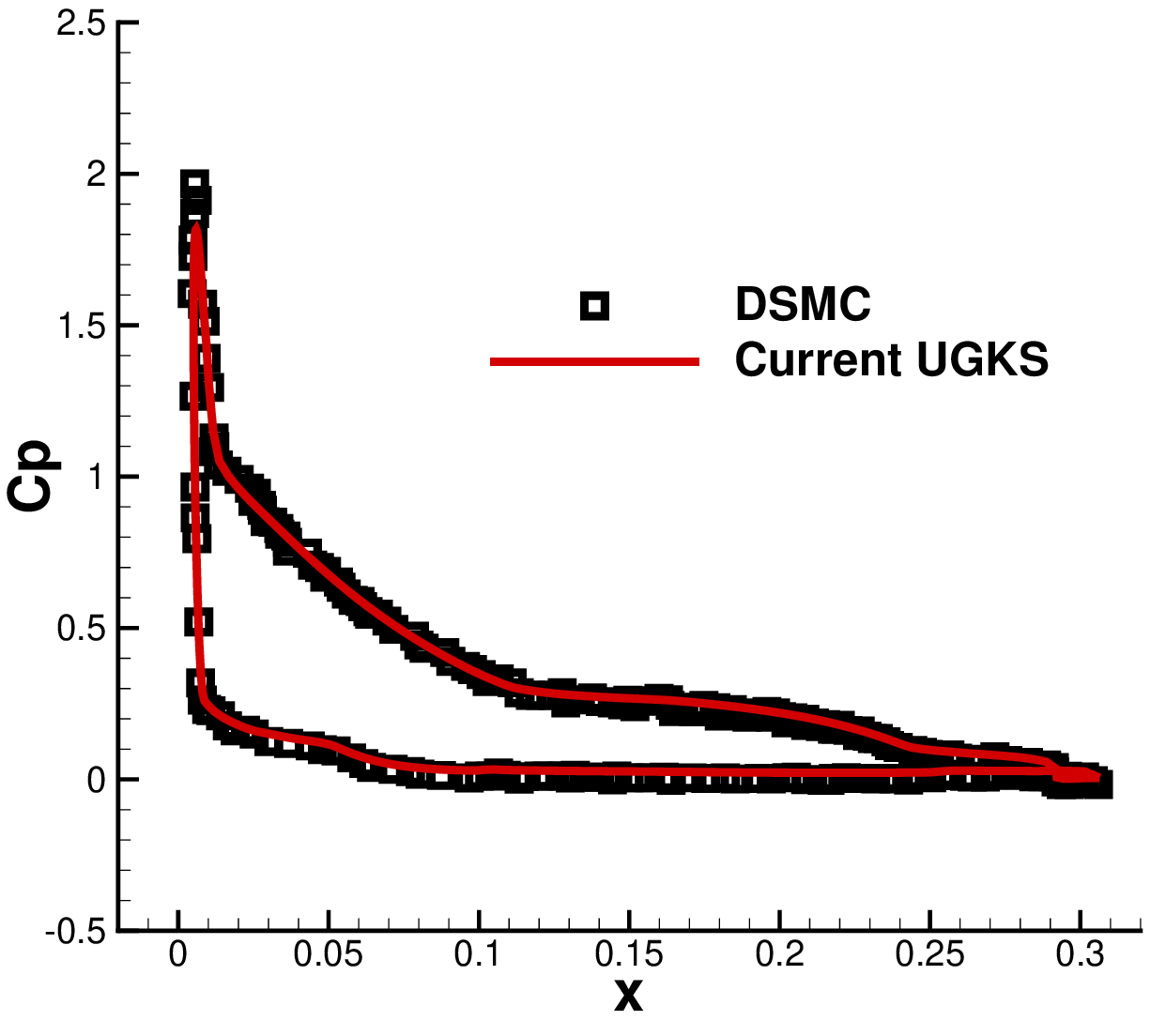}}
 \subfigure[Shear stress coefficient]{\includegraphics[width=5cm]{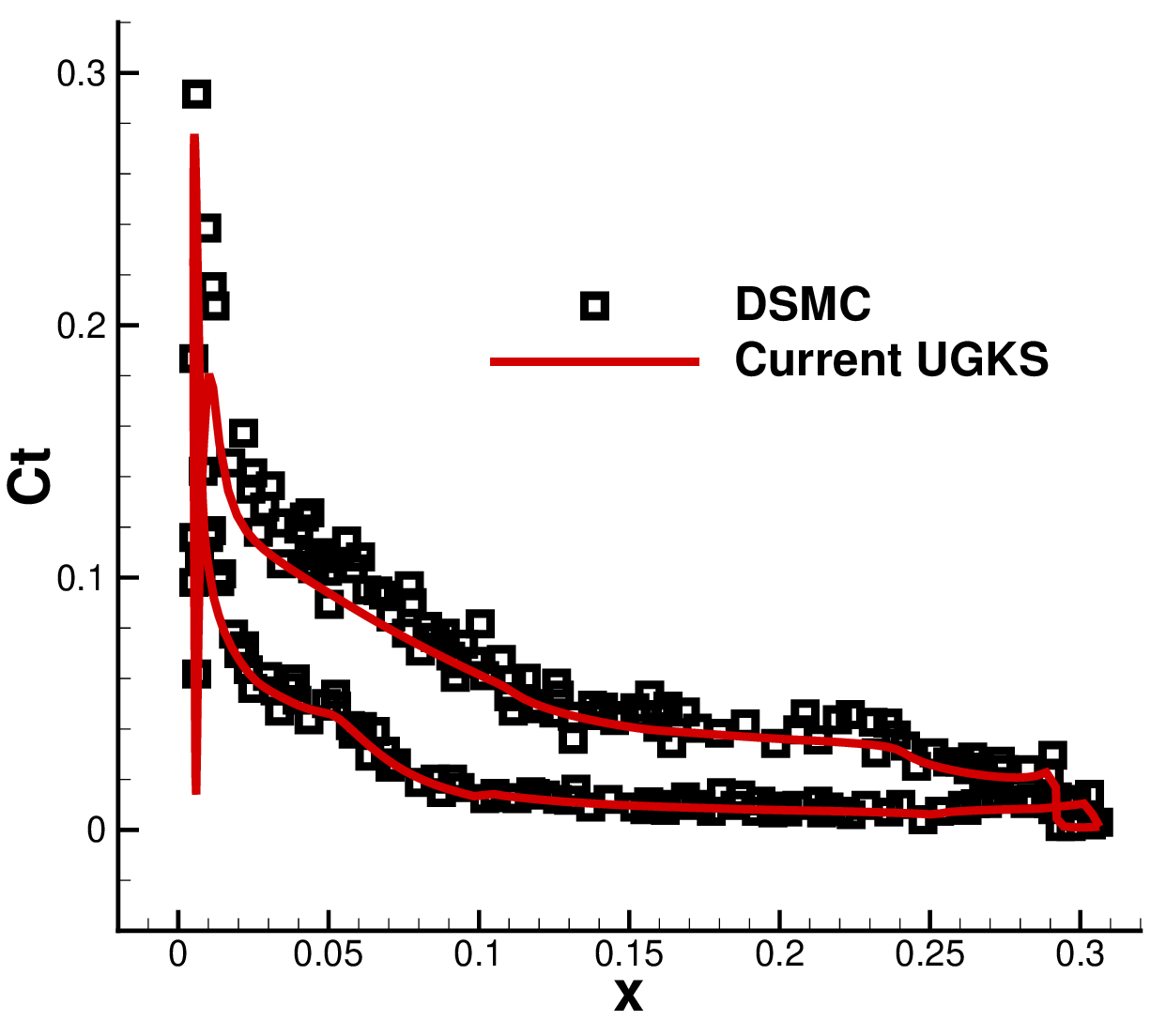}}
 \subfigure[Heat flux coefficient]{\includegraphics[width=5cm]{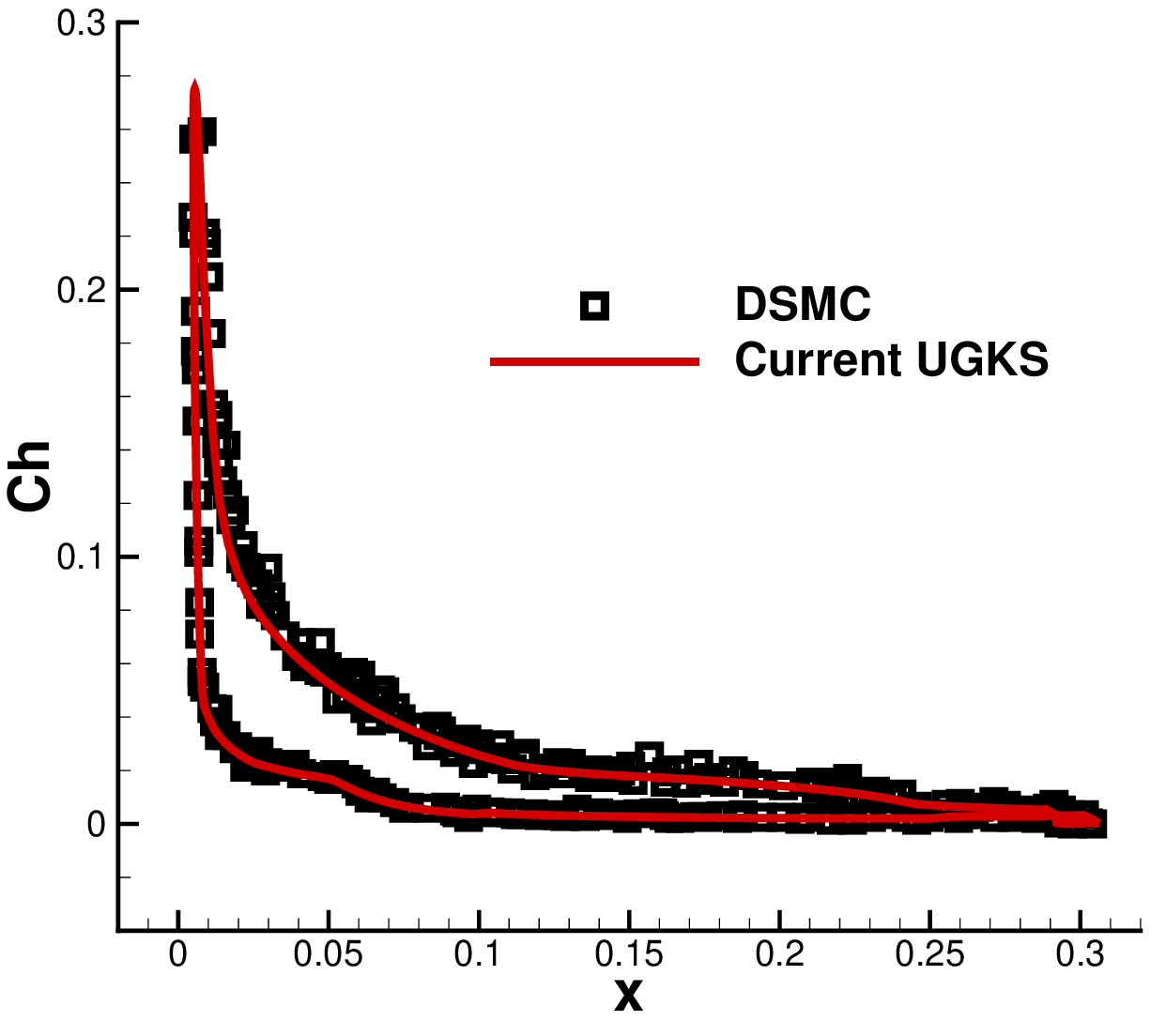}}
 \caption[]{Surface quantities  of hypersonic flow around an X38-like space vehicle at $\text{Mach}_\infty=8.0$ and $\text{Kn}_\infty=0.00275$ by the current UGKS program for argon gas compared with the DSMC method \cite{li2021kinetic}.}\label{x38surface}
\end{figure}

The contour of Mach number and temperature are shown in Figure \ref{x38contour}.
Figure \ref{x38surface} depicts the surface quantities on the symmetric cross-section perpendicular to the $y$-axis and comparisons with the DSMC data. All the coefficients predicted by our code align well with the DSMC reference \cite{li2021kinetic}.

We now discuss the parallel efficiency of our code on large-scale problems. The simulation is conducted on the SUGON computation platform using a CPU 7285 (32 cores, 2.5GHz). The code is compiled using GCC version 7.3.1 with the \texttt{-O2} optimization flag and is linked to Open MPI version 4.1.5.
To provide a more intuitive demonstration of the program's parallel efficiency, we first conducted tests using the gas-kinetic scheme (GKS)  \cite{xuGKS2001}, which has a smaller memory footprint and faster computation speed. Table \ref{GKSeff} shows the computation time and parallel efficiency for running 5,000 steps from 1 core to 512 cores. The parallel computing efficiency can be maintained above 70\% when using fewer than 128 cores.
When running the UGKS program, the computation time and parallel efficiency are shown in Table \ref{UGKSeff}. The parallel efficiency is consistent with that of the GKS method, indicating that our program does not suffer from reduced parallel efficiency due to the multiple communications required at each step of the UGKS. Furthermore, for the computation with 64 cores, our new framework requires only 112.55 GB of memory, and for the computation with 512 cores, it requires 168.1227 GB of memory, making memory consumption no longer a bottleneck for the UGKS. The increase in memory is primarily due to the need for ghost cells for communication at parallel interfaces, necessitating extra memory to store the physical quantities of the ghost cells. However, this increase in memory is not significant relative to the rise in the number of cores, making it acceptable in computations.
Compared to previous works, the GSIS  \cite{zhang2024efficient} utilizes 512 cores and requires 1.32 TB of memory, which has about $3.28\times10^{10}$ degrees of freedom with 961,080 hexahedral cells in the spatial discretization, 8,531 cells in the velocity space, and considering the vibrational freedoms. In contrast, the current UGKS program only consumes about $1/8$ of the memory with the same number of cores for $9.63\times10^{9}$ (about 29.4\%) degrees of freedom, indicating that we have greater memory usage efficiency.
  \begin{table}[htb!]
 \centering
 \caption{Wall clock time and parallel computing efficiency of the GKS solver with 5,000 iterations in hypersonic flow around an X38-like space vehicle}\label{GKSeff}
  \begin{tabular}{cccc}
    \hline
 Cores & Wall time (s)& Actual speedup & Parallel efficiency\\
    \hline
1   &10,091& 1.00   & 100.00\% \\
2   & 5,080& 1.99   & 99.32\% \\
4   & 2,632& 3.83   & 95.85\% \\
8   & 1,345& 7.50   & 93.78\% \\
16   & 701  &14.40   & 89.97\% \\
32  & 396 &25.48   & 79.63\%  \\
64   & 207  &48.75   & 76.17\% \\
128 & 112& 90.10 & 70.39\% \\
256 & 69 &146.25 &57.13\% \\
512 & 49 &205.94 &40.22\%\\
\hline
\end{tabular}
\end{table}

  \begin{table}[htb!]
 \centering
 \caption{Wall clock time and parallel computing efficiency of the UGKS solver with 20 iterations in hypersonic flow around an X38-like space vehicle by the current UGKS program}\label{UGKSeff}
  \begin{tabular}{cccc}
    \hline
 Cores & Wall time (s)& Actual speedup & Parallel efficiency \\
  \hline
64 &2,257 &1.00 &100.00\%\\
128 &1,252& 1.80& 90.14\% \\
256 &698  &3.23 &80.84\% \\
512 &463  &4.87 &60.93\% \\
\hline
\end{tabular}
\end{table}

\section{Conclusion}
In this paper, a parallel UGKS programming paradigm based on MPI is presented. Initially, the new strategy avoids storing the entire velocity space's slopes and residuals by implementing a two-step update method for the source term. Additionally, by leveraging the independent characteristics of the velocity space distribution functions, the reconstruction and update with a one-point offset are conducted, which reduce the number of MPI communications effectively through the utilization of  non-blocking MPI communication.

The accuracy of the program is validated by testing cases, such as a three-dimensional lid-driven cavity flow, and hypersonic flows over a cylinder, a sphere, and an X38-like space vehicle. The program's parallel efficiency is evaluated for both small-scale computations (three-dimensional lid-driven cavity flow) on a personal workstation and for large-scale computations (hypersonic flow around an X38-like space vehicle) on a supercomputing platform.
The results indicate that the new programming paradigm achieves low memory consumption and high parallel efficiency.
Notably, compared to previous methods like the GSIS \cite{zhang2024efficient}, the current memory consumption is significantly reduced.

The computational efficiency of this program can be further improved by implementing implicit algorithms and adaptive velocity space strategies.
Adopted with the current programming paradigm, efficient tools for simulating non-equilibrium flow can be developed in many
other deterministic frameworks, such as DUGKS and GSIS.

\section*{Acknowledgements}

We would like to thank Mr Cao Junzhe for discussion of the algorithm and the program design.
This work was supported by the National Key R\&D Program of China (Grant No. 2022YFA1004500) and the National Natural Science Foundation of China (Nos. 12172316 and 92371107), and the Hong Kong Research Grant Council (Nos. 16208021, 16301222, and 16208324).



\bibliographystyle{elsarticle-num}
\bibliography{mybibfile}

\begin{thebibliography}{10}
\expandafter\ifx\csname url\endcsname\relax
  \def\url#1{\texttt{#1}}\fi
\expandafter\ifx\csname urlprefix\endcsname\relax\def\urlprefix{URL }\fi
\expandafter\ifx\csname href\endcsname\relax
  \def\href#1#2{#2} \def\path#1{#1}\fi

\bibitem{bird1994molecular}
G.~A. Bird, Molecular gas dynamics and the direct simulation of gas flows,
  Oxford university press, 1994.

\bibitem{xu2021unified}
K.~Xu, A unified computational fluid dynamics framework from rarefied to
  continuum regimes, Cambridge University Press, 2021.

\bibitem{senturia1997simulating}
S.~D. Senturia, N.~Azuru, J.~White, Simulating the behavior of {MEMS} devices:
  computational methods and needs, IEEE Computational Science and engineering
  4~(1) (1997) 30--43.

\bibitem{wang2022investigation}
Y.~Wang, S.~Liu, C.~Zhuo, C.~Zhong, Investigation of nonlinear squeeze-film
  damping involving rarefied gas effect in micro-electro-mechanical systems,
  Computers \& Mathematics with Applications 114 (2022) 188--209.

\bibitem{alexeenko2003numerical}
A.~Alexeenko, D.~Levin, S.~Gimelshein, B.~Reed, Numerical investigation of
  physical processes in high-temperature {MEMS}-based nozzle flows, in: AIP
  conference proceedings, Vol. 663, American Institute of Physics, 2003, pp.
  760--767.

\bibitem{wang2004simulations}
M.~Wang, Z.~Li, Simulations for gas flows in microgeometries using the direct
  simulation {Monte} {Carlo} method, International Journal of Heat and Fluid
  Flow 25~(6) (2004) 975--985.

\bibitem{bird1963approach}
G.~Bird, Approach to translational equilibrium in a rigid sphere gas, Phys.
  fluids 6 (1963) 1518--1519.

\bibitem{bird1998recent}
G.~Bird, Recent advances and current challenges for {DSMC}, Computers \&
  Mathematics with Applications 35~(1-2) (1998) 1--14.

\bibitem{fan2001statistical}
J.~Fan, C.~Shen, Statistical simulation of low-speed rarefied gas flows,
  Journal of Computational Physics 167~(2) (2001) 393--412.

\bibitem{chu1965kinetic}
C.~Chu, Kinetic-theoretic description of the formation of a shock wave, The
  Physics of Fluids 8~(1) (1965) 12--22.

\bibitem{yang1995rarefied}
J.~Yang, J.~Huang, Rarefied flow computations using nonlinear model {Boltzmann}
  equations, Journal of Computational Physics 120~(2) (1995) 323--339.

\bibitem{mieussens2000discrete}
L.~Mieussens, Discrete-velocity models and numerical schemes for the
  {Boltzmann}-{BGK} equation in plane and axisymmetric geometries, Journal of
  Computational Physics 162~(2) (2000) 429--466.

\bibitem{tcheremissine2005direct}
F.~Tcheremissine, Direct numerical solution of the {Boltzmann} equation, in:
  AIP Conference Proceedings, Vol. 762, American Institute of Physics, 2005,
  pp. 677--685.

\bibitem{xu2010unified}
K.~Xu, J.-C. Huang, A unified gas-kinetic scheme for continuum and rarefied
  flows, Journal of Computational Physics 229~(20) (2010) 7747--7764.

\bibitem{juan-chen_huang_unified_2012}
{Juan-Chen Huang}, K.~Xu, P.~Yu, A unified gas-kinetic scheme for continuum and
  rarefied flows {II}: Multi-dimensional cases, Communications in Computational
  Physics 12~(3) (2012) 662--690.

\bibitem{liu2020ugkwp}
C.~Liu, Y.~Zhu, K.~Xu, Unified gas-kinetic wave-particle methods {I}: Continuum
  and rarefied gas flow, Journal of Computational Physics 401 (2020) 108977.

\bibitem{zhu2019ugkwp}
Y.~Zhu, C.~Liu, C.~Zhong, K.~Xu, Unified gas-kinetic wave-particle methods.
  {II}. multiscale simulation on unstructured mesh, Physics of Fluids 31~(6).

\bibitem{guo2013dugks}
Z.~Guo, K.~Xu, R.~Wang, Discrete unified gas kinetic scheme for all {Knudsen}
  number flows: Low-speed isothermal case, Physical Review E-Statistical,
  Nonlinear, and Soft Matter Physics 88~(3) (2013) 033305.

\bibitem{yang2023dugkwp}
L.~Yang, Z.~Li, C.~Shu, Y.~Liu, W.~Liu, J.~Wu, Discrete unified gas-kinetic
  wave-particle method for flows in all flow regimes, Physical Review E 108~(1)
  (2023) 015302.

\bibitem{xu_improved_2011}
K.~Xu, J.-C. Huang, An improved unified gas-kinetic scheme and the study of
  shock structures, IMA Journal of Applied Mathematics 76~(5) (2011) 698--711.

\bibitem{sha_liu_unified_2014}
{Sha Liu}, P.~Yu, K.~Xu, C.~Zhong, Unified gas-kinetic scheme for diatomic
  molecular simulations in all flow regimes, Journal of Computational Physics
  259 (2014) 96--113.

\bibitem{rui_zhang_unified_2023}
{Rui Zhang}, S.~Liu, C.~Zhong, C.~Zhuo, Unified gas-kinetic scheme with
  simplified multi-scale numerical flux for thermodynamic non-equilibrium flow
  in all flow regimes, Communications in Nonlinear Science and Numerical
  Simulation 119 (2023) 107079.

\bibitem{zhao_wang_unified_2017}
{Zhao Wang}, H.~Yan, Q.~Li, K.~Xu, Unified gas-kinetic scheme for diatomic
  molecular flow with translational, rotational, and vibrational modes, Journal
  of Computational Physics 350 (2017) 237--259.

\bibitem{huang_unified_2013}
J.-C. Huang, K.~Xu, P.~Yu, A unified gas-kinetic scheme for continuum and
  rarefied flows {III}: Microflow simulations, Communications in Computational
  Physics 14~(5) (2013) 1147--1173.

\bibitem{liu_unified_2020}
C.~Liu, K.~Xu, A unified gas-kinetic scheme for micro flow simulation based on
  linearized kinetic equation, Advances in Aerodynamics 2~(1) (2020) 21.

\bibitem{xu_parallel_2022}
L.~Xu, W.~Zhang, Y.~Chen, R.~Chen, A parallel discrete unified gas kinetic
  scheme on unstructured grid for inviscid high-speed compressible flow
  simulation, Physics of Fluids 34~(10) (2022) 106110.

\bibitem{xu_discrete_2023}
L.~Xu, Z.~Yan, R.~Chen, A discrete unified gas kinetic scheme on unstructured
  grids for viscid compressible flows and its parallel algorithm, AIMS
  Mathematics 8~(4) (2023) 8829--8846.

\bibitem{zhang_discrete_2019}
Y.~Zhang, L.~Zhu, P.~Wang, Z.~Guo, Discrete unified gas kinetic scheme for
  flows of binary gas mixture based on the {McCormack} model, Physics of Fluids
  31~(1) (2019) 017101.

\bibitem{wenjun_sun_unifed_2019}
W.~Sun, S.~Jiang, K.~Xu, G.~Cao, Multiscale simulation for the system of
  radiation hydrodynamics, Journal of Scientific Computing 85 (2020) 1--24.

\bibitem{jiang_song__2021}
{Jiang Song}, X.~Kun, S.~Wenjun, X.~Xiaojing, Unified gas kinetic schemes for
  the radiation transfer equations, SCIENTIA SINICA Mathematica 51~(6) (2021)
  799.

\bibitem{liu2021unified}
C.~Liu, K.~Xu, Unified gas-kinetic wave-particle methods iv: multi-species gas
  mixture and plasma transport, Advances in Aerodynamics 3 (2021) 1--31.

\bibitem{quan2024radiative}
M.~Quan, X.~Yang, Y.~Wei, K.~Xu, Radiative hydrodynamic equations with
  nonequilibrium radiative transfer, arXiv preprint arXiv:2409.01827.

\bibitem{tan_time_2020}
S.~Tan, Time implicit unified gas kinetic scheme for {3D} multi-group neutron
  transport simulation, Communications in Computational Physics 28~(3) (2020)
  1189--1218.

\bibitem{guo_discrete_2016}
Z.~Guo, K.~Xu, Discrete unified gas kinetic scheme for multiscale heat transfer
  based on the phonon {Boltzmann} transport equation, International Journal of
  Heat and Mass Transfer 102 (2016) 944--958.

\bibitem{zhang2024electron}
C.~Zhang, R.~Guo, M.~Lian, J.~Shiomi, Electron-phonon coupling and
  non-equilibrium thermal conduction in ultra-fast heating systems, Applied
  Thermal Engineering 249 (2024) 123379.

\bibitem{fei2020unified}
F.~Fei, J.~Zhang, J.~Li, Z.~Liu, A unified stochastic particle
  {Bhatnagar}--{Gross}--{Krook} method for multiscale gas flows, Journal of
  Computational Physics 400 (2020) 108972.

\bibitem{fei2021efficient}
F.~Fei, Y.~Ma, J.~Wu, J.~Zhang, An efficient algorithm of the unified
  stochastic particle {Bhatnagar}--{Gross}--{Krook} method for the simulation
  of multi-scale gas flows, Advances in Aerodynamics 3 (2021) 1--16.

\bibitem{guo2023unified}
Z.~Guo, J.~Li, K.~Xu, Unified preserving properties of kinetic schemes,
  Physical Review E 107~(2) (2023) 025301.

\bibitem{yuan2020conservative}
R.~Yuan, C.~Zhong, A conservative implicit scheme for steady state solutions of
  diatomic gas flow in all flow regimes, Computer Physics Communications 247
  (2020) 106972.

\bibitem{xiao2020velocity}
T.~Xiao, C.~Liu, K.~Xu, Q.~Cai, A velocity-space adaptive unified gas kinetic
  scheme for continuum and rarefied flows, Journal of Computational Physics 415
  (2020) 109535.

\bibitem{wei2024adaptive}
Y.~Wei, W.~Long, K.~Xu, Adaptive unified gas-kinetic scheme for diatomic gases
  with rotational and vibrational nonequilibrium, Computer Physics
  Communications 305 (2024) 109324.

\bibitem{yang2023adaptive}
L.~Yang, L.~Han, H.~Ding, Z.~Li, C.~Shu, Y.~Liu, Adaptive partitioning-based
  discrete unified gas kinetic scheme for flows in all flow regimes, Advances
  in Aerodynamics 5~(1) (2023) 15.

\bibitem{wei2023adaptive}
Y.~Wei, J.~Cao, X.~Ji, K.~Xu, Adaptive wave-particle decomposition in {UGKWP}
  method for high-speed flow simulations, Advances in Aerodynamics 5~(1) (2023)
  25.

\bibitem{zhu2016implicit}
Y.~Zhu, C.~Zhong, K.~Xu, Implicit unified gas-kinetic scheme for steady state
  solutions in all flow regimes, Journal of Computational Physics 315 (2016)
  16--38.

\bibitem{zhu2019implicit}
Y.~Zhu, C.~Zhong, K.~Xu, An implicit unified gas-kinetic scheme for unsteady
  flow in all {Knudsen} regimes, Journal of Computational Physics 386 (2019)
  190--217.

\bibitem{zhang2024conservative}
R.~Zhang, S.~Liu, J.~Chen, C.~Zhuo, C.~Zhong, A conservative implicit scheme
  for three-dimensional steady flows of diatomic gases in all flow regimes
  using unstructured meshes in the physical and velocity spaces, Physics of
  Fluids 36~(1).

\bibitem{long2024implicit}
W.~Long, Y.~Wei, K.~Xu, {An implicit adaptive unified gas-kinetic scheme for
  steady-state solutions of nonequilibrium flows}, Physics of Fluids 36~(10)
  (2024) 106114.

\bibitem{li_unified_2018}
S.~Li, Q.~Li, S.~Fu, K.~Xu, A unified gas-kinetic scheme for axisymmetric flow
  in all {Knudsen} number regimes, Journal of Computational Physics 366 (2018)
  144--169.

\bibitem{taitano2014moment}
W.~T. Taitano, D.~A. Knoll, L.~Chac{\'o}n, J.~M. Reisner, A.~K. Prinja,
  Moment-based acceleration for neutral gas kinetics with {BGK} collision
  operator, Journal of Computational and Theoretical Transport 43~(1-7) (2014)
  83--108.

\bibitem{chacon2017multiscale}
L.~Chacon, G.~Chen, D.~A. Knoll, C.~Newman, H.~Park, W.~Taitano, J.~A. Willert,
  G.~Womeldorff, Multiscale high-order/low-order {(HOLO)} algorithms and
  applications, Journal of Computational Physics 330 (2017) 21--45.

\bibitem{chen2017unified}
S.~Chen, C.~Zhang, L.~Zhu, Z.~Guo, A unified implicit scheme for kinetic model
  equations. part {I}. memory reduction technique, Science bulletin 62~(2)
  (2017) 119--129.

\bibitem{yang2018implicit}
L.~Yang, C.~Shu, W.~Yang, J.~Wu, An implicit scheme with memory reduction
  technique for steady state solutions of {DVBE} in all flow regimes, Physics
  of Fluids 30~(4).

\bibitem{mouhot2006fast}
C.~Mouhot, L.~Pareschi, Fast algorithms for computing the {Boltzmann} collision
  operator, Mathematics of computation 75~(256) (2006) 1833--1852.

\bibitem{wu2013deterministic}
L.~Wu, C.~White, T.~J. Scanlon, J.~M. Reese, Y.~Zhang, Deterministic numerical
  solutions of the {Boltzmann} equation using the fast spectral method, Journal
  of Computational Physics 250 (2013) 27--52.

\bibitem{chen2012unified}
S.~Chen, K.~Xu, C.~Lee, Q.~Cai, A unified gas kinetic scheme with moving mesh
  and velocity space adaptation, Journal of Computational Physics 231~(20)
  (2012) 6643--6664.

\bibitem{ho2019multi}
M.~T. Ho, L.~Zhu, L.~Wu, P.~Wang, Z.~Guo, Z.-H. Li, Y.~Zhang, A multi-level
  parallel solver for rarefied gas flows in porous media, Computer Physics
  Communications 234 (2019) 14--25.

\bibitem{jiang2019implicit}
D.~Jiang, M.~Mao, J.~Li, X.~Deng, An implicit parallel {UGKS} solver for flows
  covering various regimes, Advances in Aerodynamics 1 (2019) 1--24.

\bibitem{baranger2012locally}
C.~Baranger, J.~Claudel, N.~H{\'e}rouard, L.~Mieussens, Locally refined
  discrete velocity grids for deterministic rarefied flow simulations, in: AIP
  Conference Proceedings, Vol. 1501, American Institute of Physics, 2012, pp.
  389--396.

\bibitem{li2016high}
S.~Li, Q.~Li, S.~Fu, J.~Xu, The high performance parallel algorithm for unified
  gas-kinetic scheme, in: AIP Conference Proceedings, Vol. 1786, AIP
  Publishing, 2016.

\bibitem{shuang2019parallel}
T.~Shuang, S.~Wenjun, W.~Junxia, N.~Guoxi, A parallel unified gas kinetic
  scheme for three-dimensional multi-group neutron transport, Journal of
  Computational Physics 391 (2019) 37--58.

\bibitem{zhang2022unified}
Q.~Zhang, Y.~Wang, D.~Pan, J.~Chen, S.~Liu, C.~Zhuo, C.~Zhong, Unified x-space
  parallelization algorithm for conserved discrete unified gas kinetic scheme,
  Computer Physics Communications 278 (2022) 108410.

\bibitem{wang_parallel_2022}
P.~Wang, J.~Li, D.~Jiang, M.~Mao, Parallel implementation and verification of
  implicit unified gas kinetic scheme, in: 2022 6th High Performance Computing
  and Cluster Technologies Conference ({HPCCT}), ACM, Fuzhou China, 2022, pp.
  25--30.

\bibitem{zhang2024efficient}
Y.~Zhang, J.~Zeng, R.~Yuan, W.~Liu, Q.~Li, L.~Wu, Efficient parallel solver for
  rarefied gas flow using {GSIS}, Computers \& Fluids 281 (2024) 106374.

\bibitem{xu2015direct}
K.~Xu, Direct modeling for computational fluid dynamics, Acta Mechanica Sinica
  31 (2015) 303--318.

\bibitem{zhang2023slidingmesh}
Y.~Zhang, X.~Ji, K.~Xu, A high-order compact gas-kinetic scheme in a rotating
  coordinate frame and on sliding mesh, International Journal of Computational
  Fluid Dynamics (2023) 1--20.

\bibitem{yang2019improved}
L.~Yang, C.~Shu, W.~Yang, J.~Wu, An improved three-dimensional implicit
  discrete velocity method on unstructured meshes for all {Knudsen} number
  flows, Journal of Computational Physics 396 (2019) 738--760.

\bibitem{li2021kinetic}
J.~Li, D.~Jiang, X.~Geng, J.~Chen, Kinetic comparative study on aerodynamic
  characteristics of hypersonic reentry vehicle from near-continuous flow to
  free molecular flow, Advances in Aerodynamics 3 (2021) 1--10.

\bibitem{xuGKS2001}
K.~Xu, A gas-kinetic {BGK} scheme for the {Navier}--{Stokes} equations and its
  connection with artificial dissipation and {Godunov} method, Journal of
  Computational Physics 171~(1) (2001) 289--335.

\end{thebibliography}

\end{document}